\documentclass[amsmath,eqsecnum,prb] {revtex4-1}
\usepackage{bm}
\usepackage{enumerate}
\usepackage{graphicx}
\usepackage{verbatim}
\usepackage[dvips]{epsfig}
\usepackage{epsf}
\usepackage{hyperref}
\hypersetup{colorlinks=true,linkcolor=blue,anchorcolor=blue,citecolor=blue,filecolor=blue,urlcolor=blue,bookmarksnumbered=true,pdfview=FitB}
\usepackage[utf8]{inputenc}
\usepackage[english]{babel}
\usepackage{fancyhdr}
\makeatletter
\newcommand{\beq}{\begin{equation}}
\newcommand{\eeq}{\end{equation}}
\newcommand{\bea}{\begin{eqnarray}}
\newcommand{\eea}{\end{eqnarray}}
\newcommand{\e}{\varepsilon}

\newcommand{\bk}{{\bf k}}
\newcommand{\bp}{{\bf p}}
\newcommand{\bq}{{\bf q}}

\newcommand{\nn}{\nonumber}
\newcommand{\bse}{\begin{subequations}}
\newcommand{\ese}{\end{subequations}}
\newcommand{\bwt}{\begin{widetext}}
\newcommand{\ewt}{\end{widetext}}

\newcommand{\bkp}{{\bf k}'}

\newcommand{\bl}{{\bf l}}

\newcommand{\kp}{k_{||}}
\newcommand{\pp}{p_{||}}

\newcommand{\bv}{{\bf v}}

\newcommand{\I}{\mathrm{Im}}
\newcommand{\R}{\mathrm{Re}}
\newcommand{\bsu}{\begin{subequations}}
\newcommand{\esu}{\end{subequations}}

\newcommand{\lr}{\left(}
\newcommand{\rr}{\right)}
\newcommand{\ls}{\left[}
\newcommand{\rs}{\right]}

\newcommand{\bg}{{\bar g}}

\newcommand{\te}{\tau_{\mathrm{ee}}}
\newcommand{\ti}{\tau_{\mathrm{i}}}
\makeatletter
\begin{document}

\title{Optical response of correlated electron systems}
\author{Dmitrii L. Maslov$^a$ and Andrey V. Chubukov$^b$}
\affiliation{$^a$Department of Physics, University of Florida, P.O. Box 118440, Gainesville, Florida 32611-8440, USA\\
$^b$Department of Physics
 and
 William I. Fine Theoretical Physics Institute, University of Minnesota,\\ Minneapolis, Minnesota 55455, USA}
\date{\today}
\begin{abstract}
Recent progress in experimental techniques has made it possible to extract detailed information on dynamics of carriers in a correlated electron material from its optical conductivity, $\sigma(\Omega, T)$. This review consists of three parts, addressing the following three aspects of optical response: 1) the role of momentum relaxation; 2) $\Omega/T$ scaling of the optical conductivity of a Fermi-liquid metal, and 3) the optical conductivity of non-Fermi-liquid metals. In the first part (Sec. II), we analyze the interplay between the contributions to the conductivity from normal and umklapp electron-electron scattering. As a concrete example, we consider a two-band metal and show that although its optical conductivity is finite it does not obey the Drude formula. In the second part (Secs. III and IV), we re-visit the Gurzhi formula for the optical scattering rate, $1/\tau(\Omega,T)\propto\Omega^2+4\pi^2 T^2$, and show that a factor of $4\pi^2$  is the manifestation of the ``first-Matsubara-frequency rule'' for boson response, which states that $1/\tau(\Omega,T)$ must vanish upon analytic continuation to the first boson Matsubara frequency. However, recent experiments show that the coefficient $b$ in the Gurzhi-like form, $1/\tau(\Omega,T)\propto\Omega^2+b\pi^2 T^2$, differs significantly from $b=4$ in most of the cases. We suggest that the deviations from Gurzhi scaling may be due to the presence of elastic
but energy-dependent 
scattering, which decreases the value of $b$ below $4$, with $b=1$ corresponding to purely elastic scattering. In the third part (Sec. V), we consider the optical conductivity of metals near quantum phase transitions to nematic and spin-density-wave
states. In the last case, we focus on ``composite'' scattering processes, which give rise to a non-Fermi--liquid behavior of the optical conductivity
at $T=0$
: $\sigma'(\Omega)\propto \Omega^{-1/3}$ at low frequencies  and $\sigma'(\Omega)\propto \Omega^{-1}$ at higher frequencies. We also discuss $\Omega/T$ scaling of the conductivity and show that $\sigma'(\Omega,T)$ in the same model scales in a non-Fermi-liquid way, as $T^{4/3}\Omega^{-5/3}$.
\end{abstract}
\maketitle
\tableofcontents
\section{Introduction}
\label{sec:intro}
Optical spectroscopy of strongly correlated materials plays a crucial role in advancing our understanding of these complex systems.~\cite{basov:2005,basov:2011}
 Recent progress in experimental techniques
 has made it possible
 to
 extract detailed information on
    two very important quantities: the dynamical effective mass and transport scattering rate
    of conduction electrons. The frequency and temperature dependences of these two quantities give
     one
      invaluable insights into the nature of the correlated electron state.
      A wide range of frequencies employed in optical spectroscopy allows one to extract the information both on the band structure, via studying
      inter-band transitions,
      and on dynamics of charge carriers within a given band (or several occupied bands).
        In this mini-review, we focus on the latter and discuss
     three
     particular aspects of
     intra-band
     optical response, which makes it distinct from other probes, such as angle-resolved photoemission
     (ARPES) and {\em dc} transport.

    The first
   aspect
     is the role of momentum relaxation.
      Here,
      the optical probe occupies
      a special niche
       in
       between
     ARPES
      and {\em dc} transport. The width of the ARPES
       peak is given by the inverse quasiparticle relaxation time,
    which contains contributions from all kinds of scattering:
    elastic and inelastic,
      small- and large angle,
      momentum-conserving
       and
 momentum-relaxing,
 etc.
  Consequently, all sources of scattering, i.e.,
   disorder,
   electron-phonon, and electron-electron ({\em ee})  interactions, etc.
        contribute to the ARPES
        linewidth.
          The {\em dc} resistivity,
            on the other hand,
            is sensitive only to
            processes which are
            large-angle
            and
            momentum-relaxing
            at the same time.
             For {\em ee}
             interaction,
             a process of this type is umklapp scattering.
                Such
                scattering
                 is
                 effective
 for a sufficiently large Fermi surface (FS)~\cite{abrikosov} and
 not
  too
  long-range {\em ee} interaction.\cite{maslov:2011}
  However, if either of these two conditions is not satisfied,
                umklapp
                 scattering is suppressed, and the {\em dc} resistivity
                 is almost entirely determined by
                                 coupling of electrons to the ``sinks'' of momentum,
                                      such as phonons or impurities.
                                      Note that this does not mean that {\em ee} interaction is weak but rather that it is of the ``wrong" type
                                      for
                                      transport.

                       The scattering rate probed by the optical conductivity is somewhat intermediate between those probed by ARPES
                                          and  {\em dc} transport.
                         As in the {\em dc} case, only
                         large-angle
                         scattering
                          contributes to the optical scattering rate.
                           Similar to the ARPES case, however,
                          this scattering does not have to come from momentum-relaxing processes.
                           As long as Galilean invariance is broken--by lattice or spin-orbit interaction--even
                        momentum-conserving
                          scattering gives a non-zero contribution to the optical scattering rate.
                          In certain cases, this
                          makes the optical conductivity 
                           amenable to hyperscaling analysis,\cite{patel:2015,eberlein:2016} which does not
                          rely on a specific scattering mechanism.
                            Furthermore,
                            at sufficiently high frequencies
                            {\em ee}
                            scattering
                             is stronger than
                            electron-phonon
                            and electron-impurity ones,
                            and the optical conductivity
                             probes
                            directly
                             the transport rate of {\em ee}
                              scattering.
                           At the same time, the optical response at lower frequencies,
                           including the {\em dc} limit,
                            is controlled by momentum-relaxing processes, such as  umklapp
                            {\em ee}
                             scattering or phonon/impurity scattering.
                              This makes the optical response
                               an effective tool
                               for studying {\em ee} interaction even in cases when {\em dc} transport is controlled by scattering from phonons and/or impurities.
                                  A good example of the effectiveness of the optical probe is provided by classic studies of the optical conductivity of noble metals (Ag, Au, and Cu).\cite{beach:1977,parkins:1981}
                                 The temperature dependence of the {\em dc} resistivity in these metals is
                                  determined
                                  mostly
                                  by electron-phonon interaction,
                                  while
                                 the
                                 Fermi-liquid (FL),
                                  $T^2$ term due to {\em ee} interaction is virtually indiscernible. When probed at
                                                                 frequencies above the Debye frequency,
                                  the same metals show a different behavior.
                                  Indeed, the scattering rate due to phonons saturates at
                                    such frequencies, while the scattering rate due to {\em ee} interaction continues to grow with frequency all the way up to the bandwidth
                                     and
                                 determines
                                  the measured optical scattering rate.

                                   In this review, we
                                   address
                                   several
                                    topics  related
                                   to the role of momentum conservation in
                                   optical response.
                                   In particular, we discuss the interplay between contributions from normal and umklapp {\em ee}
    scattering  to the optical conductivity,  both at finite frequency and near the {\em dc} limit.
    We consider
    the
    optical conductivity of a two-band metal
    as the simplest example of a system with broken Galilean invariance.
      We also
      discuss how  the FS  geometry affects the behavior of the optical
      and
      {\em dc} conductivities.

    The second aspect is
    $\Omega/T$ scaling
     of the optical conductivity, $\sigma(\Omega, T)$.
     \cite{marel:2003b,dodge:2000,dodge:2006,nagel:2012,mirzaei:2013,stricker:2014,tytarenko:2015,reber:2015} The real part of the inverse optical conductivity
      is proportional to the optical scattering rate: $\R\sigma^{-1} (\Omega, T) \propto 1/\tau (\Omega, T)$.
      For FL, one generally expects the optical scattering rate
       due to {\em ee} interaction,
       $1/\tau_{\text{ee}} (\Omega, T)$, to scale quadratically with
       either of the variables, i.e., as
       $\max\{\Omega^2,
       T^2\}$.
        An
         explicit computation of $1/\tau_{\text{ee}} (\Omega, T)$
         was performed by Gurzhi in 1959,\cite{gurzhi:1959}
         who showed
           that the prefactors
           of
           the
           $\Omega^2$ and $T^2$ terms are
          related in a universal manner:
        $1/\tau_{\text{ee}} (\Omega, T) \propto  \Omega^2+4\pi^2 T^2$.
         It was later realized~\cite{Fowler:1965,martin:2003,chubukov:2012,maslov:2012}
           that
         a factor of
          $4\pi^2$
           in front of
           $T^2$
           appears there for a good reason: it reflects the fact
        that
         $1/\tau_{\text{ee}}(\Omega, T)$  must vanish
           upon analytic continuation
          to the first boson Matsubara frequency: $\Omega \to \pm 2i\pi T$
           (the ``first-Matsubara-frequency rule'' for boson response).
            By a similar rule, the single-particle
            scattering rate
         in a FL behaves as $\Sigma''(\omega, T)\equiv
          \I\Sigma(\omega,T)\propto \omega^2+\pi^2 T^2$ and vanishes
               upon analytic continuation to the first fermion Matsubara frequency,
         $\omega \to \pm i \pi T$.

            However, recent studies of a number of materials, which
            might have been
            classified as FLs
            due to
            both
            $\Omega^2$ and
            $T^2$ scalings of
            their {\em dc} and optical scattering rates,
           have
           found
            that
             $1/\tau(\Omega, T)$
            often
            does not
           conform to
           the
           FL
           scaling form.
           Namely,
           a different scaling form,
       $1/\tau
        (\Omega,T)\propto \Omega^2+b\pi^2 T^2$,
         with $b$ ranging from $1$ to about $6$,  has been observed.~\cite{nagel:2012,mirzaei:2013,stricker:2014,tytarenko:2015,reber:2015}
           One might be tempted to ignore this disagreement as it affects only a number. However, this number is quite important, because strong deviation of $b$ from
           $b=4$ implies that the optical scattering rate cannot be described by just
             {\em ee}
             interaction,
            although the
            overall behavior of a system resembles that of a
            canonical
             FL.

In this paper,
we
 re-visit
  the Gurzhi formula for the optical scattering rate in a FL and discuss in some detail the ``first-Matsubara--frequency rule" (FMRF), first for the fermion self-energy  and then for the optical conductivity.
Explicitly,
 FMRF
 for the fermion self-energy
states that the Matsubara
self-energy, evaluated at $\omega_m=\pm \pi T$, does not contain the FL-like, $T^2$ term for $2< D\leq 3$:
        $\Sigma (\pm
        \pi T, T) = \pm i
        \pi
        \lambda  T + 0\times T^2 + O(T^D)$. [In 2D, the self-energy does not contain the $T^2\ln T$ term but does contain the $T^2$ one: $\Sigma (\pm
        \pi T, T) = \pm i\pi \lambda  T + O(T^2)$].
        The remainder, which is generically of order $T^D$, is further reduced if a FL
        is of the
Eliashberg type.
           The rule also applies to non-Fermi liquids (NFLs), with a proviso that the coefficient $\lambda$, which is constant in a FL, now depends on $T$.
           The corresponding rule
            for the optical scattering rate
             on the Matsubara axis
              is  $1/\tau_{\text{ee}} (\pm 2\pi
            T, T) = 0\times T^2 + O(T^D)$.
              We relate these
             two
             properties to the scaling forms of the self-energy and optical conductivity on the real axis.
     We also discuss recent experiments and the phenomenological model, proposed in Ref.~\onlinecite{maslov:2012} to explain the violation of
      Gurzhi scaling. In this model, one assumes that there are two channels of scattering.
     The first one
    is
   {\em ee} interaction, which does lead to the
   usual
    FL scaling form of the self-energy. The
    second one is
    an
    elastic
    scattering process with an effective cross-section,  which depends on the electron energy but not on temperature.
    Consequently,
    this channel contributes an $\omega^2$ but no $T^2$ term to the self-energy.
       This
       changes the balance between the $\Omega^2$ and $T^2$ terms in the optical conductivity. As a result, the coefficient $b$ is not equal to $4$ but depends on the relative weights
        of the
        inelastic and
        elastic channels. In particular, the value of $b\approx 1$, observed in a number of
        rare-earth Mott insulators
         and heavy metals, implies that the elastic channel is much stronger than the {\em ee}
      one.
         As a particular example, we consider elastic scattering at resonance levels  and show that this model is in reasonable agreement with the data on URu$_2$Si$_2$.\cite{nagel:2012}

The third
aspect is
the optical conductivity
of a NFL metal.
We consider two
particular examples of a NFL, encountered in metals near
nematic (Pomeranchuk)
 and
  spin-density-wave (SDW) instabilities.
In the last case, we focus on
a clean two-dimensional (2D) metal
 near a quantum-critical point (QCP)
 separating
the paramagnetic and
 commensurate
 SDW phases.
  Critical
 magnetic fluctuations are known to
  destroy
 fermion coherence,
 but only at particular
    \lq\lq hot
  spots\rq\rq\/
 [FS
points connected by the ordering wavenumber, $(\pi,\pi)$].
On the rest of the FS,
      quasiparticles are
 strongly renormalized compared to the non-interacting case, but still display a FL behavior at the lowest energies.
  Because $1/\tau_{\text{ee}} (\Omega) \propto \Omega^2$ for coherent quasiparticles and because the optical conductivity is obtained by averaging over the FS,
  \cite{hlubina:1995}
   it
  had
   long been
   believed
   that the conductivity at the
   SDW
   QCP retains its FL form:
   $\sigma'(\Omega)\equiv \R\sigma(\Omega)
   \propto 1/\Omega^2 \tau_{\text{ee}} (\Omega) = \text{const}$.
   However,
   it has been argued recently\cite{hartnoll:2011,chubukov:2014}
  that
    this is not the case because of
    composite scattering --
    a process in which
    fermions, located away from a hot spot,
    are
     scattered
      twice by $(\pi,\pi)$ fluctuations and return to nearly the same points.
     It turns out that this scattering
      gives  a larger  contribution to
      $\sigma'(\Omega)$
       than  direct scattering by $(\pi,\pi)$  fluctuations. As a
      result, the dissipative part of the conductivity
      displays a non-FL behavior at a QCP:
      $\sigma'(\Omega)$  scales as
       $\Omega^{-1/3}$
  at
  asymptotically low
  frequencies
    and as
   $\Omega^{-1}$
   at higher frequencies,
   nominally up to the bandwidth.
  The  $1/\Omega$ scaling of $
  \sigma'(\Omega)$  is reminiscent of the behavior
  observed in the superconducting cuprates.\cite{basov:1996,puchkov:1996,basov:2005,norman:2006}
        We
        derive these
        results
        and also show that,
       at finite temperature, the optical conductivity
       is of
       the FL form
       at frequencies below
       certain $T$-dependent scale,
       but  displays NFL-like
       $\Omega$- and $T$-dependences
          at frequencies above
         this scale:
         $\sigma'(\Omega,T)\propto T^{4/3}\Omega^{-5/3}$.

  Due to limited and focused scope of this review, we do not discuss several modern approaches to
    optical response
   of correlated electron systems,
  e.g., 
    the holographic approach.
  We refer
  the
  reader to
  recent literature on this subject, e.g., Refs.~\onlinecite{horowitz:2012,vegh:2013,donos:2014,langley:2015,davison:2015,khvesh:2015}.

The
rest of the
paper is organized as follows. In  Sec.~\ref{sec:BEE},  we
    address  various aspects of momentum conservation for the optical conductivity.
     In Sec. \ref{sec:FL}, we derive the
     Gurzhi formula for the optical scattering rate in a FL.
        In Sec.~\ref{sec:rules}, we discuss the first-Matsubara-frequency rule,
        both or the fermion self-energy (Sec.~\ref{sec:1stM}) and
        optical conductivity (Sec.~\ref{sec:1stMsigma}).
           In Sec.~\ref{sec:gurzhi_exp}, we discuss recent experiments and the phenomenological model, proposed in Ref.~\onlinecite{maslov:2012} to explain the observed
           deviations from
           FL scaling.
           In Sec.~\ref{sec:pom}, we re-visit the optical conductivity of a metal near Pomeranchuk quantum criticality.
      In Sec.~\ref{sec:SDW},
      we consider
      a
      2D
      metal near
      an SDW
      quantum phase transition,
       and show that its  optical conductivity
        exhibits a non-FL behavior due to composite scattering of fermions away from hot spots.

\section{Momentum conservation in
  optical
 response}
\label{sec:BEE}
\subsection{Normal and umklapp scattering in the optical conductivity}
\label{sec:BE}
\subsubsection{Boltzmann equation and Kubo formula}
\label{sec:bek}
\paragraph{{\bf Optical conductivity of a non-Galilean--invariant system: Boltzmann equation.}}
It is well-known that the {\em dc} conductivity of
 a
 single-band
  electron system is infinite in the absence of  umklapp scattering processes and/or disorder, even if Galilean invariance is broken by
 a
 Indeed, consider a linearized Boltzmann equation
\bea
e\bv_\bk\cdot{\bf E}=-\sum_{\bk\bp\bk'\bp'}W_{\bk,\bp;\bk'\bp'} \left(g_{\bk}+g_{\bp}-g_{\bk'}-g_{\bp'} \right)~ F_{\bk\bp\bk'\bp'},
\label{be}
\eea
where $\bv_\bk=\boldsymbol{\nabla}_\bk\e_\bk$ is the group velocity of a Bloch electron with dispersion $\e_\bk$, ${\bf E}$ is the external electric field,    $g_{\bk}$ is a non-equilibrium part of the
 distribution function, defined as $f_{\bk}=f_{0\bk}+g_{\bk}\frac{\partial f_{0\bk}}{\partial \e_\bk}$, $f_{0\bk}$ is
 the
  Fermi
   function
   for free electrons,
     $W_{\bk,\bp;\bk'\bp'}$ is the
     {\em ee}
     scattering probability, and
     $F_{\bk\bp\bk'\bp'}$ is a combination of the Fermi
     functions whose explicit form is not essential for the present discussion.
     In the absence of  umklapp processes, the total momentum is conserved: $\bk+\bp=\bk'+\bp'$. Let $g_{\bk}$ be a solution of Eq.~(\ref{be}). But then
$\tilde g_{\bk}=g_{\bk}+{\bf C}\cdot \bk$ with  a $\bk$-independent but otherwise arbitrary vector ${\bf C}$ is also a solution. Since ${\bf C}$ is arbitrary, the corresponding
charge
current
\beq
{\bf j}=2 e\sum_{\bk}  \bv_{\bf k} f_{\bk} =2e \sum_{\bk}  \bv_{\bf k} g_{\bk} \frac{\partial f_{0\bk}}{\partial \e_\bk}
\eeq
 can be made arbitrarily large even
 by
  infinitesimally weak electric field, which means that the conductivity
is infinite. In the memory-matrix formalism, the same result follows from the fact that the memory matrix has a zero mode in the absence of  umklapp scattering.\cite{maebashi:1997,maebashi:1998}

However, if the electric field oscillates in time,  umklapp processes are not necessary for the conductivity to be finite at finite frequency: just  a violation of Galilean invariance suffices. \cite{gurzhi:1959}
Let's see how this
result follows from
the Boltzmann equation.
 Adding the time derivative to the left-hand-side of Eq.~(\ref{be}), we obtain
\bea
-i \Omega g_\bk+e\bv_\bk\cdot{\bf E} =-\sum_{\bk\bp\bk'\bp'}W_{\bk,\bp;\bk'\bp'} \left(g_{\bk}+g_{\bp}-g_{\bk'}-g_{\bp'} \right)~F_{\bk\bp\bk'\bp'}.
\label{be2}
\eea
Now the Boltzmann equation can be solved by iterations with respect to the collision integral:
\beq
g_\bk=
g^{(0)}_\bk+ g^{(1)}_{\bk}+g^{(2)}_{\bk}+\dots
\label{g}
\eeq
This expansion is valid in the ``collisionless'' or ``high-frequency'' regime, defined by the condition  $\Omega\te\gg 1$,
where $1/\te$ is the
 {\em ee} scattering rate obtained by some appropriate averaging of the collision integral.
  Within the semiclassical approximation, the frequency of light must be small compared to temperature;
   therefore, $\te$ in
   the
   solution of the semiclassical Boltzmann equation is a function of temperature but not frequency,
    \footnote{In the  semiclassical Boltzmann equation, the frequency of the electric field enters only the left-hand-side. Correspondingly, the result for the conductivity contains the {\em ee} relaxation
    time, $\te$, which depends on the temperature but not frequency. On the other hand, the Kubo formula treats the frequency and temperature on equal footing. The quantum Boltzmann equation remedies
    semiclassical
    limitation and the resulting conductivity coincides with the one given by the Kubo formula.}
     which is what we will be assuming here  and in Sec.~\ref{sec:two_band}.
The zeroth-order
term in the expansion, $g^{(0)}_\bk=e\bv_\bk\cdot {\bf E}/i\Omega$, is independent of $\te$ and gives the
imaginary, i.e., non-dissipative, part of the conductivity: $\sigma''(\Omega)\propto 1/\Omega$. The first-order term is given by
\bea
g^{(1)}_\bk=-\frac{e}{\Omega^2}\sum_{\bk\bp\bk'\bp'}W_{\bk,\bp;\bk'\bp'}\left( \Delta {\bf J}\cdot{\bf E}\right)~F_{\bk\bp\bk'\bp'},
\label{g1}
\eea
where
\beq \Delta {\bf J} \equiv \bv_{\bk}+\bv_{\bp}-\bv_{\bk'}-\bv_{\bp'}\label{DJ}
\eeq
 is a change
  in the
total
 particle flux
  due to an
 {\em ee}
  collision.\cite{lawrence:1973,lawrence:1975,riseborough:1983}
 The correction
 $g^{(1)}_\bk$ is real
and,
 if non-zero,
 gives rise to a non-zero
  real, i.e., dissipative, part of the conductivity: $\sigma'\propto 1/\Omega^2\te$.
 In a Galilean-invariant system,
 $\bv_\bk=\bk/m$ and thus momentum conservation implies current conservation. In this case $\Delta {\bf J}=0$, and the dissipative term in the conductivity is absent.
 (This is true for
 an
 arbitrary order in $1/\Omega\te$.)
 If
  $\bv_\bk\neq \bk/m$, which
  is the case in the presence of
 a
 lattice and/or spin-orbit interaction,
 $\Delta {\bf J}$ is non-zero, even though momentum is conserved. Consequently,
the conductivity of a non-Galilean-invariant system
contains the dissipative part.

There are two points which one needs to keep in mind, however.
First,
 the conductivity at frequencies $\Omega\gg 1/\te$ behaves as
\beq
\sigma(\Omega)=C_1\frac{i}{\Omega}+C_2\frac{1}{\Omega^2\te}
\eeq
where $C_{1,2}$ are real constants.
 This form
does coincide with the
corresponding
 limit of the Drude formula
\beq
\sigma(\Omega)= \frac{\Omega_p^2}{4\pi} \frac{1}{\te^{-1}-i\Omega},\label{drude}
\eeq
where $\Omega_p$ is the plasma frequency.
At
an
arbitrary frequency,
however, the conductivity
is not described by Eq.~(\ref{drude}).
 This follows already from the fact Eq.~(\ref{drude}) has a finite limit at $\Omega\to 0$, while we argued
at the beginning of this section that the {\em dc} conductivity is infinite in the absence of umklapp/impurity scattering. In Sec.~\ref{sec:two_band},
we discuss the actual behavior of $\sigma(\Omega)$ for a particular example of a non-Galilean-invariant system--a two-band metal.

Second, $1/\te$ in the
high-frequency
tail of the conductivity does not always coincide with the quasiparticle scattering rate which, in a FL, scales as $T^2$ for $T\gg \Omega$.
  This implies that $\sigma'(\Omega)$ does not always scale as $T^2/\Omega^2$ even for a FL.
For certain
 FS  geometries (see Sec. \ref{sec:extra}),
      the prefactor of the leading,
      $T^2$ term in $1/\te$ vanishes,  and $1/\te$ scales as $T^4$, which implies that the conductivity scales as $T^4/\Omega^2$. \cite{gurzhi:1980,gurzhi:1982,gurzhi:1987,maebashi:1998,rosch:2005,rosch:2006,maslov:2011,pal:2012,pal:2012b,briskot:2015}
\begin{figure}[h]
\includegraphics[width=0.6 \linewidth]{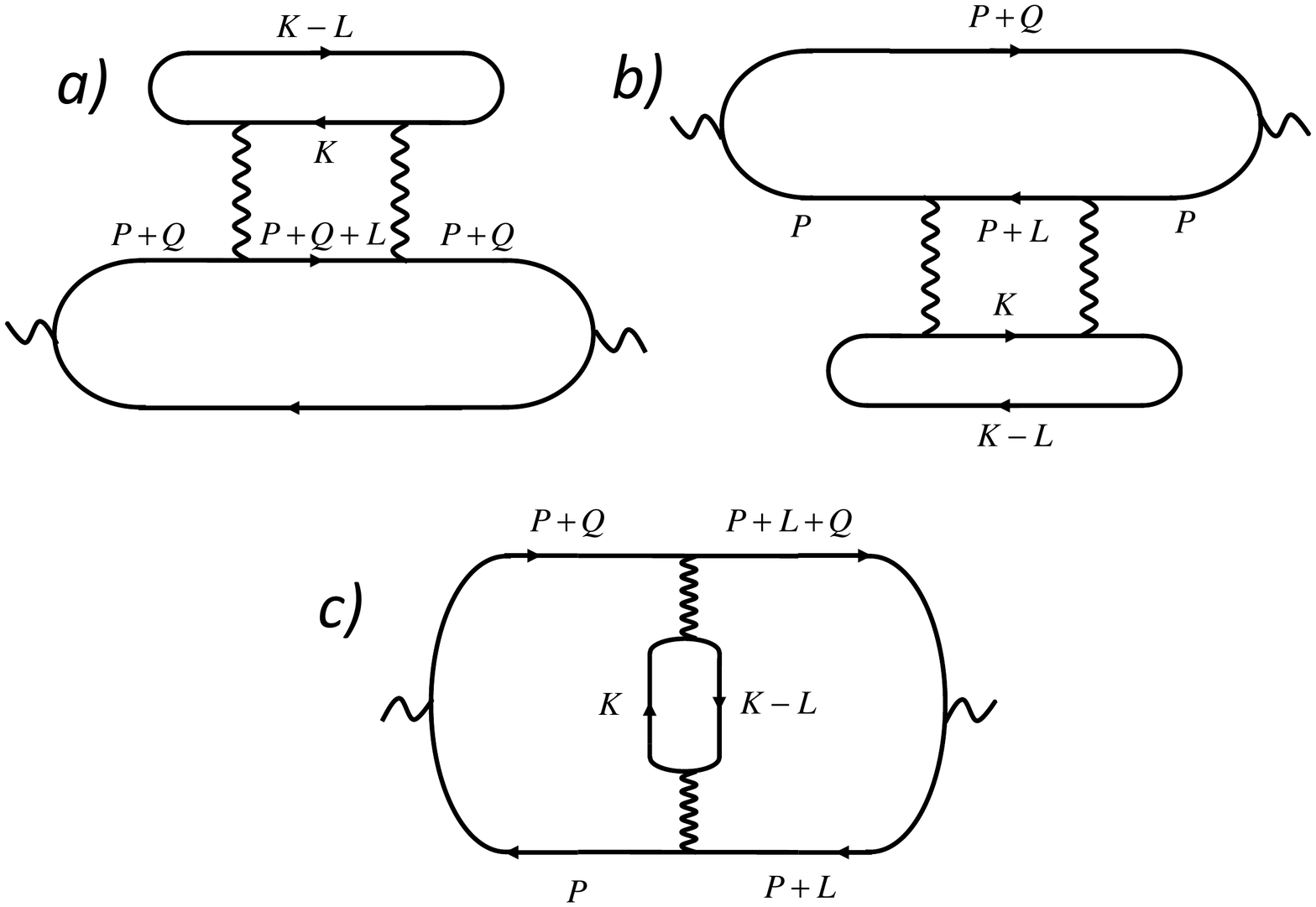}
\vspace{-1.5in}
\includegraphics[width=0.6 \linewidth]{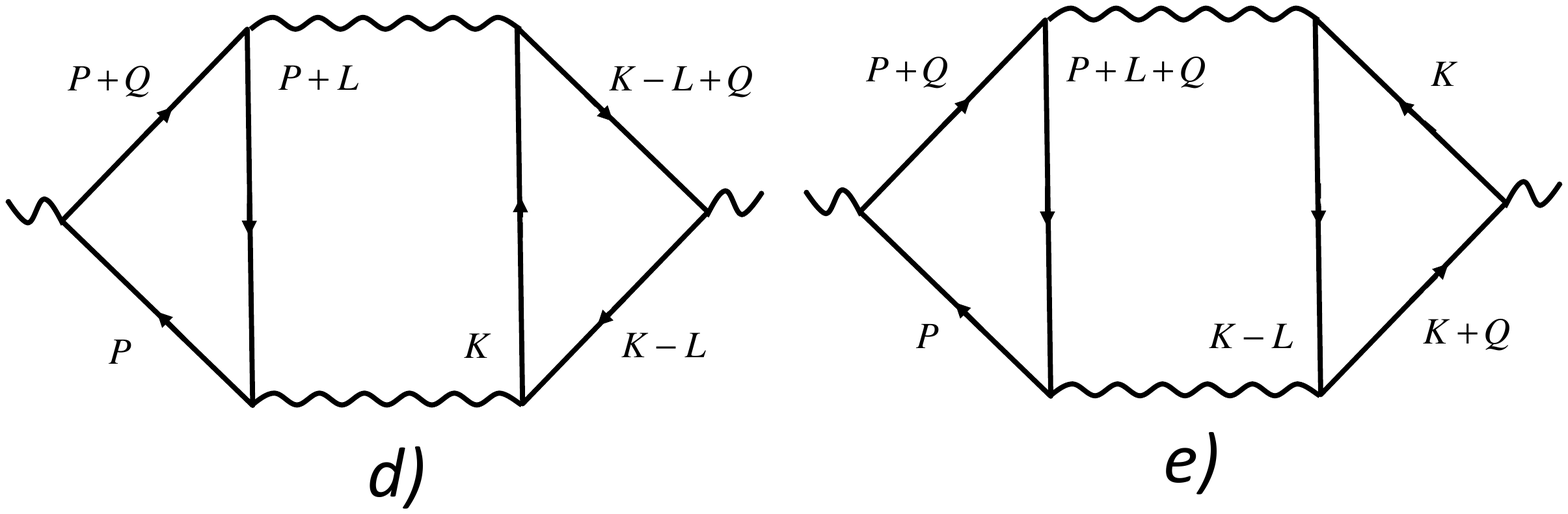}
\vspace{-0.25in}
\caption{Feynman diagrams for the optical conductivity. \label{fig:5diag}}
\end{figure}

Obviously, if umklapp scattering is allowed,
it also contributes to the optical conductivity.
There is a certain interplay between normal and umklapp scattering,
 which we discuss in Secs.~\ref{sec:ex_1} and \ref{sec:vertex}.

 \paragraph{{\bf Optical conductivity of a non-Galilean--invariant system: Kubo formula.}}
One can
 arrive at the same result--that $\sigma' (\Omega)$ vanishes in the Galilean-invariant case but $\sigma' (\Omega)\propto 1/\Omega^2 \tau_{ee}\neq 0$
if Galilean invariance is broken--by
   using
    the Kubo formula for the conductivity rather than the Boltzmann equation.
Holstein\cite{holstein:1964} and a number of authors after him\cite{riseborough:1983,maebashi:1997,maebashi:1998, gornyi:2004,farid:2006}
 demonstrated the cancellation of diagrams for the conductivity
 in the Galilean-invariant case in various physical contexts.
 Following the previous work, we show in Appendix \ref{app:diags} that
the combination of the  diagrams for the conductivity in Fig.~\ref{fig:5diag} can be
 reduced
 to a form that
 contains the same
 change in total flux, $\Delta{\bf J}$, as in Eq.~(\ref{g1}). For the current-current correlation function   to second order in the static {\em ee} interaction $U_{\bf l}$, we find
\bea
{\cal K}(iq_0)
=\frac{1}{q_0^2}\sum_{P,K,L} \lr\bv_\bp\cdot
\Delta{\bf J}\rr
U_{\bl}^2
\left(2G_PG_{P+L}-G_PG_{P+L+Q}-G_{P+Q} G_{P+L}\right)G_KG_{K-L},\label{KK}
\eea
where $P=(\bp,ip_0)$, $K=(\bk,ik_0)$, $L=(\bl,il_0)$, $Q=({\bf 0},iq_0)$, $G_P=1/(ip_0-\e_\bp)$ is the Green's function, and $\sum_K$ stands for $T\sum_{k_0}\int d^Dk/(2\pi)^D$, etc.
The dissipative part of the conductivity is obtained as $\sigma'(\Omega)=e^2\I{\cal K}(iq_0\to\Omega+i0^+)/\Omega$.
Because $\I{\cal K}(iq_0\to\Omega+i0^+)$ in Eq.~(\ref{KK}) is proportional to $1/\Omega^2$ and contains the square of the {\em ee} interaction, it gives
an ${\cal O}(1/\Omega^2\te)$ term in the conductivity.
Whether $
\sigma'(\Omega)
$  is finite or zero is determined by whether
$\Delta{\bf J}$ in Eq.~(\ref{KK}) is finite or zero, which is the same result as obtained from the  Boltzmann equation.

\subsubsection{Similarity
between the optical conductivity of a clean system and
the {\em dc} conductivity of a disordered system}
\label{sec:ex_1}
The situation with the optical conductivity is to a certain extent similar to
  that for
  the
  {\em dc}
   conductivity
    in the presence of both impurities and {\em ee} interaction.
  In the latter case,
   the impurity collision integral,
  $(f_\bk-\langle f_\bk\rangle)/\ti$ with $\langle\dots\rangle$ standing for averaging over the directions of $\bk$,
    plays the role of the time derivative in the {\em ac} case,  the zeroth-order solution is obtained in the presence of impurities only,
    and
    higher orders
are obtained by iterations with respect to the {\em ee} collision integral. In the Galilean-invariant case,
   \footnote{Although impurities violate Galilean invariance, this effect is not captured by the Boltzmann equation
   for an average (over realizations of disorder) distribution function, which is still parametrized by momentum.
   In this case, it makes sense to speak about Galilean invariance of an electron system in the presence of impurities.}
the
 Boltzmann equation predicts that {\em ee} interaction does not affect the resistivity,
 i.e.,
 the analog of $g_\bk$
 in  Eq.~(\ref{g})
 vanishes.
If Galilean invariance is broken,
  the analog of $g^{(1)}_\bk$ is non-zero, i.e., {\em ee} interaction contributes to the resistivity.
Beyond
the
 Boltzmann equation, {\em ee} interaction
 may affect the resistivity already in the Galilean-invariant case via, e.g., quantum-mechanical interference effects,\cite{altshuler:1985} superconducting fluctuations,\cite{varlamov} and finite viscosity of the electron liquid.\cite{hruska:2002,andreev:2011}

 Another similarity between the {\em dc} and optical conductivities
 is
 in
 the interplay between
 normal
  {\em ee} scattering
   (with rate $1/\te$)
    and momentum-relaxing  scattering
  (with rate $1/\ti$).
 For simplicity, we
 assume
 that the latter mechanism
 in the {\em dc} case
 is due to impurities.
   At low temperatures, when $\te\gg\ti$, the scattering rates of the two processes add up
  according to the Matthiessen rule:
   \beq
   1/\tau_{\text{eff}} = 1/\ti + 1/\te.
   \label{ch_1}
   \eeq
   The Matthiessen rule, however,
   does not hold
   at all temperatures.
 In particular, at higher temperatures, when $\te\ll\ti$,
 the {\em ee} term in the resistivity does not become the dominant one.
    Instead, the resistivity saturates at a temperature-independent value, which is proportional to
   $1/\ti$
  but, in general, differs from the
  residual resistivity at
  $T=0$. \cite{maslov:2011,pal:2012b}
The physical reason for such saturation is that normal {\em ee} collisions by themselves cannot
relax the current, no matter how frequent they are. This can done only by impurities (and/or umklapps).
All normal
 collisions can do is to modify the
  energy dependence of the
distribution function, and it is this modification that
 changes the
 resistivity compared
 to the residual one.
The ratio of the high-$T$ to low-$T$ saturation values is determined by the shape of the FS.

The interplay between
normal
 and
momentum-relaxing
 scattering in the optical conductivity is similar
  to the {\em dc} case in a sense
that  the Matthiessen
rule is, in general, also violated.
In the collisionless regime,
$\sigma'(\Omega)\propto 1/ \Omega^2\tau_{\text{eff}}$, where $1/\tau_{\text{eff}}$
is given by
 an
equation similar to Eq.~(\ref{ch_1}) but with, generally
speaking,
different
 weights
of the $1/\te$ and $1/\ti$ terms [see Eq. (\ref{resigma_imp_b}) below].
  At lower
  frequencies, however,  $1/\te$ and $1/\ti$ do not
  contribute
  additively to the frequency dependence of $\sigma' (\Omega)$.

In the next section, we analyze the optical conductivity of a
two-band metal. The results of this section are not new:
 they can be inferred from general formulas, derived, e.g, in  Ref.~\onlinecite{levinson:book}. We include this discussion for completeness and also because
 this simple case does give an idea of how the interplay between the
 normal
  and momentum-relaxing  scattering mechanisms works.

\subsection{Optical conductivity of a two-band metal}
\label{sec:two_band}
A two-band metal is the simplest example of a system with broken Galilean invariance. Even if
 each of the bands is parabolic, the system as a whole is not Galilean-invariant.
  The analysis of the conductivity in
  this model
   is
  usually associated with the name of Baber,\cite{baber:1937} who considered the effect of inter-band {\em ee} scattering. It is sometimes forgotten, however, that Baber analyzed only the
   case of a {\em compensated semi-metal}, with equal numbers of electrons and holes.
 Only in this case,
 normal {\em ee} collisions alone
 render the {\em dc} resistivity finite.
 (This
  is also true also
   for Weyl/Dirac semimetals at the charge neutrality point.\cite{*[{See, e.g.,} ]  [{, and references therein.}] hosur:2013})
 If
a
 metal is not compensated,
 one needs
   momentum-relaxing scattering to obtain a finite {\em dc} resistivity.

We now analyze how
the optical conductivity of a two-band metal  evolves as a function of frequency between the {\em dc} and
high-frequency
 limits.

\subsubsection{Momentum-conserving scattering only}
  At first, let momentum-relaxing scattering be absent.
For a
 two-band metal with parabolic bands,
the conductivity can be found by
 solving the semiclassical equations of motion
 \cite{levinson:book}
\bea
-i\Omega m_1\bv_1&=&e_1{\bf E}-\eta n_2 (\bv_1-\bv_2),\nn\\
-i\Omega m_2\bv_2&=&e_2{\bf E}-\eta n_1 (\bv_2-\bv_1),
\label{2band}
\eea
where $e_{1,2}=\pm e$, indices $1$ and $2$ denote the bands,
  and
   $n_{1,2}$ is the number density.
\footnote{By assuming that the number
density in each of the bands is finite, we exclude the case of a Weyl/Dirac semimetal at the charge neutrality point, which
is the special limit of the Baber's case with zero band overlap.}
 and $\eta>0$ parametrizes inter-band {\em ee} scattering.
 (For parabolic bands, intra-band {\em ee} scattering conserves the in-band momentum and thus does enter the equations of motion.)
\footnote{For parabolic bands, the
phenomenological equations of motion
are
 equivalent to the solution of the Boltzmann equation. What is left undetermined at the phenomenological level
 is
  an explicit
  form
   of the scattering time, which
   is
   to be found
   from the Boltzmann equation with a given collision integral.
   Solutions of the Boltzmann equation for
    two-band systems in particular physical contexts can be found in, e.g.,
 Refs.~\onlinecite{
 murzin:1998,pal:2012b}.}
Solving these equations, we find the current ${\bf j}=e_1n_1\bv_1+e_2n_2\bv_2$ and thus the conductivity
 at
 finite frequency $\Omega \neq 0$
\bea
\sigma(\Omega)=\frac{1}{4\pi}\frac{\Omega\Omega_p^2+i\te^{-1}\Omega_0^2}
{\Omega\left(\te^{-1}-i\Omega\right)}
= \frac{\Omega^2_p-\Omega^2_0}{4\pi} \frac{1}{\te^{-1}-i\Omega} + \frac{\Omega^2_0}{4\pi} \frac{
i}{\Omega},
\label{qD}
\eea
where $\Omega_p^2=\Omega_1^2+\Omega_2^2$, $\Omega_{1,2}^2=4\pi e^2n_{1,2}/m_{1,2}$
is the intra-band plasma frequency,
\beq
\Omega_0^2=4\pi \frac{\left(e_1n_1+e_2n_2\right)^2}{n_1m_1+n_2m_2}\label{O_0}
\eeq
is the ``compensation frequency", and we defined $\te$ as
\beq
1/\te\equiv \eta\left(\frac{n_1}{m_
2}+\frac{n_2}{
m_1}\right).
\eeq
As in the previous section, $\te$ depends on the temperature but not on frequency; for a FL, $1/\te\propto T^2$.

\begin{figure}[h]
\includegraphics[width=0.8 \linewidth]{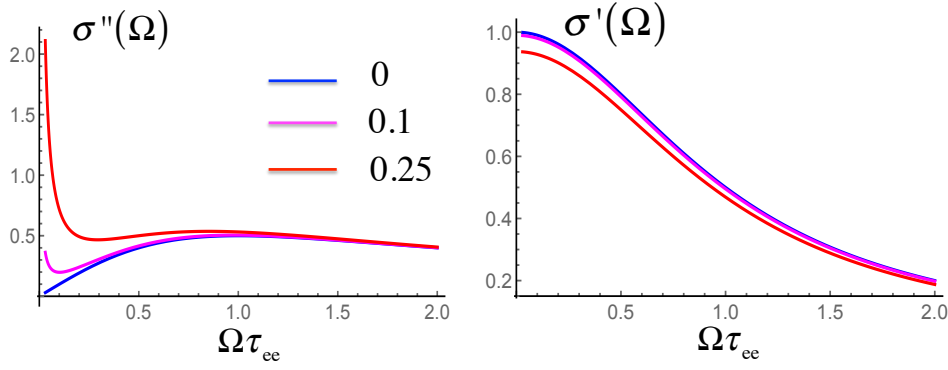}
\vspace{-2 in}
\caption{The imaginary (left) and real (right) parts of the conductivity of a two-band metal with interband {\em ee} scattering but without intra-band relaxation [Eqs.~(\ref{imsigma_2b}) and (\ref{resigma}), correspondingly]. A degree of compensation, $\Omega_0/\Omega_p$ [cf. Eq.~(\ref{O_0})], is specified in the legend. The conductivity in measured in units of $\Omega_p^2\te/4\pi$. Only in the fully compensated case ($\Omega_0=0$), both the real and imaginary parts obey the Drude formula.
Away from compensation, the real part of the conductivity 
 $\sigma'(\Omega)$ also contains a $\delta(\Omega)$ term (not shown).
\label{fig:2band}}
\end{figure}

 If the metal is compensated, i.e., $e_1n_1+e_2n_2=0$, then $\Omega_0=0$ and the conductivity is described the Drude formula [Eq.~(\ref{drude})] at all frequencies,
  including $\Omega =0$. This is the Baber's case:\cite{baber:1937}
$\sigma'$ approaches a
finite value $\sigma' = \Omega_p^2\te/4\pi$
in the limit of $\Omega\to 0$, while $\sigma''$ vanishes in this limit.
  However, if the metal is not compensated, $\sigma(\Omega)$ as a whole cannot be described by the Drude formula.
   Indeed,
    the imaginary part  of $\sigma (\Omega)$
\bea
\sigma''(\Omega)=\frac{1}{4\pi}\frac{\Omega_p^2\Omega^2\te^2+\Omega_0^2}{\Omega\left(\Omega^2\te^2+1\right)}
\label{imsigma_2b}
\eea
diverges as $1/\Omega$ at $\Omega\to 0$ (cf. Fig.~\ref{fig:2band}, left).
 This divergence is the same as
 the diamagnetic term in the conductivity
 of an ideal metal.
  The real part of the conductivity
  at $\Omega \neq 0$ is
\bea
\sigma'(\Omega)=\frac{\Omega_p^2-\Omega_0^2}{4\pi}\frac{\te}{\Omega^2\te^2+1}
\label{resigma}
\eea
is
still
of
the Drude form with a renormalized plasma frequency,
 and
 it
 remains
 finite at $\Omega\to 0$ (cf. Fig.~\ref{fig:2band}, right).
Note that $\Omega_p^2-\Omega_0^2= \frac{4\pi n_1n_2m_1m_2}{n_1m_1+n_2m_2}\left(\frac{e_1}{m_1}-\frac{e_2}{m_2}\right)^2$ is always positive.

 So far, we have completely neglected  momentum-relaxing scattering.  Infinitesimally weak momentum-relaxing scattering can be accounted for by adding a small imaginary part to $\Omega$ in the denominator of the second term in Eq.~(\ref{qD}): $\Omega\to \Omega+i0^+$. Then the Kramers-Kronig transform of this term produces an additional $\delta(\Omega)$ term in $\sigma'(\Omega)$, which is the same as in the case of an ideal metal without any scattering.
 Finite momentum-relaxing scattering smears out the delta-function into a Drude peak, which will be described in the next section. At compensation, $\Omega_0=0$ and there is no $\delta(
 \Omega)$ term even in the absence of momentum relaxation.

A non-Drude from of the conductivity
 affects the behavior of the reflection coefficient at low frequencies ($\Omega\te\ll 1$).
  For a Drude metal,  $1-R \propto \sqrt{\Omega}$ (the Hagen-Rubens relation).
 For $\sigma (\Omega)$ given by Eq.~(\ref{qD}),
 $1-R$ scales instead as $\Omega^2$:
\beq
1-R\approx
2\sqrt{2} \Omega^2\te\frac{\Omega_p^2-\Omega_0^2}
{\Omega_p^3}.
\eeq

\subsubsection{Both momentum-conserving and momentum-relaxing scattering}
Now, let's add momentum-relaxing scattering due to disorder or umklapps.
This can be modeled by adding the $-m_
j\bv_
j/\tau_
j$ terms  ($j=1,2$) to the right-hand sides of the equations of motion [Eq.~(\ref{2band})],
 where $\tau_{1,2}$ are the
momentum-relaxation times.
(The rates $1/\tau_{1,2}$ are assumed to account for both intra- and inter-band momentum-relaxing processes.)
 Even in this case, however,
the conductivity is not described by the Drude formula with the total relaxation rate given by the sum  $1/\tau_1+1/\tau_2+1/\te$ of the partial rates, because momentum-relaxing and
momentum-conserving mechanisms are not additive.  While the Drude formula has only one characteristic frequency scale ($1/\tau$), the actual conductivity in our case has three scales: the
first
 two
are given by the momentum-relaxing  and momentum-conserving scattering rates, correspondingly, and the third one (roughly a geometric mean of the two
previous ones) defines a crossover between the two regimes.
Solving the
modified
equations of motion,
 we
 obtain
\bea
\sigma'(\Omega)=\frac{1}{4\pi} \frac{ \gamma_0^2\lr \frac{\Omega_1^2}{\tau_2}+\frac{\Omega_2^2}{\tau_1}+\frac{\Omega_0^2}{\te}\rr+\Omega^2\lr \frac{\Omega_1^2}{\tau_1}+\frac{\Omega_2^2}{\tau_2}+\frac{\Omega_p^2-\Omega_0^2}{\te}\rr}{\lr\Omega^2-\gamma_0^2\rr^2+\lr\Omega/\tau_{\text{eff}}\rr^2}
\label{resigma_imp}
\eea
and
\bea
\sigma''(\Omega)=\Omega\frac{\frac{1}{\tau}\lr\frac{\Omega_1^2}{\tau_2}+\frac{\Omega_2^2}{\tau_1}+\frac{\Omega_0^2}{\te}\rr+\lr\Omega_1^2+
\Omega_2^2\rr\lr\Omega^2-\gamma_0^2\rr}{\lr\Omega^2-\gamma_0^2\rr^2+(\Omega/\tau_{\text{eff}})^2},
\label{imsigma_imp}
\eea
where
\bse
\bea
\gamma_0^2&=&\frac{1}{\tau_1\tau_2}+
\frac{1}{\te} \left(\frac{1}{\tau_1}\frac{1}{1+\frac{n_2m_2}{n_1m_1}}+\frac{1}{\tau_2}\frac{1}{1+\frac{n_1m_1}{n_2m_2}}\right)\quad\text{and}\\
\frac{1}{\tau_{\text{eff}}}&=&\frac{1}{\tau_1}+\frac{1}{\tau_2}+\frac{1}{\te}.
\eea
\ese
The high-frequency tail of Eq.~(\ref{resigma_imp}) is of the Drude form
\beq
\sigma'(\Omega)=\frac{1}{4\pi\Omega^2}\lr\frac{\Omega_1^2}{\tau_1}+\frac{\Omega_2^2}{\tau_2}+\frac{\Omega_p^2-\Omega_0^2}{\te}\rr,
\label{resigma_imp_b}
\eeq
while $\sigma''(\Omega) \propto
1/\Omega$.
In the static limit,
$\sigma'(\Omega=0) = \left(\frac{\Omega_1^2}{\tau_2}+\frac{\Omega_2^2}{\tau_1}+\frac{\Omega_0^2}{\te}\right) /4\pi \gamma_0^2 $ is finite, while $\sigma''(\Omega)$ vanishes.

In  between the high- and low-frequency
limits, however,
 the conductivity is not described by the Drude formula.  To analyze the form of conductivity at intermediate frequencies,
 we focus on the {\em hydrodynamic} regime,
when momentum-conserving scattering is stronger than momentum-relaxing one: $1/\te\gg 1/\tau_1\sim 1/\tau_2$
(here and thereafter, the $\sim$ sign means ``equal in order of magnitude").
This regime has received considerable attention recently in the context of both strongly-correlated
systems
 and Weyl/Dirac metals.
\cite{mueller:2008,andreev:2011,tomadin:2014,forcella:2014,narozhny:2015,briskot:2015,levitov:2016,bandurin:2016}
We consider a generic case when
$n_1\sim n_2$ and $m_1\sim m_2$; this implies that
$\Omega_1\sim\Omega_2\sim\Omega_0
\sim \Omega_p$,
  and $\tau_1\sim\tau_2$. In this case,
the analysis of Eq. (\ref{resigma_imp}) shows that there are three
crossover frequencies:
1) $1/\ti \sim 1/\tau_1\sim 1/\tau_2$,
2) $1/\te$, and 3) the
intermediate
scale
$1/\tau^* \sim1/\sqrt{
\ti
\te} \sim \gamma_0$.

\begin{figure}[h]
\includegraphics[width=0.8 \linewidth]{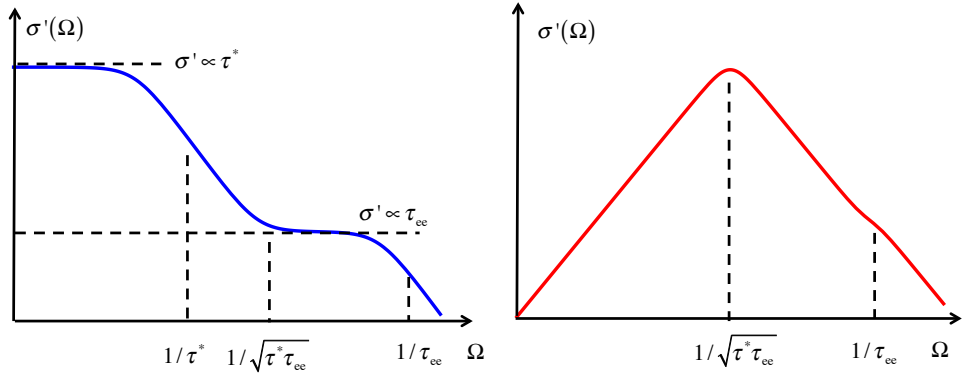}
\vspace{-2. in}
\caption{Left: $\sigma'(\Omega)$
of a two-band metal
 [Eq.~(\ref{resigma_imp})] in the hydrodynamic regime, where the momentum-relaxing scattering rate,
 $1/\ti$, is much smaller than
 the
 rate of
 momentum-conserving
 {\em ee} scattering,
 $1/\te$.  Right: the same for $\sigma''(\Omega)$ [Eq.~(\ref{imsigma_imp})]. Both $\sigma'(\Omega)$ and $\sigma''(\Omega)$ are plotted on the log-log scale.\label{fig:2band_imp}}
\end{figure}

 For $\Omega\ll 1/\tau^*$, $\sigma'(\Omega)$
 has
  a Drude
  peak
 with relaxation time
 $\ti$:
   in the {\em dc} limit,
$\sigma'(\Omega)$
 is proportional to $\ti$; for $1/\ti\ll\Omega\ll 1/\tau^*$,
  $\sigma'(\Omega)\propto 1/\Omega^2\ti. $
  This is the smeared-out delta-function peak described in the previous section.
 The high-frequency tail of this
 peak
 is cut off at $\Omega\sim 1/\tau^*$, where $\sigma'(\Omega)$ saturates at a quasi-static value proportional to $\te$.
 For
 $\Omega\gtrsim 1/\te$, $\sigma'(\Omega)$
 has
  a second Drude peak with relaxation time $\te$. The two-peak structure of $\sigma'(\Omega)$ is depicted in
 Fig.~\ref{fig:2band_imp}, left.

For $\Omega\ll1/\tau^*$, the imaginary part of $\sigma(\Omega)$  obeys the Drude formula with relaxation time $\ti$ : $\sigma''(\Omega)$ vanishes linearly with $\Omega$ for $\Omega\ll 1/\ti$
and falls off as $1/\Omega$ for $\Omega\gg 1/\ti$. In contrast to $\sigma'(\Omega)$, however, $\sigma''(\Omega)$ does not
have
 the second Drude peak for $\Omega\gg 1/\tau^*$. Instead, the $1/\Omega$ tail of the first Drude peak matches smoothly with a non-Drude form in Eq.~(\ref{imsigma_2b}), parametrized by relaxation time $\te$. Overall, $\sigma''(\Omega)$ behaves as $1/\Omega$  for $\Omega\gg 1/\ti$, with different plasma frequencies in the prefactor of the $1/\Omega$ term. As a result, $\sigma''(\Omega)$
 has
 a knee at $\Omega\sim 1/\te$, see
 Fig.~\ref{fig:2band_imp}, right.

\subsubsection{ {\em dc} limit}
\label{sec:2bdc}
We now briefly discuss
the static limit,  where Eq.~(\ref{resigma_imp})
 yields
  the {\em dc} resistivity in the following form
\beq
\rho=1/\sigma'(0)=4\pi
\frac{\gamma_0^2}{\frac{\Omega_1^2}{\tau_2}+\frac{\Omega^2_2}{\tau_1}+\frac{\Omega_0^2}{\te}}.
\eeq
Suppose that $1/\te$
is a monotonically increasing function of the temperature with $1/\te |_ {T=0}=0$, while $\tau_1$ and $\tau_2$ are $T$-independent.
In the low-$T$ regime, $1/\te\ll 1/\tau_1,\;1/\tau_2$; in the high-$T$ regime,  $1/\te\gg 1/\tau_1,\;1/\tau_2$.
The low- and high-$T$ limits of the resistivity
are
\bse
\bea
\rho|_{T=0}&=&4\pi\frac{1}{\Omega_1^2\tau_1+\Omega_2^2\tau_2},\label{rhoa}\\
 \rho|_{T\to \infty}&=&\frac{4\pi}{\Omega_0^2}
 \left(\frac{1}{\tau_1}\frac{1}{1+\frac{n_2m_2}{n_1m_1}}+\frac{1}{\tau_2}\frac{1}{1+\frac{n_1m_1}{n_2m_2}}\right).\label{rhob}
 \eea
 \ese
 These results allow for a transparent physical interpretation. At $T=0$, inter-band {\em ee} scattering is absent, the two bands conduct independently, and the total resistance is equal to  that of a circuit with two bands connected in parallel. At $T\to\infty$,  inter-band {\em ee} scattering is the strongest mechanism. Momentum gained from the electric field is  re-distributed quickly between the bands and then relaxed slowly within each band. The effective circuit for this case corresponds to two bands connected in series but weights of the two resistances that depend on  the number densities and masses.

 From Eq.~(\ref{rhob}),
we see that as long as a metal is not compensated, i.e., $\Omega_0\neq 0$,
the {\em ee} contribution to the resistivity does not grow unboundedly with temperature (as it does in the Baber case) but saturates at high temperatures.
Both the low-
  and high-$T$ limits of $\rho$ are controlled by the momentum-relaxing scattering rate.
 The ratio $ \rho|_{T\to \infty}/\rho|_{T=0}$ is determined by the ratio of the effective masses of the bands \cite{maslov:2011,pal:2012b} and can be large in transition and heavy-fermion metals, but is {\em finite}. \footnote{Explanations of a large (compared to the residual resistivity) $T$-dependent part of the resistivity of multi-band but decompensated metals, e.g., SrTiO$_3$, \cite{klimin:2012} by Baber scattering do not take this point account.}

 Recent experiments on a quantum paraelectric SrTiO$_3$\cite{kamran:2015,mikheev:2016}   have posed an interesting puzzle: the {\em dc} resistivity has been found to have a very pronounced $T^2$ term even at very low doping, when umklapps are essentially impossible and
 only one of the three conduction bands is occupied. The magnitude of this term far exceeds the theoretical predictions for a single-band non-Galilean--invariant FL.\cite{maiti:unpub}
 More work is needed, however, before one can say whether the $T^2$ term comes  from {\em ee} interactions or has a different origin, such as scattering from soft phonons modes,\cite{epifanov:1981a,epifanov:1981}
 characteristic for this material.

\subsection{Additional cancelations due to special geometry of the Fermi surface }
\label{sec:extra}
In Sec.~\ref{sec:BE}, we argued
that
since
 the change in total
 particle flux
 [Eq.~(\ref{DJ})]
 does not vanish identically
 in a non-Galilean--invariant system,
 normal {\em ee} scattering
gives rise to a non-zero dissipative part of the conductivity, $\sigma'(\Omega,T)\propto 1/\Omega^2\te$. We now relax the semiclassical condition of $\Omega$ being the smallest energy scale in the problem
and allow $\te$ to depend both on $\Omega$ and $T$.  In a FL, the (inverse) relaxation time scales as $1/\tau_{\text{sp}}\propto \max\{\Omega^2,T^2\}$.
If $\te$ entering the conductivity
were the same as $\tau_{\text{sp}}$, the optical conductivity of a non-Galilean-invariant system would behave as
\beq
\sigma'(\Omega,T)\propto \frac{ \max\{\Omega^2,T^2\}}{\Omega^2},
\label{opt2}
\eeq
while the {\em dc} resistivity in the presence of disorder would be given by
\beq
\rho=\rho_{\text{i}}+4\pi^2A_{\text{dc}}T^2,\label{dc}
\eeq
where $\rho_{\text{i}}$ is the residual resistivity and the factor of $4\pi^2$ is singled-out for further convenience.
 This is not always the case, however,
because
the geometry of the FS may lead to
 additional cancelations of the leading,
 ${\cal O}\left(\max\{\Omega^2,T^2\}\right)$ term in Eq.~(\ref{opt2}) and of the $T^2$ term in Eq.~(\ref{dc}).
 The study of such ``geometrical'' cancelations has a long history
  \cite{gurzhi:1980,gurzhi:1982,gurzhi:1987,maebashi:1998,rosch:2005,rosch:2006,maslov:2011,pal:2012,pal:2012b,briskot:2015}
   and has been reviewed recently in Ref.~\onlinecite{pal:2012b}.
   Here, we only
     list the results.
     In 2D, the cancelation of the
     leading terms occurs for any convex and simply connected FS, such as the one
      in the tight-binding model with sufficiently weak next-to-nearest neighbor hopping. In this case, momentum conservation $\bk+\bp=\bk'+\bp'$ for electrons on the FS can only be satisfied in processes
      that
      either swap the initial and final states ($\bk'=\bp$, $\bp'=\bk$) or occur in the Cooper channel ($\bk+\bp=0=\bk'+\bp'$). In both cases, $\Delta{\bf J}$ in Eq.~(\ref{DJ}) vanishes. For a circular FS, this result is almost self-evident but
      hinges on a simple geometric fact that two circles can have at most two intersection points. But then the same is true for any convex contour in 2D, and hence $\Delta{\bf J}$ vanishes for a FS of this type.  To get a non-zero result, one needs to include the states
      further away from the FS, which costs an extra factor of
      ${\cal O}(\max\{T^2,\Omega^2\}$,
       and the resulting contribution to the optical conductivity scales as
       $\max\{T^4,\Omega^4\}/\Omega^2$, while the $T^2$ term in the {\em dc} resistivity is replaced by $T^4$.
        In 3D, the restrictions are less severe: as long as one keeps quartic (and higher) terms in the dispersion, there is no cancelation. In what follows, we will assume that the FS is such that geometric cancelations
        of this kind
         do not happen.

\section{Optical conductivity of Fermi-liquid metals}
\label{sec:FL}
\subsection{Gurzhi formula}
\label{sec:gurzhi}
\subsubsection{Kubo formula without vertex corrections}
\label{sec:novert}
As we showed in the previous section,
 the optical--as opposed to {\em dc}--conductivity
of a non-Galilean-invariant system
is finite even in the absence of umklapp scattering: normal scattering
suffices.  In
the
previous section, however, we treated the
{\em ee} scattering rate,$1/\te$,
as
a
phenomenological
parameter,
 borrowing the knowledge of its $T$- and $\Omega$-dependences from the microscopic theories.
  In this section, we review the microscopic theory for the optical conductivity of a FL.
 The main result of this theory
 is that $1/\tau_{ee}$, which appears in the formula for optical conductivity at high frequencies,
 $\sigma'(\Omega,T) \propto 1/\Omega^2 \te$, scales as $\max\{\Omega^2,T^2\}$ and, moreover, the two dependences
 are described by
a universal form
  $1/\te
  \propto \Omega^2 + 4 \pi^2 T^2$.
  This result
  follows from the formula for the optical conductivity of a 3D FL,\cite{gurzhi:1959} derived by Gurzhi in 1959 from the quantum Boltzmann equation:
   \beq
\sigma'(\Omega)=\text{const}\times \frac{\Omega^2+4\pi^2 T^2}{\Omega^2} = \text{const} \left(1 + \frac{4\pi^2 T^2}{\Omega^2} \right).\label{gurzhi}
 \eeq

\begin{figure}[h]
\includegraphics[width=0.8 \linewidth]{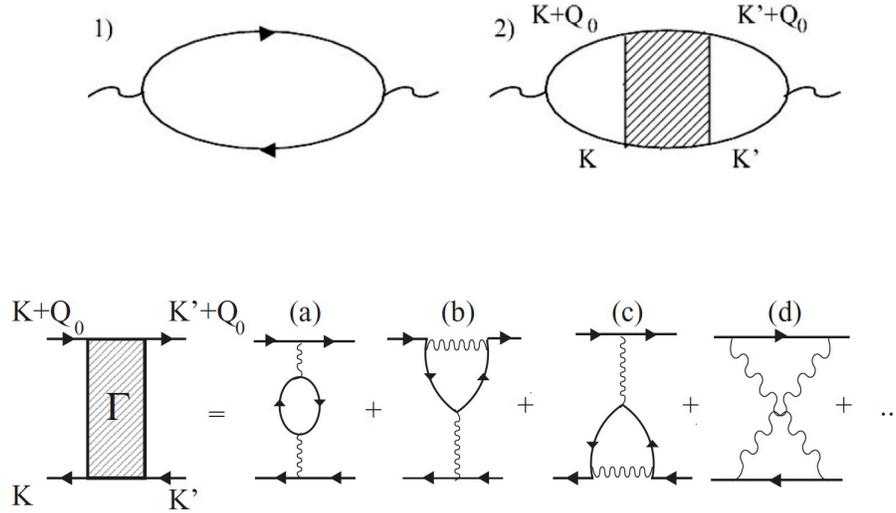}
\vspace{-0.75in}
\caption{Top: diagrammatic representation of the Kubo formula for the conductivity. Bottom: examples of vertices contributing to diagram 2 on the top.
{\em Adapted with permission from Ref.~\onlinecite{maslov:2012}. Copyrighted by the American Physical Society.}\label{fig:kubo_vertex}}
\end{figure}

To reproduce the Gurzhi result,  we use the Kubo formula, which relates the conductivity to the current-current correlation function, ${\cal K}^R(\Omega,T)$:
\cite{mahan}
\bea
\sigma'(\Omega)=\frac{e^2}{\Omega}\I{\cal K}^R(\Omega,T).
\eea
Here and thereafter, the superscript $R$ denotes a retarded version of a certain quantity.
For simplicity, we assume
 a cubic
 lattice in the $D$-dimensional space,
 so that
 the conductivity tensor, $\sigma_{ij}(\Omega)$, is diagonal and symmetric, and define
 $\sigma(\Omega)\equiv \sum_{i=1}^D\sigma_{ii}(\Omega)/D$.
Diagrammatically,
the current-current correlation function
 is
given by a fully renormalized particle-hole bubble
at zero momentum
and
finite frequency,
with
group
 velocities $
\bv_\bk$
at the vertices.
Without loss of generality, ${\cal K}^R(\Omega,T)$ can be split into parts:
with and without vertex corrections, ${\cal K}^R_
2(\Omega,T)$ and ${\cal K}^R_1(\Omega,T)$, correspondingly (see Fig.~\ref{fig:kubo_vertex}, top).  The lines in these diagrams denote full Green's functions, with self-energy corrections included.

First, we compute ${\cal K}^R_1(\Omega,T)$
and then show that the functional form of $1/\te(\Omega, T)$ does not change
 if
vertex corrections are included.
To obtain $\I{\cal K}_1^R(\Omega,T)$, we first find the current-current correlation function
 ${\cal K}_1 (\Omega_n,T)$ at the
 discrete
 Matsubara frequencies, $\Omega_n=2\pi nT$,
 and then obtain
    ${\cal K}_1^R(\Omega,T)$ by analytic continuation, $\Omega_n\to -i\Omega+i0^+$.
Along the Matsubara axis, we have
\bea
{\cal K}_1(\Omega_n,T)=-\frac{2}{(2\pi)^DD}T\sum_{\omega_m}\int d\e_\bk \oint da_\bk
v_\bk
G_{\bk}(\omega_m)G_\bk(\omega_m+\Omega_n),\label{kubo_m}
\eea
where $\omega_m=\pi (2m+1) T$
and $da_{\bk}$ is the element of an isoenergetic surface
($d^Dk=d\e_\bk da_\bk/v_\bk$).
 Note that we define the self-energy
  with
  the sign opposite to that in the traditional definition, e.g.,
   in our definition $G_{\bk}(\omega_m)=\left[i\omega_m-\e_\bk+\Sigma_\bk(\omega_m,T)\right]^{-1}$.

\begin{figure}[h]
\includegraphics[width=0.8 \linewidth]{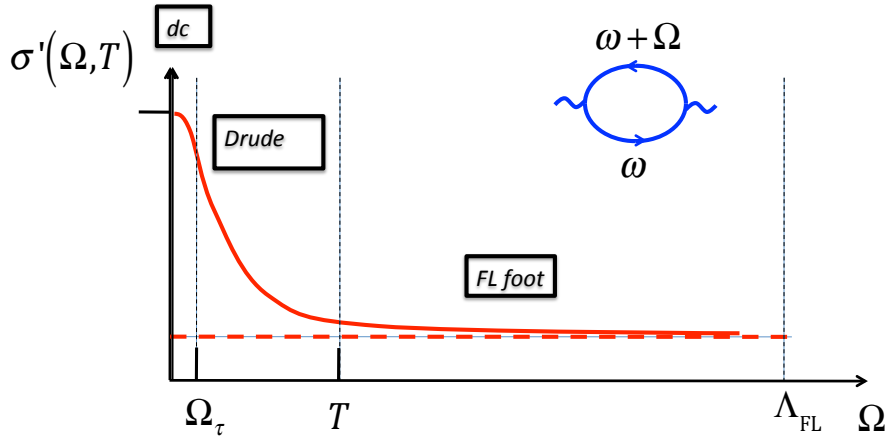}
\vspace{-1.5in}
\caption{Optical conductivity of a Fermi liquid. $\Omega_\tau\sim 1/\te(\Omega=0,T)\propto T^2$.  \label{fig:foot}}
\end{figure}

For definiteness,
we consider a 3D FL,
  the self-energy of which at low frequencies and temperatures is given by
\bea
 \R\Sigma_{\bk}^R(\omega)&=&\omega(Z_{\bk}^{-1}-1),\nn\\
\I\Sigma_{\bk}^R(\omega,T)&=&A_{\bk}(\omega^2+\pi^2 T^2),\label{FLS}
\eea
where $Z_\bk$ is the quasiparticle renormalization factor at point $\bk$ on the  FS. The real part of the self-energy
may
also contain
a term which depends
 linearly on the quasiparticle dispersion $\e_\bk$.  Such a term, however, accounts only for an overall renormalization of the conductivity, and we
 neglect it.
Transforming the Matsubara sum in Eq.~(\ref{kubo_m}) into the integral over the real axis and integrating over $\e_\bk$, we obtain
\bea
\sigma'(\Omega)=\frac{8e^2}{D(2\pi)^D}\oint da_{\bk}\; v
_{\bk}\int^\infty_{-\infty} \frac{d\omega}{\Omega}\frac{\left[n_F(\omega)-n_F(\omega+\Omega)\right]\left[\I\Sigma_{\bk}^R(\omega,T)+\I\Sigma_{\bk}^R(\omega+\Omega,T)\right]}{\left[\Omega+\R\Sigma_{\bk}^R(\omega+\Omega,T)-\R\Sigma_{\bk}^R(\omega,T)\right]^2+\left[\I\Sigma_{\bk}^R(\omega+\Omega,T)+\I\Sigma_{\bk}^R(\omega,T)\right]^2},\nn\\
\label{1b}
\eea
 where $n_F(\omega)$ is the Fermi function.
 In a FL,
$\Omega/Z_{\bk} \gg \I\Sigma_{\bk}^R(\Omega,T)$.
 Then
  one can neglect the imaginary parts of the self-energy in the denominator of Eq.~(\ref{1b}). After averaging the imaginary parts of the self-energies in the numerator of Eq.~(\ref{1b}) with the Fermi functions, we
  arrive at
\bea
\sigma'(\Omega)&=&\frac{8e^2 }{D(2\pi)^D}\oint da_\bk\; v
_{\bk} Z_{\bk}^2\int^{\infty}_{-\infty}d\omega \frac{n_F(\omega)-n_F(\omega+\Omega)}{\Omega^3}\left[\I\Sigma_{\bk}^R(\omega,T)+\I\Sigma_{\bk}^R(\omega+\Omega,T)\right]= \nonumber \\
&=& \frac{8e^2 }{D(2\pi)^D}\oint da_\bk\; v
_{\bk} Z_{\bk}^2 A_{\bk} \int^{\infty}_{-\infty}d\omega \frac{n_F(\omega)-n_F(\omega+\Omega)}{\Omega^3}\left[2 \pi^2 T^2 + \omega^2 + (\Omega + \omega)^2 \right] \nonumber\\
&=& \sigma_0\frac{\Omega^2+4\pi^2 T^2}{\Omega^2},
\label{4pi2}
\eea
where $\sigma_0=(16e^2/3D) \oint da_{\bk} v_{\bk}Z_{\bk}^2A_{\bk}$. For $\Omega\ll T$, $\sigma'(\Omega,T)$ assumes a Drude form ($\sigma'\propto 1/\Omega^2\te$) with $1/\te=2\langle\I\Sigma_{\bk}^R(0,T)\rangle
 \propto T^2$,
  where $\langle\dots\rangle$ indicates averaging over the FS.  For $\Omega\gg T$, the $\Omega^2$ term in self-energy cancels with the $\Omega^2$ term in the denominator, and $\sigma'$ saturates at a frequency-independent value (the ``FL foot", see Fig.~\ref{fig:foot}). The foot continues up to frequency $\Lambda_{\text{FL}}$
at which the FL description breaks down. At higher frequencies,  the behavior of $\sigma'$ is non-universal;\cite{berthod:2013} at even higher frequencies, comparable to the bandwidth, the situation is further complicated by interband transitions, which can sometimes mimic a non-FL behavior.\cite{dang:2015}  The FL foot is seen, for example, in the optical conductivity of a heavy-fermion material UPd$_2$Al$_3$,\cite{scheffler:2005,berthod:2013} and organic conductors  $\beta$-(BEDT-TTF)$_2$AuI$_2$/I$_2$Br\cite{jacobsen:1987} and $\kappa$-(BEDT-TTF)$_2$Cu[N(CN)$_2$]Br$_x$Cl$_{1-x}$.\cite{dumm:2009,dressel:2011}

\subsubsection{Vertex corrections}
\label{sec:vertex}
\begin{figure}[h]
\includegraphics[width=0.8 \linewidth]{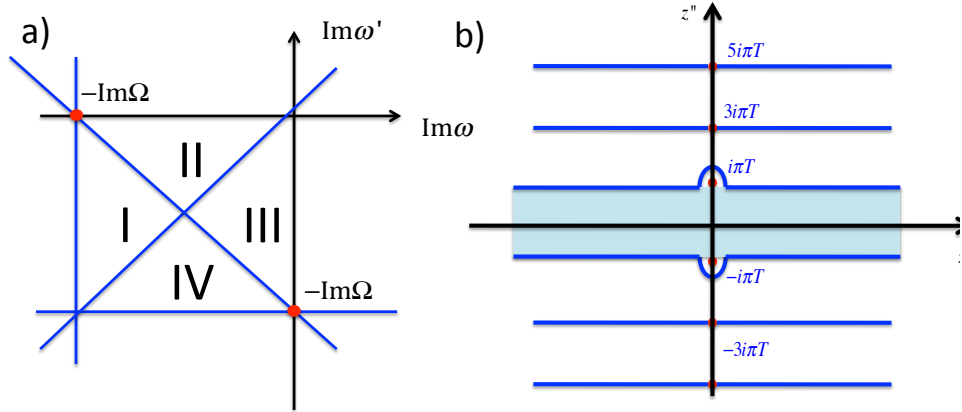}
\vspace{-1.5 in}
\caption{a) Regions in the complex plane in which the vertex functions $\Gamma_{\bk,\bk'}^{\text{I-IV}}$ are defined.  b) Analytic structure of the imaginary part of the self-energy continued to the complex plane. {\em Adapted with permission from Refs.~\onlinecite{chubukov:2012} and \onlinecite{maslov:2012}. Copyrighted by the American Physical Society.}\label{fig:cuts}}
\end{figure}

\paragraph{{\bf Vertex-correction diagrams.}}
While neglecting the vertex corrections simplifies the derivation of the Gurzhi formula, it is by no means a necessary condition for its validity. Originally, the Gurzhi formula was derived from the quantum Boltzmann equation, which takes into account vertex corrections automatically.\cite{gurzhi:1959}
In the diagrammatic approach,
 one can
 show that any vertex-correction diagram for the conductivity, such as the ones in the bottom panel of Fig.~\ref{fig:kubo_vertex}, produces a contribution of the same form as in Eq.~(\ref{1b}). Indeed, diagram 2 in Fig.~\ref{fig:kubo_vertex}, top) reads
\bea
{\cal K}_2(\Omega_n,T)=\frac{2}{D(2\pi)^{2D}}
 T^2\sum_{\omega_m,\omega_{m'}}
\oint \frac{da_\bk}{v_\bk}\oint \frac{da_{\bkp}}{v_{\bk'}}&&\bv_{\bk}\cdot\bv_{\bk'}G_{\bk}(\omega_m) G _{\bk}(\omega_m+\Omega_n)G_{\bk'}(\omega_{m'})
 G_{\bk'}(\omega_{m'}+\Omega_n)\nn\\
 &&\times\Gamma_{\bk,\bk'}(\omega_m,\omega_{m'},\Omega_n),
 \label{16}
 \eea
where $\Gamma_{\bk,\bk'}(\omega_m,\omega_{m'},\Omega_n)$ is the four-leg vertex, which we assume to depend on the directions of $\bk$ and $\bk'$ but not on their magnitudes.  Employing
the  Eliashberg procedure of analytic continuation,\cite{eliashberg:1962} integrating over $\e_\bk$ and $\e_{\bkp}$ with the assumption of a local self-energy, and neglecting the imaginary parts of the self-energies in the numerators of the resulting integrals
[this corresponds to the same
assumption
 that we used to arrive at Eq.~(\ref{4pi2})], we obtain for the vertex part of the conductivity
 \beq
\sigma_V'(\Omega)=\text{const}\times \oint da_\bk \oint da_{\bk'}\frac{\bv_{\bk}\cdot\bv_{\bk'}}{v_\bk v_{\bk'}}Z_\bk Z_{\bk'}
\int^\infty_{-\infty} \frac{d\omega}{\Omega^3}\int^\infty_{-\infty} d\omega'\left[n_F(\omega)-n_F(\omega+\Omega)\right]
\I\Gamma_{\bk,\bk'}(\omega,\omega',\Omega),\label{vert4}
\eeq
where
 \bea
 \Gamma_{\bk,\bkp}(\omega,\omega',\Omega)=\coth\frac{\omega'-\omega}{2T}\left(\Gamma^{\mathrm{II}}_{\bk,\bkp}-\Gamma^{\mathrm{III}}_{\bk\bkp}\right)+\coth\frac{\omega+\omega'+\Omega}{2T}\left(\Gamma_{\bk,\bkp}^{\mathrm{III}}-\Gamma_{\bk,\bkp}^{\mathrm{IV}}\right)-\tanh\frac{\omega'}{2T}\Gamma_{\bk,\bkp}^{\mathrm{II}}+\tanh\frac{\omega'+\Omega}{2T}\Gamma_{\bk,\bkp}^{\mathrm{IV}}.\nn\\
 \label{vert1}
 \eea
 Vertices $\Gamma_{\bk,\bkp}^{\text{II-IV}}$ (which are functions of $\omega$, $\omega'$, and $\Omega$) are obtained by analytically continuing the Matsubara vertex into the corresponding regions of the $(\I\omega,\I \omega')$ plane for $\I\Omega>0$, as shown in Fig.~\ref{fig:cuts}a. As an example, diagram {\em a} in the bottom panel of Fig.~\ref{fig:kubo_vertex} gives
 \bea
 \I\Gamma_{\bk\bkp}(\omega,\omega',\Omega)
 =-2 U_{\bk-\bk'}^2 \I \Pi^R_{\bk-\bk'}(\omega-\omega')\left[2n_B(\omega'-\omega)+n_F(\omega')+n_F(\omega'+\Omega)\right],
 \label{vert3}
  \eea
  where $U_{\bq}$ is the (static) interaction corresponding to the wavy line, $n_B(\omega)$ is the Bose function and
  \bea
  \Pi^R_{\bq}(\Omega)=\frac{2}{(2\pi)^D}\oint \frac{da_\bk}{v_\bk} \int d\e_\bk\frac{n_F(\e_\bk)-n_F(\e_{\bk+\bq})}{\Omega-\e_{\bk+\bq}+\e_{\bk}+i0^+}
  \eea
is the particle-hole polarization bubble. Now we recall that $\I\Pi^R_{\bq}(\Omega)\propto \Omega$ for $\Omega\ll v_Fq$, relabel $\Omega'=\omega'-\omega$, and rewrite the double integral over $\omega$ and $\omega'$ in Eq.~(\ref{vert4}) as
\bea
\int^\infty_{-\infty} d\omega \left[n_F(\omega)-n_F(\omega+\Omega)\right]\left[S(\omega)+S(\omega+\Omega)\right],
\eea
where
\bea
S(\epsilon)\equiv \int^\infty_{-\infty} d\Omega'\Omega'\left[n_B(\Omega')+n_F(\epsilon+\Omega')\right]=\frac{1}{2}\left(\epsilon^2+\pi^2T^2\right)
\eea
is the same integral that appears in the imaginary part of the self-energy. Therefore, the contribution from diagram {\em a} has the same dependence on $\Omega$ and $T$ as Eq.~(\ref{4pi2}). The remaining diagrams can also be shown to give the same contribution.\cite{maslov:2012}
Following the Eliashberg's proof that an arbitrary order diagram for the self-energy produces  the combination of $\omega^2+\pi T^2$,\cite{eliashberg:1962b} one can also prove that any diagram for the conductivity scales with $\Omega$ and $T$ as specified by Eq.~(\ref{4pi2}).\cite{maslov:2012}

\paragraph{{\bf Physical consequences of vertex corrections.}}
 Because
 vertex corrections do not seem to
  affect the scaling form of the conductivity,  it is widely accepted \cite{dressel:book}
  that to get a reasonably accurate (within a ``factor of two'')  description of the conductivity,
  one can consider
   only the bare bubble diagram for the current-current correlator
   (diagram 1 in Fig.~\ref{fig:kubo_vertex}).
 This is generally true, but there are
  two caveats, which
   we discuss below.

First,
we recall that
$\I\Sigma^R\propto (\omega^2+\pi^2T^2) \ln \left(
\omega
^2 + \pi^2 T^2\right)$ in 2D
 contains an extra logarithm compared to the expression for $\I\Sigma^R$ in 3D.
This extra
 logarithm
   comes from a special subset of scattering processes, which are essentially one-dimensional (1D): \cite{chubukov:2003,suhas:2005b,chubukov:2005b} the two interacting fermions move on either almost parallel or antiparallel trajectories both before and after the collision. However, a change in total current due to such processes [$\Delta{\bf J}$ in Eq.~(\ref{DJ})]
is negligibly small, and thus
  these
  almost
  1D processes
  should not contribute to the conductivity. It turns out
  that
  it
   is the vertex corrections that cancel the contribution to $\sigma' (\Omega)$ from  1D processes.\cite{maslov:2012}
   As a result, $\sigma'(\Omega)$ of a 2D FL still scales
   as
   $\Omega
   ^2$ and $T
   ^2$,
   as
   in the Gurzhi formula, Eq.~(\ref{gurzhi}), without extra logarithmic factors.

Second, Eq.~(\ref{1b}) formally predicts a finite conductivity with the relaxation time $1/2\langle\I\Sigma_{\bk}^R(0,T)\rangle$ even at $\Omega=0$. This is
 an artifact of neglecting the vertex corrections. If only normal processes are allowed.  The deviations from the behavior predicted by Eq.~(\ref{1b})
 occur at $\Omega\lesssim\Omega_\tau$, where $\Omega_\tau\sim\langle\I\Sigma_{\bk}^R(0,T)\rangle\propto T^2$.
 (In Sec.~\ref{sec:two_band} we discussed these deviations for the case of a two-band metal.) In the presence of umklapp scattering, however, the {\em dc} conductivity is finite, and then Eq.~(\ref{1b}) is qualitatively correct  at all frequencies, provided that  $\I\Sigma_{\bk}^R(\omega,T)$ is replaced by the transport scattering rate. One needs to keep in mind, however, that the coupling constants of {\em e-e} interaction entering the conductivity in the {\em dc} ($\Omega\ll\Omega_\tau$) and high-frequency ($\Omega\gg \Omega_\tau$) limits are, in general, different.
Indeed, the {\em dc} conductivity is finite only in the presence of umklapp processes but, once they are allowed, normal processes can also contribute.\cite{maebashi:1998} The Matthiessen rule in this regime in violated because the total scattering rate is not the sum of the umklapp and normal contributions.  The {\em dc} resistivity can be written as $\rho_{\text{dc}}=A_{\text{dc}} 4\pi^2 T^2$, where $A_{\text{dc}}$ depends on the effective coupling constants for normal and umklapp scattering ($g_{\text{n}}$ and $g_{\text{u}}$, correspondingly) as
\beq
A_{\text{dc}}=\text{const}\times g_{\text{u}} f\left(\frac{g_{\text{n}}}{g_{\text{u}}}\right),
\eeq
where $f(x)$ takes constant and, in general, different values at $x\to 0$ and $x\to \infty$.
On the contrary,
the high-frequency conductivity contains contributions of both umklapp and normal scattering processes, which add up in a Matthiessen-like way, so that the optical resistivity
\beq
\rho_{\text{opt}}(\Omega,T)\equiv \R{\sigma^{-1}(\Omega,T)}
\eeq
can be written as
\beq
\rho_{\text{opt}}(\Omega,T)=A_{\text{opt}}(\Omega^2+4\pi^2 T^2)
\label{rho_opt}
\eeq
with $A_{\text{opt}}=\text{const}\times\max\{(g_{\text{n}},g_{\text{u}}\}$.
This means that the prefactors of the $T^2$ terms in the {\em dc} and optical resistivities, in general,  are supposed to be different.
In some cases, e.g., in Nd$_{1-x}$TiO$_3$\cite{yang:2006} and CaRuO$_3$,  \cite{schneider:2014}
the slopes of the $T^2$ terms in the {\em dc} and optical resistivities are indeed markedly different.
However, this difference may be partially due by a systematic error in the absolute value of the {\em dc} resistivity,
related to uncertainties in the sample size and shape.\cite{timusk:unpub}

\section{First-Matsubara-frequency rules}
\label{sec:rules}
Comparing the scaling forms of the self-energy and optical resistivity [Eqs.~(\ref{FLS}) and (\ref{rho_opt})], one notices that they are {\em not} identical: the universal numerical prefactor of the $T^2$ term
in the self-energy is $\pi^2$, whereas  that  in the optical resistivity is $(2\pi)^2=4\pi^2$.
 In this section we argue
  that this
 difference is not coincidental.
Specifically, we show
 that the $T$-dependent part of $\I\Sigma_{\bk}^R(\omega,T)$, which  measures the decay rate of fermion
  quasiparticles,
 contains the square  of first fermion Matsubara frequency ($=\pi T$),
while the $T$-dependent part of
$\rho_{\text{opt}}(\Omega, T)$, which  measures the decay rate of current fluctuations, which are bosons,
 contains the square of the  first non-zero boson Matsubara frequency ($=2\pi T$).
 Also
  not coincidentally, Eq.~(\ref{rho_opt}) is of the same form as the sound absorption rate in a FL.~\cite{physkin}

 In more general terms, we argue that
 the scaling forms,
 given by Eqs.~(\ref{FLS}) and (\ref{rho_opt}),
  are manifestations of quite general
  ``first-Matsubara-frequency rules" (FMFR)
    that must be satisfied by the self-energy and optical conductivity of any electron system, not necessarily of a FL.
   \cite{martin:2003,chubukov:2012,maslov:2012}
  In the context of electron-phonon interaction, such a rule for the fermion self-energy
  is known as  the ``Fowler-Prange theorem''. \cite{Fowler:1965,engelsberg:1970,engelsberg:1978, wasserman:1996}
 In Secs.~\ref{sec:1stM} and \ref{sec:1stMsigma}, we analyze FMFR  from the theoretical point of view.
    In Sec.~\ref{sec:gurzhi_exp},
 we discuss the
  experimental status of the Gurzhi formula for the optical conductivity,
  which is a consequence of FMFR.

\subsection{First-Matsubara-frequency rule for the self-energy}
\label{sec:1stM}
\subsubsection{Self-energy on the Matsubara axis}
\paragraph{{\bf Formulation of the rule.}}
For a FL in $D>2$,
the
FMFR
states that
 the
 Matsubara self-energy,
 evaluated at the
first fermion frequency
$\omega
 = \pm \pi T$,
 contains
  a linear-in-$T$ term
   but no $T^2$ term:
 \beq
\Sigma_{\bk}(\pm\pi T,T)=\pm i
\pi
\lambda_{\bk} T+
0
\times
T^2 +
R(T),
\label{rule1}
\eeq
where $\lambda_\bk$ is related to the quasiparticle residue as $Z_{\bk}=1/(1+\lambda_{\bk})$.
In general,
the remainder $R(T)$
scales as $T^D$, but
the
prefactor of the $T^D$  term vanishes
in the
``local
approximation'',\cite{chubukov:2012}
in which
the fully renormalized interaction between quasiparticles
is replaced
by its value for the initial
and final fermion states right on the Fermi surface.
In $D=2$, the rule is modified to the extent that $\Sigma_\bk(\pm
\pi
T,T)$ contains a $T^2$ term but no $T^2\ln T$ term.

For all but the first
Matsubara frequencies,
$\Sigma_{\bk}(
\omega_m, T)$
contains
a
 $T^2$ term
 in $D >2$ and
 a
 $T^2 \ln T$ term
 in $D=2$.
  Equation
   (\ref{rule1}) is also applicable to
  NFLs, the only difference is that
   the coefficient $\lambda_\bk$
    in a NFL
  itself depends on the temperature,
  and the
  form
  of
  the
  remainder changes.
    Two particular cases of NFLs, a marginal FL\cite{varma:1989,varma:2002}  and a Hertz-Millis-Moriya quantum-critical metal,\cite{hertz:1976,millis:1993,moriya:1985}
 are discussed later in this section and in Appendix~\ref{sec:remainder}, correspondingly.

In dimensions $1<D <2$,
FMFR  is not satisfied even for FLs:
the next-to-linear term in
$\Sigma_{\bk}(\pm\pi T,T)$  is of the same order as
in $\Sigma_{\bk}(
\omega_m,T)$ with $
m\neq 0,-1$.
\begin{figure}[h]
\includegraphics[width=0.8 \linewidth]{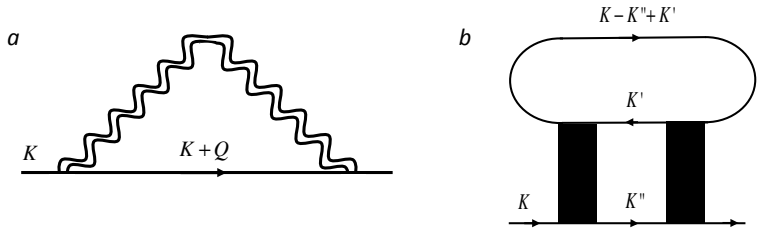}
\vspace{-2.5 in}
\caption{{\em a} One-loop self-energy in Eq.~(\ref{sigmaM}).  {\em b}) A general self-energy diagram with $K=(\bk,\omega_m)$, $K'=(\bk,\omega_{m'})$, and $K''=(\bk'',\omega_{m}+\Omega_n)$. \label{fig:selfenergy}}
\end{figure}

\paragraph{{\bf One-loop order.}}
The analysis of the remainder $R(T)$ requires special care and is presented in
Appendix~\ref{sec:remainder},  but the derivation of the leading term
in Eq.~(\ref{rule1})
  is quite straightforward. For example, a one-loop diagram for the Matsubara self-energy (Fig.~\ref{fig:selfenergy}a)
 reads
\bea
\Sigma_\bk(\omega_m,T)=T\sum_{\Omega_n} \int \frac{d^{D} q}{(2\pi)^D}G_{\bk+\bq}(\omega_m+\Omega_n) \chi(\bq,\Omega_n),\label{sigmaM}
\eea
where $\chi(\bq,\Omega_n)$ is some dynamic interaction (double wavy line), which can represent, e.g., a phonon, screened Coulomb potential, spin fluctuation, etc.
 Suppose that
$\chi$ decreases dramatically for $q$ above some scale $\Lambda$,
 then typical momentum transfers are $\lesssim\Lambda$. We are interested in the low-energy dynamics, when $\max\{\omega,T\}\ll v_F\Lambda$. Then the projection of $\bq$ on the normal to the FS, $q_{\perp}$, is on the order of $ \max\{\omega,T\}/v_F$, and is thus smaller that than the projection on a plane tangential to the FS, $q_{||}\sim \Lambda$. Since $q_{||}\gg q_{\perp}$,
 the leading term in  the self-energy
can be obtaining by
 replacing
 $q$ by $q_{||}$ in $\chi(\bq,\Omega_n)$. Consequently, the integral over $\bq$ factorizes into a one-dimensional integral over $q_{\perp}$ and a  $D-1$-dimensional integral over the  tangential plane.  The 1D integral involves only the Green's function and gives
\beq
\int \frac{dq_{\perp}}{2\pi} G_{\bk+\bq}(\omega_m+\Omega_n)=\int \frac{dq_{\perp}}{2\pi} \frac{1}{i(\omega_m+\Omega_n)-\e_\bk-v_F q_{\perp}-q_{||}^2/2m^*}=-\frac {i}{2v_F} \text{sgn}(\omega_m+\Omega_n),\label{qpar}
\eeq
where $v_F$ and $m^*$ are the
  Fermi velocity and effective mass at point $\bk$, correspondingly.
The $D-1$-dimensional integral over $q_{||}$ gives a local form of the interaction:
\beq
\chi_{\text{loc}}(\Omega_n)=\int \frac{d^{D-1}q_{||}} {(2\pi)^{D-1}}\chi(\bq_{||},\Omega_n).\label{vloc}
\eeq
 We thus arrive at
\bea
\Sigma_\bk(\omega_m,T)=-\frac{i}{2v_F}T\sum_{\Omega_n}\text{sgn}(\omega_m+\Omega_n) \chi_{\text{loc}}(\Omega_n).
\eea
For the remainder of the proof it matters only that $\chi_{\text{loc}}$ is an even function of $\Omega_n$. Using this property and singling out the $\Omega_n=0$ term, we re-arrange the sum in the equation above as
\bea
\Sigma_\bk(\omega_m,T)=
-
\frac{i}{2v_F}\text{sgn}\omega_m T\left[\chi_{\text{loc}}(0)+2\sum_{\Omega_n=2\pi T}^{|\omega_m|-\pi T} \chi_{\text{loc}}(\Omega_n)\right].\label{sum}
\eea
For $\omega_m=\pm \pi T$,  the sum vanishes and
\beq\Sigma_\bk(\pm \pi T,T)=\mp \frac{i}{2v_F}T\chi_{\text{loc}}(0),
\label{sum1}
\eeq
which yields
 the leading term in Eq.~(\ref{rule1}) with
 $\lambda_\bk=-\chi_{\text{loc}}(0)/2
 \pi
 v_F$.   For other $\omega_m$, the sum in Eq.~(\ref{sum}) does not vanish and gives rise to additional terms in $\Sigma_\bk(\omega_m,T)$ besides
 the
 $
 \pm i
 \lambda_\bk T$ one.

\paragraph{{\bf Examples.}}
FMRF
 can be verified for a number of
 specific models
  with
   different forms of $\chi_{\mathrm{loc}}(\Omega_n)$.
   For example,
    if the boson in Eq.~(\ref{sigmaM}) corresponds to an Einstein phonon with frequency $\Omega_0$, then $\chi_{\mathrm{loc}}(\Omega_n)=\text{const}/(\Omega_n^2+\Omega_0^2)$.
(In this and other examples, we assume that the FS
is
 isotropic, and suppress the dependence of the self-energy on the direction of $\bk$.)  The Matsubara sum
 in Eq.~(\ref{sum})
 in this case can be calculated exactly: \cite{allen:1982}
\bea
\Sigma(\omega_m,T)&=&
\text{const}\times\text{sgn}\omega_m\left[-i\pi\coth\frac{\Omega_0}{2T}+\Psi\left(\frac{1}{2}+i\frac{\Omega_0}{2\pi T}+\frac{|\omega_m|}{\pi T}\right)-\Psi\left(\frac{1}{2}-i\frac{\Omega_0}{2\pi T}+\frac{|\omega_m|}{\pi T}\right)\right]\nn\\
&&=\text{const}\times\text{sgn}\omega_m\left[\frac{2\pi T}{|\omega_m|-\pi T+i\Omega_0}+\Psi\left(i\frac{\Omega_0}{2\pi T}+\frac{|\omega_m|-\pi T}{2\pi T}\right)-\Psi\left(i\frac{\Omega_0}{2\pi T}-\frac{|\omega_m|-\pi T}{2\pi T}\right)\right],\nn\\
\eea
where $\Psi(x)=d \ln \Gamma(x)/dx$ is the digamma function. [We used the identity $\Psi(1-z)=\Psi(z)+\pi\cot(\pi z)$ to obtain the result shown in the second line].  For $|\omega_m|=\pi T$, the two last terms cancel each other, whereas the first term reduces to a linear $T$ dependence.

Another familiar example is a FL with {\em ee} interaction, where bosons correspond to particle-hole excitations. The frequency dependence of $\chi$ comes from Landau damping: $\chi_{\mathrm{loc}}(\Omega_n)=\text{const}_1+\text{const}_2\times|\Omega_n|$. The first term ($\text{const}_1$)  gives an $\omega_m$ term in
the
self-energy,
 which is non-zero for all $\omega_m$.
The second term gives a
$T^2, \omega^2_m$  contribution, which vanishes for $\omega_m=\pm \pi T$:
\beq
\Sigma(\omega_m,T)=i\lambda\omega_m-\frac{i\text{const}_2}{2v_F}T\sum^{|\omega_m|-\pi T}_{\Omega_n=2\pi T}|\Omega_n|=i\lambda\omega_m-i\frac{\text{const}_2}{4\pi v_F}(\omega_m^2-\pi^2T^2).\label{flm}
\eeq

\paragraph{{\bf Beyond the one-loop order.}}
In fact, the FL case allows for a more rigorous treatment than the one presented above just for the one-loop diagram. Following the arguments by Luttinger\cite{luttinger:1961} and Eliashberg, \cite{eliashberg:1962b}one can show that any diagram for the self-energy gives a contribution that scales with $\omega_m$ and $T$
as
 specified by Eq.~(\ref{flm}). Details of the derivation are given in Ref.~\onlinecite{chubukov:2012}; here we just outline the main idea. First, one realizes that a FL contribution to the self-energy comes from any diagram that contains three internal fermions with energies close to the Fermi energy. All other fermions, which are away from the Fermi energy, renormalize the interaction between these three low-energy fermions. All diagrams of this type can be represented by diagram {\em b} in Fig.~\ref{fig:selfenergy}, in which the shaded rectangles denote the exact interaction vertices. Since the
 dynamics in the problem is
 already
 coming from the three Green's functions, the vertices
 can
  be taken as static.
 After this
 simplification, diagram {\em b}
 is reduced to
\bea
\Sigma(\omega_m,T)=T^2\sum_{\omega_{m'},\Omega_n}  \int \frac{d^Dk'}{(2\pi)^D}\int \frac{d^Dk''}{(2\pi)^D} G(\bk',\omega_{m'})G(\bk'',\omega_m+\Omega_n) G(\bk-\bk''+\bk',\omega_{m'}-\Omega_n)\Gamma^2(\bk,\bk',\bk'').\nn\\
\eea
Dispersions $\e_{\bk'}$, $\e_{\bk''}$, and $\e_{\bk-\bk''+\bk'}$ are assumed to be small, of order $\omega_m$ or $T$.
Within the same local approximation already employed for diagram {\em a},
the vertices
 can be then assumed to depend only on the directions of the incoming and outgoing momenta but not on their magnitudes.
The condition of $\e_{\bk-\bk''+\bk'}$ being small imposes a geometrical constraint on the angles between the momenta, e.g., on the angle between $\bk'$ and $\bk-\bk''$. We thus have three ``infrared'' integrals--over $\e_{\bk'}$, $\e_{\bk''}$, and over the angle--which are carried out in infinite limits and produce signs of the corresponding Matsubara frequencies, similar to the integral over $q_{\perp}$ in Eq.~(\ref{qpar}). Integrating over $\e_{\bk'}$ and over the angle, we obtain
 a combination
$\text{sgn}\omega_{m'}\text{sgn}(\omega_{m'}-\Omega_n)$.
Upon summation over $\omega_{m'}$,
 this combination
 produces a
  term, which
  can
  absorbed into the linear-in-
  $\omega_m$
   part of the self-energy, and an
    $|\Omega_n|$
     term.
      The
       latter  comes in a combination
\beq
\int d\theta_{\bk''}\left[T\sum_{\Omega_n}|\Omega_n|\int d\e_{\bk''}G(\bk'',\omega_m+\Omega_n)\right]\Gamma^2(\bk,\bk'\{\bk,\bk''\},\bk''),
\eeq
where the notation $\bk'\{\bk,\bk''\}$ means that the constraint on the angles has already been taken into account.
 The square brackets
in the equation above give a factor of
$\omega_m^2-\pi T^2$, which means that
 this
part of the self-energy vanishes at $\omega_m=\pm \pi T$, and thus the FMFR
  is satisfied.

\paragraph{{\bf Partial self-energy.}}
Notice that the combination $\omega_m^2-\pi T^2$ occurs {\em before} integrating over $\theta_{\bk''}$. This means that
 FMFR
can also be formulated for a {\em partial self-energy},  defined in the same way as the usual  self-energy but without the last angular integration:
\bea
{\cal S}_{\bk,\bk''}(\omega_m,T)=N_F T^2\sum_{\omega_{m'},\Omega_n} \int d\e_{\bk{''}}\int \frac{d^Dk'}{(2\pi)^D} G(\bk',\omega_{m'})G(\bk'',\omega_m+\Omega_n) G(\bk-\bk''+\bk',\omega_{m'}-\Omega_n)\Gamma^2(\bk,\bk',\bk'')\nn\\
\label{part2}
\eea
where the density of states ($N_F$) was introduced for ${\cal S}_{\bk,\bk''}(\omega_m,T)$ to have units of energy. The usual self-energy is obtained by averaging the partial one over the direction of $\bk''$. Restoring anisotropy of  FS, the relation between the partial and usual self-energies can be written as
\beq
\Sigma_{\bk}(\omega_m,T)=\frac{1}{N_F(2\pi)^D}\int \frac{da_{\bk''}}{v_{\bk''}}{\cal S}_{\bk,\bk''}(\omega_m,T).\label{part3}
\eeq
FMFR for the partial self-energy implies that
\beq
{\cal S}_{\bk,\bp}(\pm \pi T)=\pm i
 T \mu_{\bk,\bp} +
0\times T^2.
\label{part}
\eeq
Averaging  $\mu_{\bk,\bp}$
 over the direction of $\bp$ gives $\lambda_\bk$
 in Eq.~(\ref{rule1}).
FMRF for the partial self-energy will be important for deriving the analogous rule for the optical conductivity,
see Sec.~\ref{sec:1stMsigma}.

\paragraph{{\bf Physical consequences.}}
In physical terms, FMFR  says that $\Sigma_\bk(\pm \pi T,T)$ does not contain a part which, if continued to real frequencies, would correspond to damping of quasiparticles.
 This, by itself, has important physical consequences,
 especially for
    thermodynamic quantities,
    which can be calculated
    entirely in the Matsubara representation.
     For example, the effect of many-body interactions on the amplitude of de Haas-van Alphen (dHvA) oscillations,
     $A(T)$,
     is encapsulated by the Matsubara self-energy:
\beq
A(T)=\frac{4\pi^2T}{\omega_c}\sum
^\infty_{\omega_m
=\pi T}\exp\left[-2\pi \frac{\omega_m-i\Sigma(\omega_m,T)}{\omega_c}\right],
\eeq
where $\omega_c$ is the cyclotron frequency.
 If $T\gtrsim\omega_c$, one can keep only the first term in the sum which, thanks to Eq.~(\ref{rule1}), is reduced to
 \beq
A(T)=\frac{4\pi^2T}{\omega_c}\exp\left[-2\pi \frac{\pi T-i\Sigma(\pi T,T)}{\omega_c}\right]=\frac{4\pi^2T}{\omega_c}\exp\left(-2\pi^2\frac{T}{\omega^*_c}\right),\label{LK}
\eeq
where $\omega_c^*=\omega_c/(1+\lambda)$.
 Therefore,  the effect of interactions on dHvA amounts only  to mass renormalization while damping disappears.\cite{Fowler:1965,engelsberg:1970,engelsberg:1978, wasserman:1996,martin:2003,adamov:2006}
 (Damping by disorder is still present because the corresponding self-energy $\Sigma(\omega_m,T)=\text{const} \times i\text{sgn}~\omega_m$ is not a subject to FMFR.)

 Another example is
superconductivity
mediated by soft boson modes
 near a quantum phase transition in $D <3$.
 In this case, fermions are strongly scattered by the same near-critical
 bosons that provide the glue for superconductivity.
It might seem that
this scattering
would impede superconductivity.
As in the previous example, however, the effect of
  fermion
  damping on $T_c$
 is embodied by the Matsubara self-energy, which does not contain the damping part at $\omega_m=\pm \pi T$.
  It turns out\cite{wang:2016}
  that the vanishing of the damping
  part of the self-energy
  at these two frequencies render $T_c$ finite,
   even
    if
     damping
    at other Matsubara frequencies is very strong.

 \subsubsection{Self-energy on the real axis}
 \label{sec:retarded}
 \paragraph{{\bf Analytic structure of the self-energy in the complex plane.}}
 For
  other observables, such as photoemission
  intensity
  and optical conductivity, one would like to know the constraints on the real and imaginary parts of the self-energy. For positive (negative) Matsubara frequencies, the Matsubara self-energy can be obtained by analytic continuation of the retarded (advanced) self-energy from the real axis to the complex plane (which is opposite to usual analytic continuation,
  in which
  the Matsubara self-energy is continued to the real axis). Let $S_R(z,T)$ and $S_I(z,T)$ be analytic continuations of the real and imaginary parts of the
  retarded
  self-energy into the complex plane, correspondingly.  For $\omega_m>0$,
   FMRF
    implies that
\beq
\Sigma(\pi T,T)=
S_R(i\pi T,T)+iS_I(i\pi T,T)=
i
 \lambda T.
\label{rule2}
\eeq
In  the FL regime, the real part of the self-energy can be written as
$\R\Sigma^R(\omega,T)=\lambda\omega+\R\tilde \Sigma^R(\omega,T)$, where $\R\tilde \Sigma^R(\omega,T)$ contains higher-order terms in $\omega$ and $T$.  The function $S_R(z,T)$ can be likewise separated into the linear part and the remainder: $S_R(z,T)=\lambda z+\tilde S_R(z,T)$.
 After that,
Eq.~(\ref{rule2})
 is reduced to
\beq
\tilde S_R(i\pi T,T)+iS_I(i\pi T,T)=0.
\label{rule3}
\eeq
Equation (\ref{rule3}) does not imply that $S_R(i\pi T,T)$ and $S_I(i\pi T,T)$ must vanish separately. However, they do
in all
particular cases
 that we know of.
The best-known case is the conventional FL, where
\beq
\I\Sigma^R(\omega,T)=\text{const}\times\left(\omega^2+\pi^2T^2\right)
\label{FL}
\eeq
on the real axis.
Obviously, $\I\Sigma^R(\pm i\pi T,T)=0$.

One might argue that the vanishing of $\I\Sigma^R(\pm i\pi T,T)$ is a general property.
 Indeed, within the local approximation $\I\Sigma^R(\omega,T)$ can be written as an integral
 \bea
\I\Sigma^R(\omega,T)=-\frac{1}{2\pi v_F}\int^\infty_{-\infty} d\Omega\left[n_B(\Omega)+n_F(\omega+\Omega)\right] \I \chi^R_{\text{loc}}(\Omega),\label{is}
\eea
where $\chi^R_{\text{loc}}(\Omega)$ is obtained by analytic continuation of Eq.~(\ref{vloc}) and $n_B(\Omega)$ is the Bose function.   Substituting $\omega=\pm i\pi T$ directly into Fermi function in Eq.~(\ref{is}) and noting that
\beq
n_F(\Omega\pm i\pi T)=-n_B(\Omega),
\label{nfnb}
\eeq
we see that $\I\Sigma^R(\pm i\pi T,T)=0$.

The argument leaves unclear, however, why
$\I\Sigma^R(\omega,T)$ vanishes only at the first but not all Matsubara frequencies, whereas the identity
\beq
n_F(\Omega+i\omega_m)=-n_B(\Omega),
\label{nfnb_2}
\eeq
holds for any $\omega_m$. To understand why analytic continuation of $\I\Sigma^R(\omega,T)$ does not vanish at all Matsubara frequencies,  one needs to look more carefully into the analytic structure
of  the function $f(z)\equiv \I\Sigma^R(z,T)$  in the complex plane. In fact, this function is multi-valued,  and analytic continuation of $\I\Sigma^R(z,T)$,
 i.e., function
 $S_I(z,T)$,
 corresponds to a
 particular
 branch that coincides with $ \I\Sigma^R(\omega,T)$ on the real axis. In the FL case, the result for $f(z)$,
 which is
  valid for any complex $z$,  is especially simple:
\beq
f(z)=\text{const}\times T^2\left[{\pi^2}+(\ln e^{z/T})^2\right].
\label{fz}
\eeq
On the real axis, Eq.~(\ref{fz}) gives obviously the same result as Eq.~(\ref{FL}). In the complex plane, however,
$\ln\exp(z)$ is a multi-valued function: $\ln\exp(z)=z-2\pi i n$, with $n=0,\pm 1,\dots$ Single-valued branches are selected by
cutting the plane by horizontal lines
at $\I z=i\pi (2m+1)T$, that is, exactly at the Matsubara frequencies (see Fig.~\ref{fig:cuts}b). Analytic continuation of Eq.~(\ref{FL}) is achieved by choosing the branch of $\ln\exp(z)$ with $n=0$, which coincides with Eq.~(\ref{FL}) on the real axis: $f(z)=\text{const}\times \left[{\pi^2}T^2+z^2\right]$. This branch obviously vanishes only at $z=\pm i\pi T$ but not  at any other Matsubara frequency.
 Therefore, the first Matsubara rule for the retarded the self-energy can be formulated as follows: the analytic continuation of $\I\Sigma^R(\omega,T)$ vanishes at  $\omega\to \pm i\pi T$. Equation (\ref{rule3}) implies then that the same holds for analytic continuation of $\R\tilde\Sigma^R(\omega,T)=\R\Sigma^R(\omega,T)-\lambda\omega$.

\paragraph{{\bf Example: marginal Fermi liquid.}
}As an example when the implementation of FMFR leads to non-trivial results,
 we
 consider
the
marginal Fermi liquid (MFL) model.\cite{varma:1989,varma:2002}
The MFL model was introduced phenomenologically to
 explain
 the ubiquitously observed linear scaling of the scattering rate with temperature
 and
 frequency. In this model,  the scattering rate is identified with the imaginary part of the self-energy. The required behavior of $\I\Sigma^R(\omega,T)$, i.e., $\I\Sigma^R(\omega,T)=\text{const}\times\max\{\omega,T\}$, is achieved by choosing $\I \chi^R_{\text{loc}}(\Omega,T)$ to be a scaling function of
$\Omega/T$ with the limits given by $\I \chi^R_{\text{loc}}(\Omega,T)\propto \text{sgn}\Omega$ for $T\ll \Omega
 \leq \Omega^*$ and $\I \chi^R_{\text{loc}}(\Omega,T)\propto\Omega/T$ for $\Omega\ll T$, where $\Omega^*$ is the high-energy cutoff of the model.  A convenient choice that satisfies these conditions is
\beq
\I \chi^R_{\text{loc}}(\Omega,T)=\text{const}\times \tanh(\Omega/T).\label{mfl_chi}
\eeq
 With this choice, Eq.~(\ref{is}) gives
\beq
\I\Sigma^R(\omega,T)=\text{const}\times T F\left(\frac{\omega}{T}\right),
\eeq
where
\beq
 F(x) = \frac{\pi}{2}\left(1+\frac{1}{\cosh x}\right)+x\tanh x.\label{mfl}
\eeq
(Note that this
 result is different from the often
used empirical form $\I\Sigma^R(\omega,T)=\text{const}\times\sqrt{\omega^2+\pi^2T^2}$.)

 From the mathematical point of view, Eq.~(\ref{mfl}) is different from the corresponding result for a FL [Eq.~(\ref{FL})] not only in the overall linear vs quadratic scaling with $\omega$ and $T$, but also in that $\I\Sigma^R(\omega,T)$ of a MFL vanishes not only at the first but also at any Matsubara frequency,
 if
 one formally sets $\omega = i\pi T (2m+1)$ in  Eq.~(\ref{mfl}).
 Nevertheless, one can show that
   the correct self-energy on the Matsubara axis, obtained by 
  analytic continuation of $\I\Sigma^R(\omega,T)$,
  vanishes only at the first Matsubara frequency,
   in agreement with FMRF.
The Matsubara self-energy
 at $\omega_m = \pm \pi T$
 is given by Eq.~(\ref{sum1}) with $\chi_\text{loc}(0)$ related to $\I \chi^R_{\text{loc}}(\Omega,T)$ by the spectral representation
\beq
\chi_{\text{loc}}(0)=\frac{2}{\pi} \int^
{\Omega^*}
\frac{d\Omega}{\Omega}\I \chi^R_{\text{loc}}(\Omega,T).\label{chiloc_mfl}
\eeq
Using the MFL form of $\I \chi^R_{\text{loc}}(\Omega,T)$ [Eq.~(\ref{mfl_chi})], we obtain
\beq
\Sigma(\pm\pi T,T)=\pm i\;\text{const} \times T\ln\left(\frac{4e^\gamma
\Omega^*}{\pi T}\right),
\eeq
where $\gamma=0.577\dots$ is the Euler constant. A $T\ln T$-scaling of the Matsubara self-energy (as opposed to a $T$-scaling in  a FL)  is what makes this model ``marginal".

Other examples of the same kind are encountered, for example, for electrons interacting with both optical and acoustic phonons, with the corresponding self-energies given by
 \bse
 \bea
\text{Opt. phonons:}\;\I\Sigma^R(\omega,T)&=&\text{const}\times \left [n_B(\Omega_0)+n_F(\omega+\Omega_0)-n_B(-\Omega_0)-n_F(\omega-\Omega_0)\right],\label{opt}\\
\text{Ac. phonons:}\; \I\Sigma^R(\omega,T)&=&\text{const}\times T^3\left[2\zeta(3)-\text{Li}_3\left(-e^{\omega/T}\right)-\text{Li}_3\left(-e^{-\omega/T}\right)\right],
\label{acph}
\eea
\ese
where $\zeta(x)$ is the Riemann $\zeta$-function and $\text{Li}_n(x)=\sum_{k=1}^\infty x^k/k^n$ the polylogarithmic function.
In
both
cases,
$\I\Sigma^R(\omega,T)$
also
 vanishes
 at $\omega = i \pi T (2m+1)$  with any integer $m$. However the correct $\Sigma (\omega_m, T)$
vanishes only at the first Matsubara frequency.

Recent photoemission study of a HTC cuprate (Bi$_2$Sr$_2$CaCu$_2$O$_{8+\delta}$) \cite{reber:2015} used a phenomenological form  $\I\Sigma^R(\omega,T)=\text{const}\left(\omega^2+\beta^2 T^2\right)^\alpha$ (the ``power-law liquid") to describe the observed scaling of the momentum distribution peak with $\omega$ and $T$ over a wide range of doping. The coefficient $\beta$ was found
to be
near
$\pi$ without a systematic variation with doping, which means that
FMFR
 is satisfied. On the other hand, the exponent $\alpha$ was found to vary across the phase diagram from $\alpha\approx 0.3$ (underdoped) to $\alpha\approx 0.6$ (overdoped). To explain the values of $\alpha<0.5$ within the MFL model,  one would need to modify the Ansatz for the local susceptibility as
$\I\chi_{\text{loc}}(\Omega,T)=\text{const}\times|\Omega|^{-(1-2\alpha)}\tanh(\Omega/T)$.

\subsection{First-Matsubara-frequency rule for the optical conductivity}
\label{sec:1stMsigma}
\subsubsection{Optical conductivity on the Matsubara axis}
Similarly to the self-energy considered in the previous section, the optical conductivity also satisfies FMRF, which can be formulated both on the Matsubara and real axes. First, we consider the Matsubara axis and define the ``Matsubara conductivity'' as
\beq
\sigma(\Omega_n,T)=-\frac{e^2}{\Omega_n}{\cal K}(\Omega_n,T).
\eeq
Analytic continuation of $\sigma(\Omega_n,T)$ gives the usual conductivity on the real axis. The first Matsubara rule for $\sigma(\Omega_n,T)$ can be formulated as
\beq
\sigma(\pm 2\pi T,T)=\frac{{\cal D}}{T}+R_\sigma(T).\label{rule1_s}
\eeq
In a FL, the ``Drude weight'' ${\cal D}$ is independent of the temperature.  The remainder $R_\sigma(T)$
 is related to $R(T)$
 in Eq.~(\ref{rule1})
 and  will be discussed
below.

We split the proof of Eq.~(\ref{rule1_s}) into two steps. At the first step, we consider only  the bare bubble diagram for the conductivity (diagram 1 in Fig.~\ref{fig:kubo_vertex}, top). In this case, Eq.~(\ref{rule1_s}) follows directly from FMRF for the self-energy, Eq.~(\ref{rule1}). To see this,
we assume again that the self-energy is local and recall that $\mathrm{sgn}\Sigma_{\bk}(\omega_m)=\mathrm{sgn}\omega_m$. Integrating over $\e_\bk$ in Eq.~(\ref{kubo_m}), we then obtain for $\Omega_n>0$
\bea
{\cal K}_1(\Omega_n,T)=-\frac{2iT}{D(2\pi)^{D-1}}\oint da_{\bk}v_{\bk} \sum^{\omega_m=-\pi T}_{\omega_m=-\Omega_n+\pi T}\frac{1}{i\Omega_n+\Sigma_{\bk}(\omega_m+\Omega_n)-\Sigma_{\bk}(\omega_m)}.\label{K4}
\eea
For $\Omega_n=2\pi T$, only the $\omega_m=-\pi T$ term survives in the sum and
\bea
{\cal K}_1(2\pi T,T)=
-\frac{2iT}{D(2\pi)^{D}}\oint da_{\bk}v_{\bk}\frac{1}{2\pi i T+\Sigma_{\bk}(\pi T)-\Sigma_{\bk}(-\pi T)}.
\eea
Thanks to FMFR  for the self-energy [Eq.~(\ref{rule1})], we have $\Sigma_{\bk}(\pi T)-\Sigma_{\bk}(-\pi T)=2\pi i T\lambda_\bk +{\cal O}(T^D)$. Therefore, we arrive at Eq.~(\ref{rule1_s}) with ${\cal D}=-[2/D(2\pi)^D]\oint da_k v_\bk Z_\bk$ and
$R_\sigma (T)
={\cal O}(T^{D-2})$.

At the second step, we consider the vertex corrections to the conductivity. This step is more involved and requires
the use of the first Matsubara rule for the {\em partial} self-energy, Eq.~(\ref{part}).  Although there is no Ward identity for the current vertex in
in the absence of Galilean invariance,
some Ward-type relations between the vertices and partial self-energies can still be derived within the local approximation. This procedure is described in Appendix \ref{sec:vertex_sigma}.
The result is that the vertex part of the conductivity also obeys
 FMRF in Eq.~(\ref{rule1_s}).

\subsubsection{Optical conductivity on the real axis}
On the real axis, FMRF for the optical conductivity
 states that its real part vanishes at $\Omega=\pm 2\pi i T$ up to subleading terms:
\bea
\R\sigma(\pm 2\pi i T,T)=0+\tilde R_\sigma(T),
\label{rule_res}
\eea
with $\tilde R_\sigma(T)={\cal O}(T^D)$.
This rule follows from substituting $\Omega=\pm 2\pi i T$ directly into the self-energy [Eq.~(\ref{4pi2})] and vertex [Eq.~(\ref{vert4})]   parts of the conductivity and noticing that $n_F(\omega \pm 2\pi i T)=n_F(\omega)$. As in the case of the self-energy (see Sec.~\ref{sec:retarded}), the functions defined by Eqs.~(\ref{4pi2}) and (\ref{vert4}) have branch-cut singularities in the complex plane, and thus  analytic continuation is possible only to the first (boson) frequency but not to the higher ones.

As we mentioned in Sec.~\ref{sec:retarded}, gauge invariance makes the conductivity to be more robust with respect to infrared singularities, which modify the canonical FL scaling in $D\leq 2$. In 2D, for example, only
${\cal O}(T^2\ln T)$  terms in the self-energy vanish at $\omega=\pm i\pi  T$, while
${\cal O}(T^2)$ terms survive. This is not the case for the conductivity, which does not have
 ${\cal O}(T^2\ln T )$ terms.

\subsection{Experimental verification of the Gurzhi formula}
\label{sec:gurzhi_exp}
\subsubsection{Review of recent experiments}
In this section, we discuss the experimental status of the Gurzhi formula,
  Eq.~(\ref{gurzhi}),  for the optical conductivity of a Fermi liquid.
Although it seems to be most natural to verify this formula in conventional metals,
 which are expected to obey  the FL-theory predictions,
it is very hard to
detect
  the FL, $T^2$ term
 in $\tau^{-1}(\Omega,T)$
 because it is masked by the electron-phonon interaction at any temperatures except for very low ones.
On the other hand,
 the FL
 part
 of
 the
  $\Omega$-dependence
 can be verified because the electron-phonon contribution to
 $\tau^{-1}(\Omega,T)$
saturates at frequencies above the characteristic phonon scale
(``Debye frequency"),
whereas the {\em ee} contribution continues to grow as $\Omega^2$ until interband transitions become important.
Indeed,
the $\Omega^2$ dependence of $\tau^{-1}(\Omega,T)$ was convincingly demonstrated for a number of conventional metals (Au,Ag, and Cu).\cite{beach:1977,parkins:1981}
 As expected, however,
the $T$ dependence of  $\tau^{-1}(\Omega,T)$ was found to result from the electron-phonon rather than {\em ee} interaction.
To the best of our knowledge, the $\Omega^2+4\pi^2T^2$ scaling form has not been verified in conventional metals.

On the other hand,  $\Omega/T$ scaling of the optical conductivity in strongly correlated metals has been studied quite extensively in the past, \cite{sulewski:1988,katsufuji:1999,yang:2006,dumm:2009,dressel:2011} and the interest to this subject has been reignited recently by
a detailed study of the optical conductivity of a ``hidden order'' heavy-fermion compound URu$_2$Si$_2$. \cite{nagel:2012}    By now, the Gurzhi formula has been checked for several classes of materials: heavy fermions, \cite{sulewski:1988,nagel:2012} doped Mott insulators, \cite{katsufuji:1999,yang:2006} organic materials of the BEDT-TTF  family
\cite{dumm:2009,dressel:2011}, underdoped cuprates \cite{mirzaei:2013}, Sr$_2$RuO$_4$, \cite{stricker:2014} and iron-based superconductors. \cite{tytarenko:2015}

It is customary to use a particular parametrization of $\sigma (\Omega, T)$, called an extended Drude formula:\cite{basov:2005,basov:2011}
\beq
\sigma(\Omega,T) = \frac{\Omega_p^2}{4\pi} \frac{1}{1/\tau (\Omega, T) - i \Omega \left[1+ \lambda_{\text{tr}}(\Omega, T)\right]},
\label{ac_n1}
\eeq
where $\Omega_p$ is the plasma frequency, $1/\tau (\Omega, T)$ is the optical scattering rate, and
 $1+ \lambda_{\text{tr}} (\Omega, T)$ is the ratio of the ``transport effective mass'' to the band one.
 In general, the transport mass differs from the quasiparticle one. This can be seen already from the fact that, in a Galilean-invariant interacting system, the former coincides with the bare mass
  but the latter does not.
    In this parametrization,
\beq
\tau^{-1} (\Omega, T) \equiv \frac{\Omega_p^2}{4\pi}
\R\sigma^{-1}(\Omega,T).
\label{tau_exp}
\eeq
The Gurzhi formula implies that
\beq
\tau^{-1} (\Omega, T) \propto \Omega^2+4\pi^2 T^2.
\eeq
In an experiment,  one  fits the data to a phenomenological form
\beq
\frac{1}{\tau(\Omega,T)}
=\mathrm{const}\times\left(\Omega^2+b\pi^2 T^2\right),
\label{ge}
\eeq
where $b$ is treated as a fitting parameter. The theoretical value is $b=4$.
The results from a number of studies are summarized in Table~\ref{table:b}, which is an updated version of Table I in Ref.~\onlinecite{nagel:2012}.
As one can see, the discrepancy between the experimental and theoretical values is quite significant in most cases, except for
Sr$_2$RuO$_4$.\cite{stricker:2014} The discrepancy is especially pronounced in
 doped
 rare-earth Mott insulators and heavy-fermion materials,
where $b$
 is
 smaller than $2$
 and
 remarkably close to $1$ in some cases, e,g.,
in URu$_2$Si$_2$. \cite{nagel:2012}
 The cuprates and iron pnictides occupy an intermediate niche with $2<b<3$.

   \begin{table*}[ht]
\caption{\label{table:b}Measured values of parameter $b$ in the Gurzhi formula for the optical scattering rate [Eq.~(\ref{ge})] and phenomenological parameter $a$, defined by Eq.~(\ref{pheno_se_3}) and related to $b$ via Eq.~(\ref{b_def}). Parameter $a$ measures
the relative strength of an elastic contribution to the self-energy [Eq.~(\ref{pheno_se_3})].
$^*$ indicates  that the slope of the $T^2$ term was taken from the {\em dc} resistivity, which may not be an accurate procedure due to a difference in the normal and umklapp contributions to the {\em dc} and optical resistivities, see Sec.~\ref{sec:vertex}.
}
\centering
\begin{tabular}{|c|c|c|c|c|}
\hline
Material & $b$ &Reference for raw data  & Reference for  $b$ & $a$ \\
\hline
 UPt$_3$ & $<1$& [\onlinecite{sulewski:1988}] & [\onlinecite{sulewski:1988}]  &$\to\infty$\footnote{$a\to\infty$ indicates that the elastic contribution exceeds by far the inelastic one. A precise determination of $a$ is impossible given the accuracy of the reported values of $b$.}\\ 
\hline
CePd$_3$ &$1.3^*$ &[\onlinecite{sulewski:1988}] & [\onlinecite{nagel:2012}] &$9.0$\\
\hline
doped CeTiO$_3$ &$1.8^*$
& [\onlinecite{katsufuji:1999}] &  [\onlinecite{katsufuji:1999}] & $2.8$\\
\hline
doped NdTiO$_3$ & $1.0$ & [\onlinecite{yang:2006}] &  [\onlinecite{yang:2006}]  &$\to\infty$\\
\hline
$\kappa$-BEDT-TTF & $5.6^*$ &[\onlinecite{dumm:2009,dressel:2011}]& [\onlinecite{nagel:2012,berthod:2013}] &$-0.3$\\
\hline
URu$_2$Si$_2$ & $1.0\pm 0.1$ & [\onlinecite{nagel:2012}] &  [\onlinecite{nagel:2012}] & $\to\infty$\\
\hline
Hg1201, YBCO, Bi2201&$2.3$& [\onlinecite{mirzaei:2013}]& [\onlinecite{mirzaei:2013}]&$1.3$\\
\hline
Sr$_2$RuO$_4$ & $4$\footnote{Extrapolated value.} &[\onlinecite{stricker:2014}]& [\onlinecite{stricker:2014}]& $0$\\
\hline
Co-doped BaFe$_2$As$_2$ & $2.16$ & [\onlinecite{tytarenko:2015}]&[\onlinecite{tytarenko:2015}]&$1.58$\\
\hline
\end{tabular}
\end{table*}

\subsubsection{Phenomenological model: elastic and inelastic scattering}
A quantitative analysis of $\Omega/T$  scaling of the optical conductivity is a difficult task, as one needs to make sure that the data are taken over regions of comparable
 $\Omega$ and $T$.\cite{scheffer:2013}
  In addition, Gurzhi scaling
    is expected to work only at the lowest frequencies and temperatures, which is not necessarily the case in all of the experiments cited in Table \ref{table:b}. Nevertheless, it is interesting to ask whether there are fundamental reasons for the coefficient $b$ to deviate from the theoretical value of $4$.
If the $\pi^2 T^2$ and $\Omega^2$ terms in the scattering rate come from the same mechanism, namely, from {\em ee} interaction of any sort, the relative weight of these terms should  be equal exactly to $4$.
However, if there are other, non-electron mechanisms which contribute extra $
 T^2$ and $\Omega^2$ terms, the relative weight of these terms in the total scattering rate may be different
 from $4$.

 One example of such a situation is the low-temperature (Nozi{\`e}res) regime of the Kondo effect,\cite{hewson:book}  where electrons interact with screened magnetic impurities and also with electrons forming the screening clouds. The first mechanism is elastic, which means that the corresponding part of $\I\Sigma^R$ does not depend on temperature but it does depend on frequency, as $\omega^2$ at the lowest $\omega$. The second mechanism is inelastic and contributes
 a FL-like, $\omega^2+\pi^2T^2$ combination to $\I\Sigma^R$.
   In the unitary limit,
  the prefactor of the elastic
  $\omega^2$ term is twice larger than that of the inelastic one,
    \cite{hewson:book} so
    the
    total  $\I\Sigma^R$ is given by
\bea
\I\Sigma^R(\omega,T)=\text{const}
+\frac{2}{3}\text{const}'
\left(\omega^2+\frac{1}{2}\left[\omega^2+\pi^2T^2\right]\right)=\text{const}
+
\text{const}'\left(\omega^2+\frac{1}{3}\pi^2T^2\right),\label{sek}
\eea
where $\text{const}
 >0$ is the $\omega$-independent part of the elastic contribution and $\text{const}'
<
0$
in the dilute limit.
Substituting  Eq.~(\ref{sek}) into Kubo formula (\ref{4pi2}),
we obtain for the optical scattering rate
 \beq
\frac{1}{\tau(\Omega,T)}
=\text{const}
+
\frac{2}{3}\text{const}'\left(\Omega^2+2\pi ^2T^2\right).
\label{tauk}
\eeq
A reduced weight of  the $T^2$ term
in $\I\Sigma^R$
is reflected in the corresponding reduction
of the $T^2$ term
 in the
optical scattering rate, where now the coefficient $b$ is equal to $2$ rather than $4$.
Thus the local (Kondo) FL belongs to a different universality class compared to the usual (itinerant) FL. Of course, the Kondo model (in the dilute limit) cannot explain the experiment because, first of all,
 $1/\tau(\Omega,T)$ in Eq.~(\ref{tauk}) {\em decreases} with $\Omega$ and $T$, while it increases with $\Omega$ and $T$ in the experiment.
However, this example gives one an idea to address the issue phenomenologically, by asking what form of the self-energy would produce the measured scattering rate.

Suppose that
there are
 two scattering channels in a system.
   The first
    one
   is the conventional channel of  inelastic {\em ee}
      scattering, which gives an $\omega^2 + \pi^2 T^2$ contribution to $ \I\Sigma^R(\omega,T)$ with a positive prefactor,
     and
     the
     second one
      is the elastic channel that contributes an $\omega^2$ but not $T^2$ term to the self-energy.
         The  total self-energy is then
         given by
     \beq
 \I\Sigma^R(\omega,T)=\text{const}+\text{const}
 ''
 \left[a\omega^2+\left(\omega^2+\pi^2T^2\right)\right]
 \label{pheno_se_3}
 \eeq
 with $\text{const}''>0$.
 Substituting this form into the Kubo formula, we find the corresponding optical rate
\beq
\frac{1}{\tau(\Omega,T)}=\text{const}+\text{const}''(a+1)\left[\Omega^2+b\pi^2T^2\right]
\label{tau}
\eeq
with
   \beq
   b=\frac{a+4}{a+1}.
   \label{b_def}
   \eeq

   The  (itinerant) FL value of $b=4$ corresponds to $a=0$.  The opposite limit of $a\to \infty$ (and thus of $b\to 1$) corresponds to purely elastic scattering.
   The range $0 <a< \infty$ yields
     $1<b<4$. This  corresponds to a mixture of elastic and inelastic mechanisms,
    with  a ``metallic'' sign
    ($\partial_\omega\I\Sigma^R>0$) of the elastic
   contribution to the self-energy.
  For $-1<a<0$, the  elastic contribution to the self-energy has a  ``non-metallic'' sign ($\partial_\omega\I\Sigma^R<0$), although the $\Omega$ and $T$ dependences of $1/\tau(\Omega,T)$ are still metallic-like.
  This interval of $a$ corresponds to
  $b >4$.
  The special case of $a=-1$ corresponds to $\I\Sigma^R(\omega,T)$ that depends only on $T$ but not on $\Omega$.
 For $a <-1$, the $\Omega$ dependence of $1/\tau (\Omega, T)$ has a non-metallic sign.
 Such a behavior
 was
 not
  observed in the experiments discussed here.

  According to
  this classification scheme,
    the value of $b\approx 1$ observed in U, Ce, and Nd-based compounds\cite{sulewski:1988,katsufuji:1999,yang:2006,nagel:2012} indicates
    that
    a purely elastic scattering mechanism
    is the dominant one
     in these materials
     ($a\to\infty$).
    The values of $b\approx 2.3$ and $2.16$
    observed
    in the cuprates \cite{mirzaei:2013} and pnictides\cite{tytarenko:2015}, correspondingly,  points at a mixture of elastic and inelastic mechanisms with a metallic sign of the elastic contribution. Finally,
    $b\approx 5.6$ in BEDT-TTF~\cite{dumm:2009,dressel:2011} also corresponds to a mixture of  two mechanisms
    with a non-metallic sign of the elastic contribution
    to the self-energy.

\begin{figure}[h]
 \vspace{-0.5in}
\includegraphics[width=1.0
\textwidth]{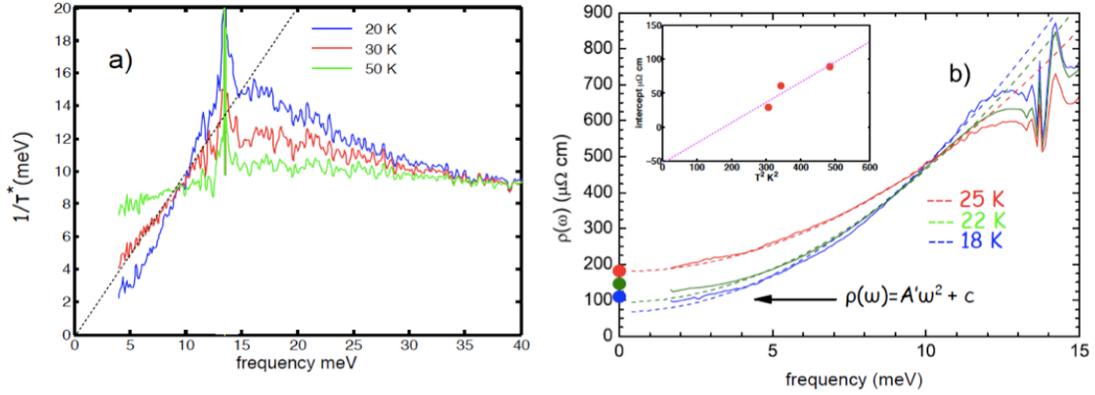}
\vspace{-2.5in}
\caption{(color on-line). Experimental results for  URu$_2$Si$_2$ from Ref.~\onlinecite{nagel:2012}.
a) Optical scattering rate $1/\tau(\Omega,T)$.
b) The optical resistivity at lower frequencies from the refined reflectivity. {\em Reproduced with permission from Ref.~\onlinecite{nagel:2012}.  Copyrighted by U. Nagel et al.}}
\label{fig:tom_data}
\end{figure}

Elastic scattering from resonant levels is one example where $b
\approx 1$.\cite{maslov:2012}  If the resonant level is described by a Lorentzian centered at $\omega=\e_F+\omega_0$ and of width $\gamma$, the imaginary part of the self-energy also has a Lorentzian form:
\beq
\I\Sigma^R_{\mathrm{el}}(\omega)
=\text{const}\times \frac{\gamma}{(\omega-\omega_0)^2+\gamma^2}.
\label{ser1}
\eeq
Expanding Eq.~(\ref{ser1}) to ${\cal O}(\omega^2)$ and substituting the result into the Kubo formula, we obtain for the $\Omega$- and $T$- dependent parts of the optical scattering rate:
\beq
\frac{1}{\tau(\Omega,T)}=\mathrm{const}\times\frac{\gamma\left(3\omega_0^2-\gamma^2\right)}{\left(\omega_0^2+\gamma^2\right)^3}\left(\Omega^2+\pi^2T^2\right),
\label{ose1}
\eeq
which corresponds to $b=1$.
 \begin{figure}[h]
\includegraphics[width=0.4\textwidth, angle=90]{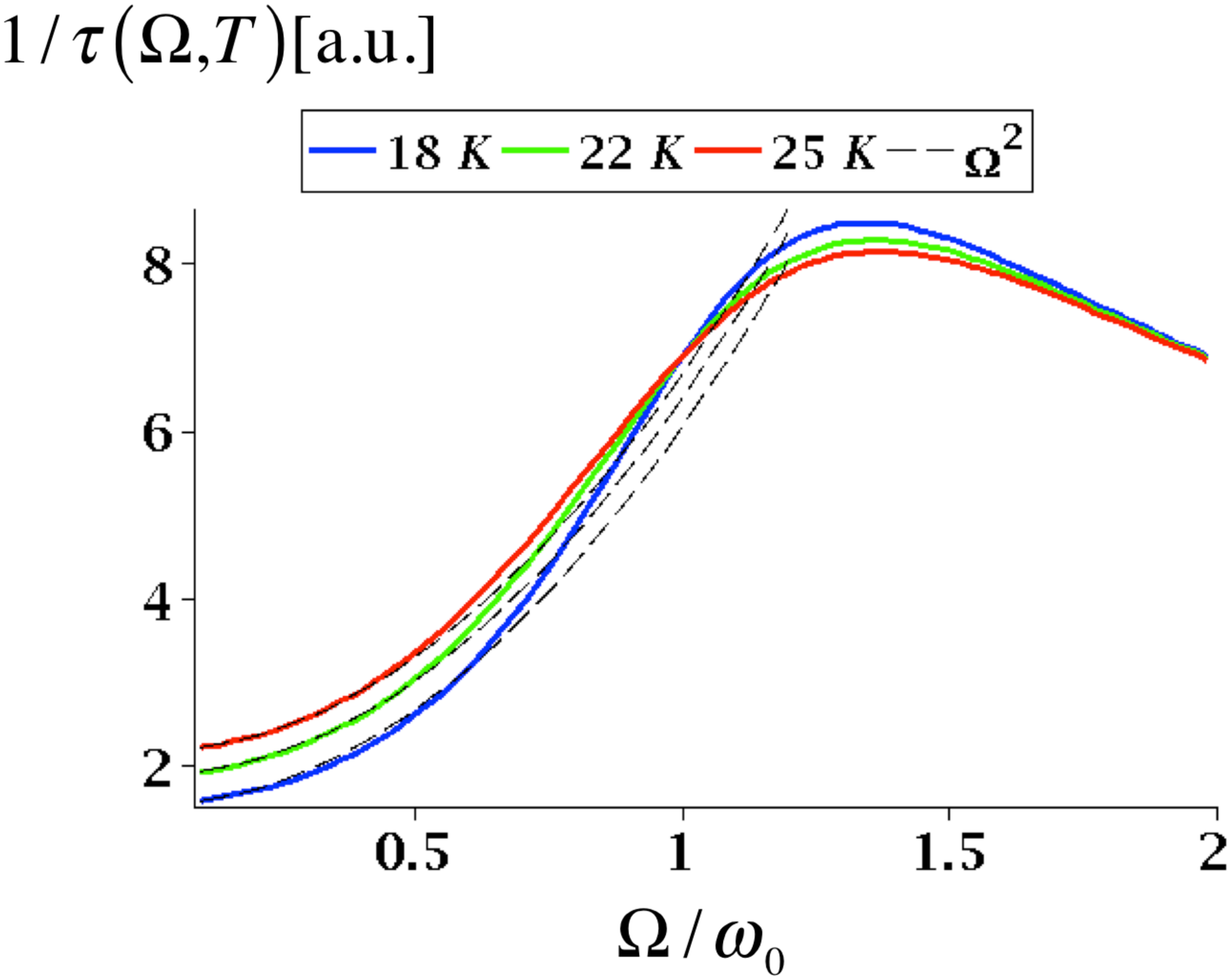}
\includegraphics[width=0.3
\textwidth]{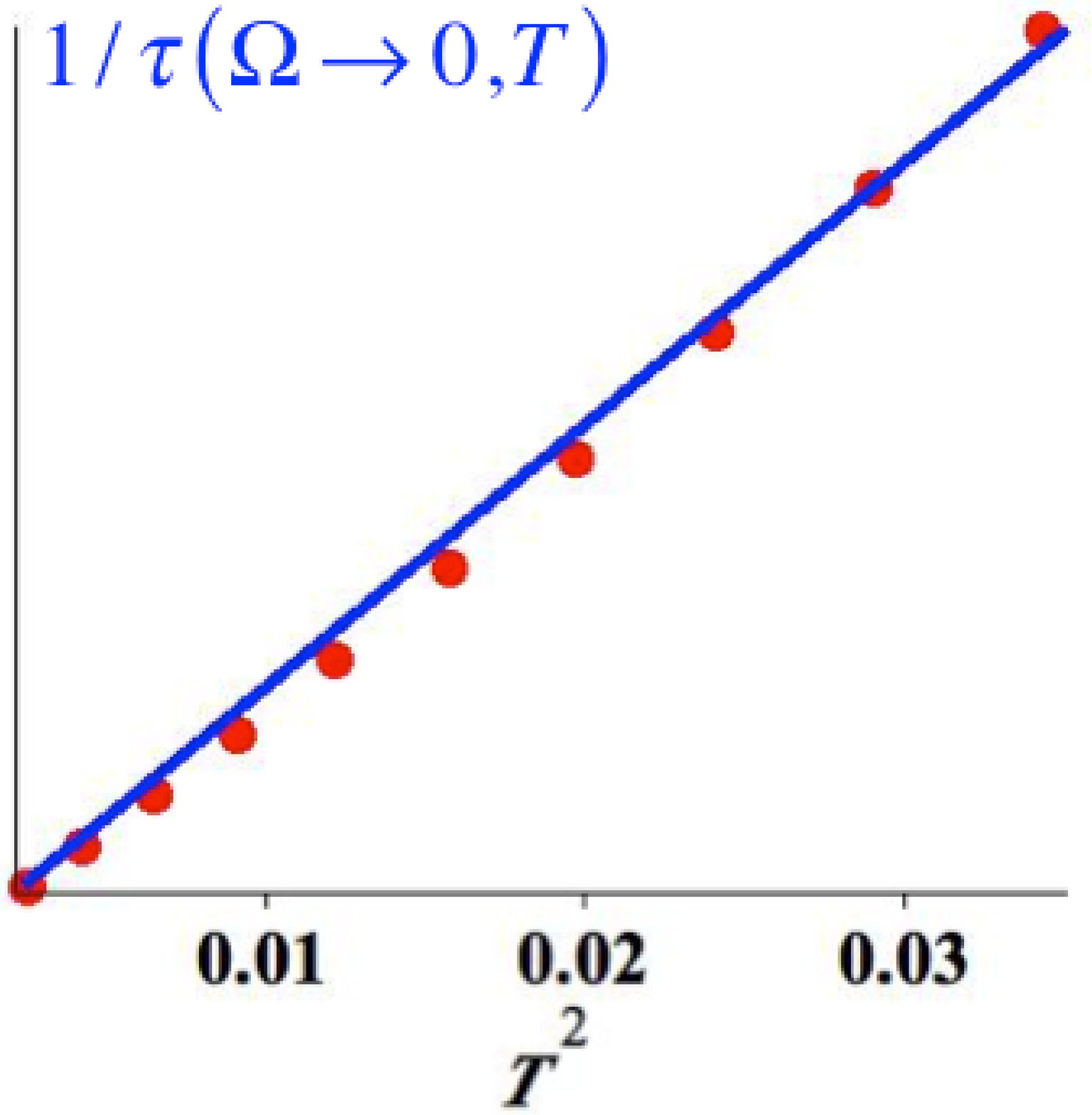}
\caption{(color on-line). Left: Optical scattering rate in the resonant-impurity model as a function of frequency at several temperatures. The absolute scale of  temperature is fixed by choosing $\omega_0=12.5$ meV and $\gamma=0.2\omega_0$. The dashed lines are the $\Omega^2$ fits of the actual dependences. Right: The
zero-frequency
 intercept, $1/\tau(\Omega\to 0,T)$, in the resonant-impurity model as a function of $(T/\omega_0)^2$. {\em Adapted with permission from Refs.~\onlinecite{maslov:2012}. Copyrighted by the American Physical Society.}}
\label{fig:res_omega}
\end{figure}
The
full
 optical scattering rate,
  obtained from the
  self-energy in Eq.~(\ref{ser1})
    without
    expanding
     in $\omega$,
    is shown in Fig.~\ref{fig:res_omega}, left. The quadratic $\Omega$ dependence at the lowest $\Omega$ is followed by a maximum at $\Omega\approx\omega_0$. The $T$-dependence of $1/\tau(\Omega,T)$ at the lowest $\Omega$ is  almost quadratic, see  Fig.~\ref{fig:res_omega}, right. The overall shape of the $\Omega$ and $T$ dependences in this model reproduce some subtle features of the experimental data for  URu$_2$Si$_2$\cite{nagel:2012} (Fig.~\ref{fig:tom_data}), including an isosbestic point at $\Omega\approx 10$ meV.

The microscopic origin of resonant levels is not clear at the moment.
It is very unlikely, however,  that clean samples of materials with $b\approx 1$ contain considerable amounts of {\em extrinsic} resonant impurities. Therefore, resonant states must be {\em intrinsic} to these compounds. We surmise that localized $f$-electrons of U, Ce, and Nd atoms
  may
  play the role of {\em incoherent} resonant levels at sufficiently high energy scales  probed in optical measurements. On the other hand, materials with itinerant $d$-orbitals (cuprates and pnictides) tend to have $b$ closer to the theoretical value of $4$. The $124$ strontium ruthenate, which exhibits a robust FL behavior at low enough temperatures (below $30$ K
  and above the superconducting transition at $\approx 1$ K),\cite{mackenzie:2003}
   comes also the closest to the theoretical prediction in its optical conductivity.
    Apparently, more data need to be accumulated before one can say that
    the materials with $b$ in the intermediate range between $1$ and $4$ represent a new universality class.

\section{Optical conductivity of non-Fermi liquids}
\label{sec:sigmaNFL}
\subsection{Pomeranchuk criticality}
\label{sec:pom}
\subsubsection{Model}
A Pomeranchuk instability is a quantum phase transition that breaks rotational but not translational symmetries of a Fermi liquid.\cite{pomeranchuk:1958} The most common example of a Pomeranchuk instability is a ferromagnetic phase transition in the $\ell=0$ angular momentum channel. Nematic
instability
in channels with $\ell\geq 1$
 \cite{fradkin:2010}  is currently a subject of considerable interest stimulated, in part, by observations of nematic phases in Sr$_3$Ru$_2$O$_7$,\cite{borzi:2007}
   iron-based superconductors,\cite{hu:2012,Fernandes:2014,Gallais:2016}
  and  high-$T_c$ cuprates.
   \cite{*[{See, e.g.,} ]  [{, and references therein.}] louis:2015}

  The order parameter in the ordered phase is spatially uniform ($q=0$). At the critical point,
the
 propagator of
 order-parameter fluctuations (the dynamical susceptibility) is assumed to be Landau overdamped
 and, at weak coupling, has the form
\beq
\chi(\bq,\Omega)=\frac{\bar g}{q^2+\gamma
d(\bq)\frac{|\Omega|}{q}},
\label{HMb}
\eeq
where  $\bar g$
is the coupling constant
of
 the effective 4-fermion interaction, mediated by critical bosons,
 and $d(\bq)$ is the form-factor,
 determined by the angular momentum of the critical channel. Below, we will focus of the simplest case of $\ell=0$ instability, when $d(\bq)=1$.
Because typical $\Omega$ in Eq.~(\ref{HMb}) scale as $q^3$,
Pomeranchuk criticality belongs to the Hertz-Millis-Moriya class\cite{hertz:1976,millis:1993,moriya:1985}  with
the
 dynamical exponent $z=3$.
 In the single-band case,
the damping parameter $\gamma$
 is related to
$\bar g$ via $\gamma\sim \bar g N_F/v_F$,\cite{abanov:2003} where $N_F$ is the density of states.

  A singular form of the critical propagator leads to a breakdown of the FL, which is manifested by a NFL behavior of the self-energy. In 2D,
\beq\Sigma'(\omega)\sim\Sigma''(\omega)=\omega_0^{1/3}\omega^{2/3},
\eeq
where $\omega_0\sim \bar g^2/\e_F$
(in this case,
$\bar g$ has units of energy). In 3D,
\bea
\Sigma'(\omega)\sim\frac{\bar g}{v_F}\omega\ln\frac{\e_F}{\omega},\;\;
\Sigma''(\omega)\sim\frac{\bar g}{v_F}\omega
\eea
(in this case, $\bar g$ has units of velocity). A necessary (but not sufficient) condition for
a controllable theory with the interaction of this type is
the requirement that the coupling is weak,
i.e.,
$\bar g\ll \e_F$ in 2D and $\bar g\ll v_F$ in 3D. This ensures that
 the low-energy sector, where the
 behavior is universal,  and the high-energy sector, where it is not,
 are not mixed
  by the interaction.

  \subsubsection{Optical conductivity}
\begin{figure}[h]
\vspace{-0.3in}
\includegraphics[width=0.6 \linewidth]{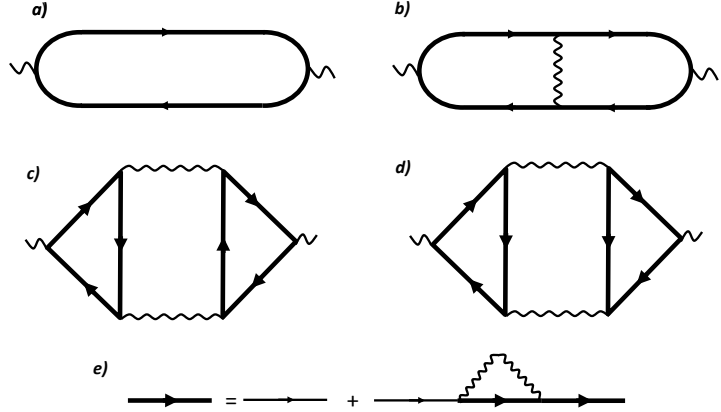}
\vspace{-0.5in}
\caption{
 {\em e-d)}
 Feynman diagrams for the optical conductivity in the
boson-fermion
model. The wavy line denotes
the
dynamic
susceptibility
of the order-parameter fluctuations, Eq.~(\ref{HMb}).
{\em e)}
The Dyson equation for the Green's function.
 \label{fig:3diag}}
\end{figure}

  As is typical for a $q=0$ criticality, characteristic momentum transfers $q\sim (\gamma\omega)^{1/3}$
  near Pomeranchuk criticality
  are small compared to $k_F$. Correspondingly, the transport time is much longer than the quasiparticle lifetime, given by $1/\Sigma''(\omega)$. Yet, the real part of the self-energy is larger than the frequency below certain energy scale [$\omega_0$ in 2D and $\e_F\exp(-v_F/\bar g)$ in 3D], which marks the onset of the NFL behavior. To account for the effect of small-angle scattering quantitatively, one needs to consider the set of diagrams shown in Fig.~\ref{fig:3diag}.\cite{riseborough:1983} This set is similar to the one considered in Sec.~\ref{sec:BE} and Appendix \ref{app:diags} (Fig.~\ref{fig:5diag}), except for that now the Green's functions, depicted by thick solid lines in Fig.~\ref{fig:3diag}, in are not free but satisfy the Dyson equation with one-loop self-energy,  Fig.~\ref{fig:3diag}{\em e}. After analytic continuation, one can expand the Green's functions in diagram {\em a} to first order in the {\em imaginary} part of the self-energy.
  In the rest of the diagrams, which contain additional interaction lines, one can neglect the imaginary parts of the self-energy in the Green's functions. The sum of the four diagrams treated in this way contains the
  square of
    the change in total current
      due to collisions, $(\Delta{\bf J})^2=(\bv_\bk+\bv_\bp-\bv_{\bk-\bq}-\bv_{\bp+\bq})^2$,
see
       Eq.~(\ref{DJ}).
        Since $q\ll k_F$, we have $(\Delta{\bf J})^2\propto q^2$. This is an analog of the ``transport factor'' encountered in the context of impurity scattering.\cite{mahan}  Qualitatively, the effect can be captured by introducing the
          ``transport self-energy'',
           whose imaginary part
               differs from the usual one by an extra factor of $q^2$ under the integral:
\bea
\Sigma''_{\text{tr}}(\omega,T)
=\int\frac{d^Dq}{(2\pi)^D} \int^\infty_{-\infty} \frac{d\Omega}{\pi}\left[n_B(\Omega)+n_F(\omega+\Omega)\right] \I G^R(\bp+\bq,\omega+\Omega)
\I \chi^R(\bq,\Omega)\frac{q^2}{2k_F^2}.
\eea
At $T=0$, this gives\cite{mathon:1968,kim:1994,schofield:1999}
\bea
\Sigma''_{\text{tr}}(\omega)=\left\{\begin{array}{cl}
\omega_1^{-1/3}\omega^{4/3},\;\text{(2D)}\\
\omega_1^{-2/3}\omega^{5/3},\;\text{(3D)}
\end{array}
\label{transp}
\right.
\eea
where $\omega_1\sim
 \e^5_F/\bar g^4$
  and $\omega_1\sim \e_F(v_F/\bar g)^{5/2}$, in 2D and 3D correspondingly. Since $\omega_{1}\gg \e_F$ in the weak-coupling regime, $\Sigma''_{\text{tr}}(\omega)\ll\omega$ for $\omega\ll \e_F$.
  {\em A posteriori},
   this
  justifies an
   expansion in the imaginary part of the self-energy.
More specifically,
each diagram in Fig.~\ref{fig:3diag} is proportional to $\Sigma''$, which is not small compared to $\max\{\omega,\Sigma'\}$,
 but
 their sum is proportional to $\Sigma''_{\text{tr}}$, which is small.
 \footnote{The series for the self-energy does not converge in 2D due to so-called planar diagrams, which are not small in
 $1/N$, where $N$
  is the number of fermion flavors.\cite{lee:2009} It can be surmised, however,  that planar diagrams, which correspond to almost 1D scattering processes,\cite{chubukov:2010v} should not affect the conductivity\cite{paul:2016}}
    If  the FS satisfies the conditions specified
in Sec.~\ref{sec:extra}, the conductivity can be described by the
 Drude formula
 with $1/\tau (\Omega, 0) = \Sigma''_{\text{tr}}(\Omega)$:
\beq
\sigma'(\Omega)=\frac{\Omega_p^2}{4\pi}\frac{\Sigma''_{\text{tr}}(\Omega)}{
\Omega^2 +\left[\Sigma''_{\text{tr}}(\Omega)\right]^2}\approx \frac{\Omega_p^2}{4\pi}\frac{\Sigma''_{\text{tr}}(\Omega)
}{\Omega^2}.\label{rs}
\eeq
In accord with the argument given above, we neglected  $\Sigma''_{\text{tr}}(\Omega)$ in the denominator at the last step.
It is tempting to include mass renormalization into this procedure, i.e., to replace $\Omega$ in the denominator of Eq.~(\ref{rs}) by $\Omega/Z(\Omega)$, but
 it can be shown 
 that, near a nematic QCP, the
 $Z$-factor 
 is canceled out
 by vertex corrections.\cite{chm_unpub}
  
 In a FL,
     $\Sigma''_{\text{tr}}(\Omega)\sim \Sigma''(\Omega)\propto \Omega^2$, and $\sigma'(\Omega)$ does not depend on $\Omega$.
  This is the FL foot discussed in Sec.~\ref{sec:novert}. The result for a NFL is different
     because of NFL scaling of  $\Sigma''_{\text{tr}}$ with $\Omega$.
Substituting the 2D form
 of $\Sigma''_{\text{tr}}(\Omega) \propto \Omega^{4/3}$ into Eq.~(\ref{rs}), one obtains\cite{kim:1994,eberlein:2016}
 \bea
\sigma'(\Omega)\propto \frac{1}{\Omega^{2/3}}
\label{rs_2}
\eea
This behavior of $\sigma'(\Omega)$
was interpreted in Ref.~\onlinecite{eberlein:2016} as a violation of hyperscaling,
because hyperscaling
 implies
that the conductivity 
is independent of $\Omega$
  in 2D.
 We
 note
  in passing that
 hyperscaling would be satisfied if one would replace
 $\Omega$ by $\Omega/Z(\Omega)$ in Eq. (\ref{rs}) and use the fact that  $Z(\Omega) \propto \Omega^{1/3}$ at a nematic QCP in 2D.

 In 3D,
the same reasoning gives
\bea
\sigma'(\Omega)
\propto\frac{1}{\Omega^{1/3}}.
\label{rs3d}
\eea

\subsubsection{{\em dc} limit}

The
$T$-dependence of
{\em dc} resistivity of quantum-critical metals had long been believed to arise from scattering of electrons from fluctuations of the order
parameter.\cite{mills:1966,rice:1967,schindler:1967,mathon:1968,kaiser:1970,riseborough:1983} Since such a mechanism by itself cannot provide current relaxation, it had been assumed implicitly that umklapp processes quickly relax the momentum gained by
  small-$q$
  fluctuations.
  It was shown in Refs.~\onlinecite{maslov:2011,pal:2012b},
  however,
  that
  umklapp scattering is
   suppressed near a $q=0$ criticality,\cite{maslov:2011,pal:2012b} where scattering is predominantly of the small-angle type,
   while umklapp processes require momentum transfers comparable to the reciprocal lattice constant.
       In this case, the only effective mechanism of current relaxation in scattering by impurities. As was discussed in Sec.~\ref{sec:BE}, normal (as opposed to umklapp) scattering by fluctuations of the order parameter modifies the electron distribution function but in such a way that the momentum is still conserved. Nevertheless, if the FS satisfies certain conditions specified in Sec.~\ref{sec:extra}, normal scattering does give rise to a temperature-dependent
       (inelastic)
       term in the resistivity. In the single-band case, such a term
       cannot exceed substantially
       the residual resistivity,
       but it can happen if
       a  metal
       has several bands
        with
       substantially
        different effective masses
        (cf. Sec.~\ref{sec:2bdc}).

         The $T$-dependence of the inelastic term
         in the resistivity
is obtained by substituting $\omega$ by $T$ in the transport self-energy [Eq.~(\ref{transp})], which leads to
 $\rho\propto T^{4/3}$ and $\rho\propto T^{5/3}$ in 2D and 3D, correspondingly. The $5/3$ scaling (or numerically close $3/2$ one) was observed in a number of weak ferromagnets close to the quantum critical point, e.g., UGe$_2$,\cite{oomi:1998}Ni$_x$Pd$_{1-x}$, \cite{nicklas:1999} Ni$_3$Al,\cite{niklowitz:2005} and NbFe$_2$\cite{brando:2008} (for a more complete list, see Ref.~\onlinecite{brando:2016}).\footnote{In some materials, e.g., helimagnetic MnSi\cite{pfleiderer:2001}  and ferromagnetic ZrZn$_2$,\cite{smith:2008b}  the $5/3$ ($3/2$) scalings are observed in wide regions of the phase diagrams, and are therefore not likely to manifest quantum-critical physics \cite{brando:2016}} Additional evidence for the quantum-critical nature of this scaling comes from the concomitant logarithmic divergence of the specific heat coefficient. To the best of our knowledge, however, quantum-critical scaling of the optical conductivity
  in 3D
  [Eq.~(\ref{rs3d})]  has never been verified experimentally. \footnote{NFL scaling of the optical conductivity in SrRuO$_3$ and CaRuO$_3$\cite{dodge:2000,lee:2002,dodge:2006,geiger:2016}  is not consistent with the predictions of the quantum-critical theory. Recently, this behavior has been attributed to minigaps produced by rotational and tilting distortions of the lattice\cite{dang:2015}}
Such an experiment is very much desirable.
In 2D, the situation is not clear even as far as {\em dc}
resistivity
is concerned: to the best of our knowledge,   $4/3$ scaling
of $\rho$
 has
 not
 been yet
 observed experimentally.

\subsection{
Quantum phase transition at  finite $q$}
\label{sec:SDW}

\subsubsection{Hot and cold regions}

As was pointed out several times in this paper, a conventional FL is characterized by two properties: i) the $Z$-factor is finite and independent of the frequency and ii) the scattering rate scales as
$\max\{\omega^2,T^2\}$. The combination of these two properties result in a FL foot (cf. Fig.~\ref{fig:foot}): a wide region of frequencies in which the real part of the conductivity is independent of the frequency.

However, this is not what the experiment shows in many cases. Of particular importance is  a violation of the expected FL behavior in under- and optimally doped superconducting cuprates, where $\sigma'(\Omega)$ scales as  $1/\Omega^{d}$ with $d>0$. The exponent $d$ was found to be close to $0.7$ in the intermediate frequency range ($\Omega \sim100-500$ meV)\cite{azrak:1994,marel:2003} and close to $1$
in
 a wider frequency range (from $
 100$ meV to about $1$ eV).\cite{basov:1996,puchkov:1996,basov:2005,norman:2006} At the same time, the optical scattering rate scales roughly linearly with $\Omega$ at sufficiently high frequency, both in the cuprates\cite{azrak:1994,basov:2005} and iron-based superconductors.\cite{qazilbash:2009} This scaling appears to be dual to enigmatic linear scaling of the {\em dc} resistivity with temperature, observed in many classes of strongly correlated materials.\cite{bruin:2013}

 Phenomenologically, one can argue that both linear dependences represent two limiting cases of the same scattering rate. This assumption is employed in the marginal FL model,\cite{varma:1989,varma:2002} which postulates that the transport scattering rate is identical to the single-particle self-energy $\Sigma''(\omega,T)=\text{const}\times\max\{\omega,T\}$, while the latter does not vary significantly over the FS.  However, it is
  generally problematic
  to justify these assumptions within a microscopic theory, which considers explicitly coupling between fermions and boson degrees of freedom. Indeed, if bosons represent long-wavelength fluctuations of the incipient order parameter, the transport scattering rate is reduced compared to the single-particle self-energy, and
  its
  $\Omega$ and $T$ dependences differ, in general, from those of $\Sigma''(\omega,T)$
 (this case was discussed in Sec.~\ref{sec:pom}). If bosons represent fluctuations of a spin- or charge-density-wave order parameter,  one runs into a different kind of problem known as the ``Hlubina-Rice conundrum'':\cite{hlubina:1995} the self-energy acquires a NFL form only near the special points of the FS  (``hot spots'') that are connected by the nesting wavenumber,  while the rest of the FS stays cold. The {\em dc} conductivity contains an average of the scattering time over the FS. This average is dominated by the cold regions of the FS, where the scattering time is long, and the conductivity retains its FL form: $\sigma(T)\propto 1/T^2$. A number of ways out of the Hlubina-Rice conundrum were proposed in the past, either by introducing anisotropic scattering rates phenomenologically \cite{hussey:2008,hussey:2011} or by invoking impurity scattering, which effectively smears out the boundaries between the hot and cold regions.\cite{rosch:1999}

 In the optical conductivity, one averages the scattering rate instead of the scattering time, which
 at
 first glance implies that the optical conductivity should be dominated by hot regions, where the scattering rate is the highest.
  However, a small size of hot spots diminishes their contribution.
 The hot-spot contribution to the optical conductivity near an SDW QCP in 2D has been calculated
 explicitly
  in several recent papers,\cite{hartnoll:2011,patel:2015}
 where it was found that $\sigma'(\Omega)=\text{const}$.
 The same result can be obtained from the extended Drude formula, which does take into account mass renormalization.
  Indeed, generalizing Eq.~(\ref{ac_n1}) for an anisotropic FS, replacing $1/\tau$ by $\Sigma''(\Omega,\bk)$ and $1+\lambda_{\rm{tr}}$ by $\approx 1/Z(\Omega,\bk)$, and expanding in $\Sigma''$, we obtain   \beq
\sigma'(\Omega)=\frac{2e^2}{D(2\pi)^D\Omega^2}\oint da_\bk v_{\bk}\Sigma''(\Omega,\bk)Z^2(\Omega,\bk).\label{rsan}
\eeq
In 2D,
  both
  $\Sigma''(\Omega,\bk)$ and $Z(\Omega,\bk)$ at the hot spot
  scale as
  $\Omega^{1/2}$, while the width of the hot spot scales as $\Omega^{1/2}$. This gives $\sigma'(\Omega)=\text{const}$, which coincides with the hyperscaling prediction.\cite{patel:2015}
  As we mentioned in the previous section, this may be another indication of the relation between hyperscaling and renormalization of the optical conductivity  by the real part of the self-energy, i.e.,
  by the $Z$-factor. Regardless of an interpretation, however, what matters is that the hot-spot contribution in the 2D case, which is relevant to cuprate and iron-based superconductors, is
  indistinguishable from the cold-region contribution, which remains of the FL-type even right at the QCP.

 In 3D, $\Sigma''(\Omega,\bk) \propto \Omega$  and $Z(\Omega,\bk_{\text{hs}}) = 1/|\ln \Omega|$,  while the width of the hot line still scales as $\Omega^{1/2}$.
This implies that
$\sigma'(\Omega)\propto 1/\Omega^{1/2}$ in 3D (modulo logarithmic renormalization).

 \subsubsection{Composite scattering, $T=0$}

  A new microscopic approach to the optical conductivity near SDW criticality
 has recently been put forward by
 Hartnoll,
 Hofman, Metlitski, and Sachdev (HHMS),\cite{hartnoll:2011} who noticed that the interaction between ``lukewarm fermions'', occupying an intermediate portion of the FS in between the hot and cold regions, may affect the optical conductivity significantly.
 This paper and
 the subsequent one
 by the two of us and Yudson \cite{chubukov:2014}
 considered
  a particular case of finite-$q$ instability--
 a $(\pi,\pi)$  spin-density-wave (SDW) on a 2D square lattice
  --and
 we focus on this case below.

   \begin{figure}[h]
\includegraphics[scale=0.5]{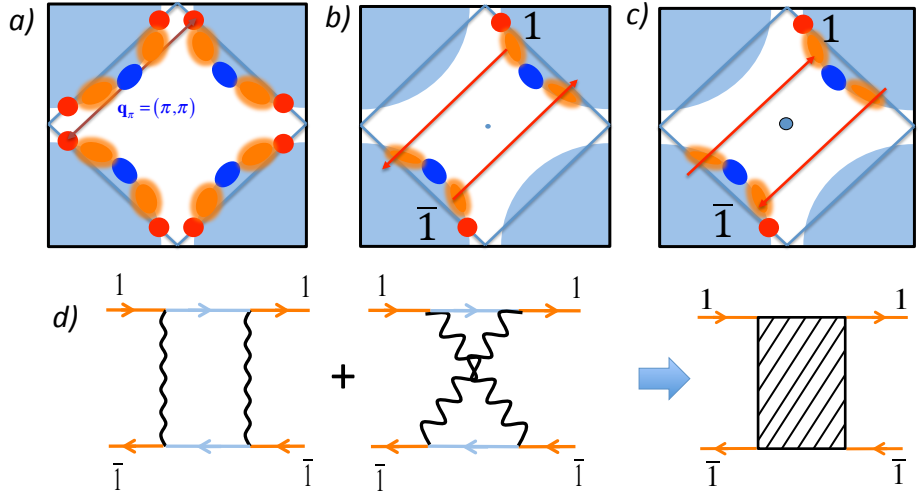}
\vspace{-1.in}
\caption{a) A cuprate-like FS  with hot (red), cold (blue), and lukewarm (orange) regions.  Hot spots are connected by the spin-density-waver ordering vector $\bq_\pi=(\pi,\pi)$. b) The first step of the composite scattering process: a lukewarm fermion near hot spot $1$ is scattered by $q_{\pi}$ from its original position and lands away from the FS; simultaneously, another lukewarm fermion, located near the mirror image of hot spot $1$ ($\bar 1$), is scattered into a state also away from the FS. c) The second step of the composite process: lukewarm fermions return to their original positions. d) Diagrammatic representation of the composite scattered process. The internal Green's functions (solid blue lines) correspond to off-shell intermediate states. The wave line is the spin-density wave propagator, Eq.~(\ref{vd}).  }
\label{fig:comp}
\end{figure}
 An
  SDW
 instability occurs
  in systems where  the FS
 crosses the magnetic Brillouin zone boundary.  The intersection points--the hot spots--are connected by the SDW ordering wavenumber, $\bq_\pi=(\pi,\pi)$ (the lattice constant is set to unity).
 For a FS in Fig.~\ref{fig:comp}, there are eight such spots (red circles). The interaction
 between fermions located in the vicinity of hot spots
 is mediated by exchange of SDW fluctuations with a propagator
  \beq
 \chi(\bq, \Omega) =\frac{\bar g}{\xi^{-2}+
 (\bq-\bq_\pi)^2  - i \gamma \Omega},
  \label{vd}
  \eeq
  where $\gamma=4\bar g/\pi v_F^2$.\cite{abanov:2003}
    At criticality, i.e., at $\xi=\infty$, the one-loop self-energy of a fermion located at point $k_{||}$ on the FS (measured from  the hot spot) is\cite{abanov:2003}
\bea
\Sigma^{(1)}_\bk(\omega)  =
i \frac{3 {\bar g}}{2\pi v_F \gamma} \left(\sqrt{-i\gamma \omega + k_{||}^{2}} - |k_{||}|\right).
\label{1.3}
\eea
The asymptotic limits of Eq.~(\ref{1.3}) allows one to identify different regions of the FS.
For $\omega\ll \bar g$, each quadrant of the FS can be partitioned into hot, lukewarm, and cold regions (depicted by the red, orange, and blue areas in Fig.~\ref{fig:comp}{\em a}, correspondingly.)
The characteristic scales for $k_{||}$ are
\beq
k_1=\sqrt{\bar g\omega}/v_F\;\text{and}\;k_2=\bar g/v_F\label{k12}
\eeq
(the assumption of $\omega\ll \bar g$ ensures that $k_1\ll k_2$). The hot
  region
  corresponds to $|k_{||}|\ll k_1$. In this region, the self-energy is of a NFL form: $\R\Sigma^{(1)}_\bk (\Omega)=\I\Sigma^{(1)}_\bk(\omega)\sim \sqrt{\bar g\omega}$
   and $Z_\bk(\omega) \propto (
   \omega/\bar g)^{1/2}
   \ll 1
   $.
    The cold region is the farthest one from the hot spot: $|k_{||}|\gg k_2$.  In there, we have a weakly renormalized FL with the $Z$-factor close to unity and
\beq
\I\Sigma^{(1)}_\bk(\omega)
\sim \bar g^2\omega^2/(v_F |k_{||}|)^3.\label{imslw}\eeq
The lukewarm region occupies the intermediate range $k_1\ll |k_{||}|\ll k_2$. In there, we have a strongly renormalized FL with the $Z$-factor that scales linearly with $k_{||}$
\beq
Z_\bk\sim v_F|k_{||}|/\bar g\label{Zlw}
\eeq
and approaches zero and unity at the two opposite
ends
of the
 lukewarm
region, correspondingly.
 At the same time, $\I\Sigma^{(1)}_\bk(\Omega)$ is still given by Eq.~(\ref{imslw}).
 If $\omega\gg \bar g$ (but still smaller than the bandwidth), there is no lukewarm region: the hot are cold region are adjacent to each other.

\begin{figure}[h]
\includegraphics[scale=0.5]{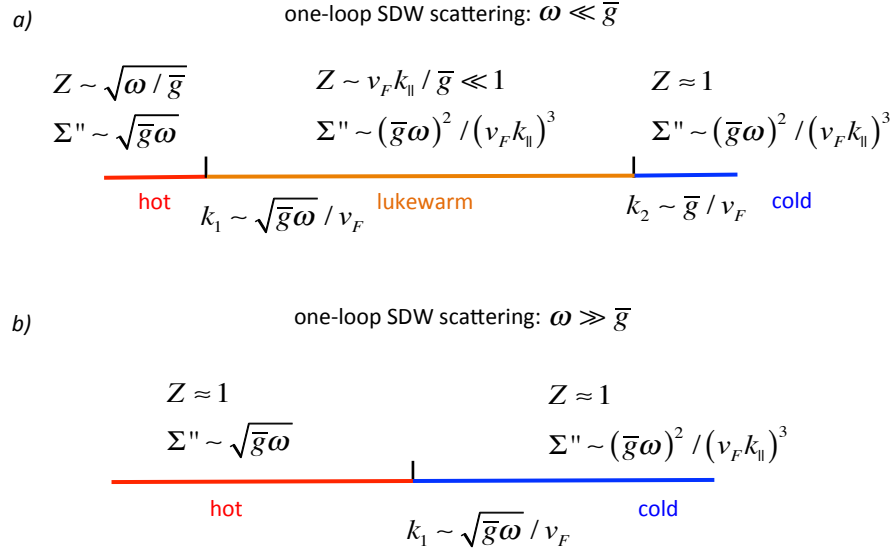}
\vspace{-1.in}
\caption{Partitioning of the FS  into hot, lukewarm, and cold regions at one-loop order in SDW scattering.}
\label{fig:regions_1}
\end{figure}

    Another
  characteristic energy scale  is set by the curvature of the fermion dispersion $\e_\bk = v_Fk_\perp +\kp^2/(2m^*)$, where $m^*$ is inversely proportional to the local curvature of the FS at point $\kp$.
  This  scale can be deduced from comparing different parts of the Green's function of a lukewarm fermion with the $Z$-factor from Eq.~(\ref{Zlw}):
\beq
G_\bk(\omega)=\left(\frac{\bar g\omega}{v_F|\kp|}-v_Fk_\perp-\frac{\kp^2}{2m^*}\right)^{-1}.\label{gf}
\eeq
 If the last term in the denominator of Eq.~(\ref{gf}) is larger than the first one, curvature is important, and the FS must be treated as a 2D object. In the opposite case, the last term can be neglected and the FS becomes essentially 1D. As one moves along the FS away from the hot spot, one first enters the 1D region and then the 2D region. The crossover between the two occurs at $|\kp|\sim k_3$, where
\beq
k_3=(\bar g\e_F^*\omega)^{1/3}/v_F\label{k3}
\eeq
and $\e_F^*=m^*v_F^2/2$. Depending on $\omega$, $k_3$ can be
either
 smaller or larger than the other two crossover scales ($k_1$ and $k_2$), and this complicates the partitioning scheme in Fig.~\ref{fig:regions_1} significantly. A detailed description of the self-energy with
 in
 both 1D and 2D
 regions
  taken into account can be found in Ref.~\onlinecite{chubukov:2014}.

To one-loop order in SDW scattering,
 the optical conductivity of both cold and lukewarm fermions is independent of frequency (the FL foot). This is not immediately obvious for lukewarm fermions, as $\sigma'(\Omega)$ in Eq.~(\ref{rsan})
with $\Sigma_{\text{tr}}$ replaced by the self-energy from Eq.~(\ref{imslw})  and with the $Z$-factor from Eq.~(\ref{Zlw}) appears to have a logarithmic dependence on $\Omega$: $\sigma'(\Omega)
\propto \Omega^{-2}\int_{k_1} d\kp \kp^2 \Omega^2/\kp^3\sim \ln\Omega$. However, a correction to the current vertex cancels the logarithmic singularity,\cite{hartnoll:2011} and the conductivity reduces a constant, which is what one would expect in a FL regime.

Going beyond one-loop order, HHMS considered a composite process depicted in panels {\em b} and {\em c} of  Fig.~\ref{fig:comp}. At the first step, two lukewarm fermions located near diametrically opposite hot spots ($1$ and $\bar 1$) are scattered by SWD fluctuations with ordering wavenumber $\bq_{\pi}$ (panel {\em b}). Because lukewarm fermions are not located at hot spots, the final states after scattering are away from the FS.  At the second step, fermions are scattered again by SDW fluctuations are return close to where they started from (panel {\em c}). The corresponding diagrams for the vertex are shown in panel {\em d}. The  intermediate states (light blue lines) are severely off-shell, and therefore their Green's functions can be approximated by the inverse dispersions. For initial states at momenta $k_{||}$ and $p_{||}$ away from the corresponding hot spots, the product of the two Greens function then reduces to $1/v_Fk_{||}p_{||}$. The rest of the diagram contains a product of two SDW propagators integrated over one of the two sets of the boson energy and momentum (the other set gives the energy and momentum transfers of incoming fermions). This integral depends logarithmically on the distance from the hot spots. Collecting everything together, we obtain the composite vertex
  \beq
  \Gamma^{c}
  (k_{||}, p_{||}; \Omega_m,q)
  =
 \frac{\bar g}{16\pi}\frac{1}{|k_{||} p_{||}|} \ln\frac{\Lambda^2}{
  q^2
  +\gamma |\Omega_m|},
  \label{1.7}
  \eeq
  where $\Lambda=\min\{k_{||},p_{||}\}$. The most important feature of this result is its strong dependence on the distance from the hot spot: for $k_{||}\sim p_{||}$, $\Gamma^c\propto k_{||}^{-2}$. This dependence continues down to the boundary between the lukewarm and hot regions located at $k_{||}\sim k_1$ [Eq.~(\ref{k12})].
   \begin{figure}[h]
\includegraphics[scale=0.5]{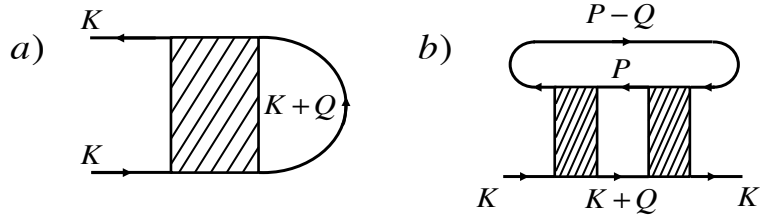}
\vspace{-2.5in}
\caption{Self-energy at one-loop ({\em a}) and two-loop ({\em b}) orders in composite scattering. The shaded box is defined in Fig.~\ref{fig:comp}{\em d}.}
\label{fig:comp_se}
\end{figure}
Treating $\Gamma^c$ as a new effective interaction of the theory, one can consider self-energy at one- and two-loop orders in this interaction (Fig.~\ref{fig:comp_se}{\em a} and {\em b}, correspondingly).

Already at one-loop order, one gets something interesting: the self-energy of lukewarm fermions acquires a non-analytic
 frequency dependence
          \bea
 \I\Sigma^{(1c)}_\bk(\omega)\sim (\bar g
\omega)^{3/2}/(v_Fk_{||})^2, \label{1.8}
  \eea
which exceeds the $\omega^2$ term from one-loop SDW scattering [Eq.~(\ref{imslw})].  This does not mean a FL breakdown: the quasiparticles are still well-defined in a sense that $\omega+\R\Sigma\gg\I\Sigma$, i.e.,  the quasiparticle energy is still larger than its linewidth. One can say that the lukewarm FL is a FL of the
unconventional
type because $\I\Sigma$ decreases with $\omega$ slower than $\omega^2$
but still faster than $\omega$.
  By naive power counting, the corresponding optical conductivity should scale as $\Omega^{-1/2}$, which would signal a strong deviation from the FL picture. However, this does not happen, again because of the vertex correction. Note that double scattering by (large) SDW momentum is subsumed into the composite vertex which, by itself, corresponds to  scattering by small momentum: $q \sim \sqrt{\gamma\Omega}$. The correct expression for the conductivity must include the current imbalance factor, $(\Delta{\bf J})^2=\left(\bv_\bk+\bv_\bp-\bv_{\bk+\bq}-\bv_{\bp-\bq}\right)^2$, which is of order $q^2\propto \Omega$. This additional factor of $\Omega$ leads to a $\sqrt{\Omega}$ term in the conductivity, which is a subleading to the FL (constant) term from one-loop SDW scattering.

A more interesting effect occurs at two-loop order in composite scattering (Fig~\ref{fig:comp_se}{\em b})
-- this process gives rise to another non-analytic frequency dependence of $\I\Sigma_\bk (\omega)$, which does
lead
to
a NFL behavior
of the optical conductivity.
 This behavior is different in 1D and 2D regimes,
 which are defined by whether characteristic momenta $\kp$
 (proportional to
 the frequency of light)
 are smaller or larger than $k_3$ in Eq.~(\ref{k3}), respectively.

In the 2D regime, the self-energy is the same as in a 2D FL liquid with an effective interaction given by Eq.~(\ref{1.7}):
 \bea
 \I\Sigma^{(2c)}_\bk(\omega) \sim
 \frac{{\bar g}^2}{(v_F| \kp|)^3} \frac{\e^*_F}{v_F |\kp|}  \omega^2\ln^3{\frac{Z_{\bk} v_F |\kp|}{\omega}}.
 \label{sigma2D}
  \eea
  The prefactor of $\kp^{-4}$ came from the square of the vertex in Eq.~(\ref{1.7})--this is the most important part of the result as it will lead to a NFL behavior of the optical conductivity. The origin of the $\log^3\omega$ factor is also easy to trace down: two out of three logs came again from the square of the vertex while the third one is the conventional feature of a 2D FL. Depending on whether one is in the lukewarm or cold region, the $Z$-factor under the log is either given by Eq.~(\ref{Zlw}) or almost equal to $1$.

  The 1D regime requires a more detailed analysis.\cite{chubukov:2014} In a true 1D system, the self-energy is a highly singular function of the ``distance" to the mass shell, $\zeta\equiv \omega\mp v_F(k\mp k_F)$, where $\mp$ corresponds to right/left-moving  fermions: the self-energy
 from
  forward scattering has a pole at $\zeta=0$ while that
  from
  backscattering vanishes at $\zeta=0$. But our system is not really 1D in a sense that even if
  the $\kp^2/2m^*$ term in the Green's function (\ref{gf})
  is
   neglected,
  the information about the 2D nature of the FS still enters through the $\kp$-dependence of the $Z$-factor. As a result, none of the two 1D singularities occur in our case but the self-energy is still of the 1D type,  in a sense that it scales linearly with $\omega$:
  \beq
\I\Sigma^{(2c)}_\bk(\omega)\sim \lr \frac{\bar g }{v_F\kp}\rr^2Z_\bk \omega.
\label{se2_11}\eeq

A crossover between the 1D and 2D regimes occurs at $|\kp|\sim k_3$ [Eq.~(\ref{k3})].  For $\omega\ll \bar g^2/\e^*_F$, $k_3$ divides the lukewarm region (from $k_1$ to $k_2$) into two parts, such that the 1D part is closer to the hot spot while the 2D part is closer to the cold region. For $\omega\gg \bar g^2/\e^*_F$, the 1D part extends over the entire lukewarm region. Notice that that 1D regime
can be classified as a MFL: since $\I\Sigma$ scales linearly with $\omega$, quasiparticles are just barely defined, in a sense that $\omega+\R\Sigma\sim \I\Sigma$. However, in contrast to the traditional MFL phenomenology,\cite{varma:1989,varma:2002} only a fraction of the FS exhibits such a behavior. The 2D regime corresponds to a conventional but strongly renormalized and anisotropic FL.

We now turn to the optical conductivity. For $\omega\ll \bar g^2/\e^*_F$, the integral over $k_{||}$ in Eq.~(\ref{rsan}) is controlled by $|\kp|\sim k_3$, which means that either 1D or 2D forms of the self-energy can be used to estimate the conductivity. Using the 2D form [Eq.~(\ref{sigma2D})] and setting the lower limit of the integral at $\kp\sim k_3$, we obtain
\bea
\sigma'_\Sigma(\Omega)\sim\sigma_0\ls\frac{(\e_F^*)^2}{\bg\Omega}\rs^{1/3}
\ln^3\frac{(E_F^*)^2}{\bg\Omega}.
\label{sse5}
\eea
Here $\sigma_0=e^2N_{\text{hs}}/4\pi^2 c$ sets the overall scale of the conductivity ($N_{\text{hs}}$ is the total  number of hot spots and $c$ is the lattice constant along the $c$-axis) and the subscript $\Sigma$ indicates that we have not taken into account the vertex corrections yet. At higher frequencies, the 1D form of the self-energy  [Eq.~(\ref{se2_11})] dominates the integral, which yields
\beq
\sigma'_\Sigma(\Omega)\sim \sigma_0\frac{\bar g}{\Omega}.
\label{s1d}
\eeq

At first glance,  the vertex corrections may modify the results for the optical conductivity significantly. Indeed, fermions involved in composed scattering belong to the vicinities of either the same or diametrically opposite hot spots, and are displaced only a little along the FS in the process of the scattering. This seems to make the
 change in total current
 [Eq.~(\ref{DJ})]
 small. However, the velocities entering $\Delta{\bf J}$ are the {\em renormalized} rather than the bare ones: $v_\bk=Z_\bk v^0_\bk$. While
the bare Fermi velocities may be assumed to vary slowly along the FS, the renormalized ones vary rapidly, following the rapid variation of the $Z$-factor. Replacing the slowly varying bare Fermi velocity by a constant ($=v_F$) and using the lukewarm form of the $Z$-factor [Eq.~(\ref{Zlw})], we obtain for the current imbalance factor
\beq
(\Delta{\bf J})^2=(v_F^2/\bar g)\left(|\kp|+|\pp|-|\kp+q_{||}|-|\pp-q_{||}|\right)^2.
\eeq
The momenta in the equation above are small compared to the size of the Brillouin zone.
 However, typical $q_{||}$ are {\em not} small compared to $\kp$ and $\pp$, in fact $q_{||}\sim\kp\sim\pp$.\footnote{Note that the situation here is different compared to one-loop composite scattering, where
 typical
 $q_{||}$
 are
 small compared to $k_{||}$.}
The overall smallness of $|\Delta{\bf J}|$ reflects the smallness of the $Z$-factor,
 which has been already taken into account when deriving Eqs.~(\ref{sse5}) and (\ref{s1d}).  The vanishing of $(\Delta {\bf J})^2$ at $q_{||} =0$ has only one effect: it regularizes the infrared singularity which gave rise to the logarithmic factors in Eq.~(\ref{sse5}), and the correct expression for the conductivity in the 2D regime does have these factors. In the 1D regime, the only effect of this vanishing is a change in the numerical prefactor. However, the power-law dependences in Eqs.~(\ref{sse5}) and (\ref{s1d}) remain intact.

\begin{figure}[h]
\includegraphics[scale=0.6]{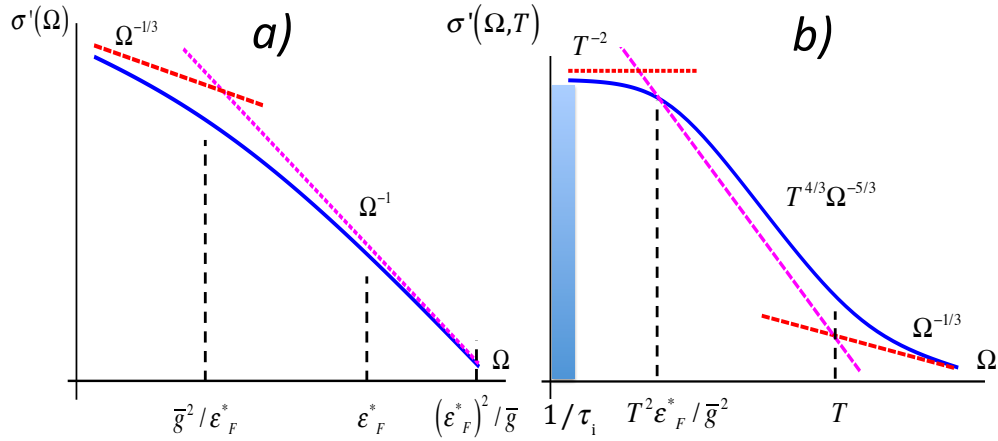}
\vspace{-2.in}
\caption{Schematic:
real part of
the conductivity (on a double-log scale) of a 2D metal at a spin-density-wave critical point. a) $T=0$. b)
 same at finite $T$ and
in the hydrodynamic regime, where (momentum-conserving) composite scattering is stronger than (momentum-relaxing) umklapp or impurity scattering. The temperature is finite but sufficiently low:  $T\ll \bar g^2/\e_F^*$. The high-frequency range ($\Omega\gg \bar g^2/\e^*_F$) is not shown here.
$1/\ti
$ is the momentum-relaxation rate. The shaded region corresponds to frequencies $\lesssim\gamma$, where momentum-relaxing scattering needs to be considered explicitly. }
\label{fig:13to1}
\end{figure}

These
hand-waving
 arguments are confirmed by considering a full set of the diagrams for the conductivity--which is the same set as in Fig.~\ref{fig:5diag} but with the wavy lines replaced by the composite vertices.\cite{chubukov:2014} The boundaries of the 2D and 1D regimes are identified by comparing the regions of validity of the approximations made in the process of deriving the corresponding formulas. The final result for the optical conductivity reads
\bea
\sigma'(\Omega)\sim \sigma_0\times\left\{
\begin{array}{l}
\ls(\e_F^*)^2/\bg\Omega\rs^{1/3},~~\mathrm{for}\;\Omega\ll\bg^2/\e_F^*;
\\
\bg/\Omega,\quad\quad\quad\quad\quad\mathrm{for}\; \bg^2/\e^*_F\ll \Omega\ll \e_F^*.
\label{sigma_Sigma}
\end{array}
\right.
\eea
 The first (second) equation corresponds to the 2D (1D) regime.
[Formally, the upper limit of 1D scaling is $(\e_F^*)^2/\bar g\gg \e_F^*$ but we replaced it by $\e_F^*$ as the model
 considered here cannot be trusted at energies above $\e_F^*$.] In addition to $\Omega^{-1/3}$ scaling at  lower frequencies, the conductivity also exhibits $1/\Omega$ scaling in a parametrically wide interval at higher frequencies (cf. Fig.~\ref{fig:13to1}{\em a}). The latter is reminiscent of scaling observed in the cuprates. \cite{azrak:1994,marel:2003,basov:1996,puchkov:1996,basov:2005,norman:2006}

The final result [Eq.~(\ref{sigma_Sigma})] should be taken with a number of warnings. First, the exponents $1/3$ and $1$
are evaluated at
  two-loop
  order in composite scattering.
   Contributions from higher-loop
   orders are likely to change these
   values.
   \cite{hartnoll:2011}
    One can show that higher-loop terms  are on the same order as the one-loop result in the 2D regime and are larger by a $\ln\Omega$ factor in the 1D regime.~\cite{chubukov:2014} At best, one can hope to have two regions of scaling with exponents smaller than one (2D regime) and closer to one (1D regime). In addition, the low-frequency part of the 2D regime may be expected to be cut by either charge-density-wave or superconducting instabilities arising in the same model.
    Finally, our analysis was based on the assumption of $\bar g$ being smaller than $\e_F^*$, while in reality these two energies are on the same order, and thus one can only hope
that a fortunate game of numbers  will separate the low- and high-frequency regimes in the conductivity. Nevertheless, it is still encouraging to have a microscopic model that, with all the limitations described above, predicts a NFL behavior of the optical conductivity.

\begin{figure}[h]
\includegraphics[scale=0.5]{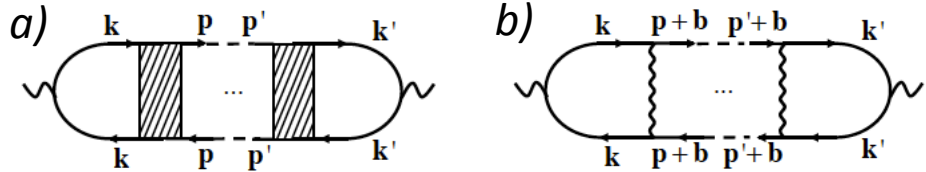}
\vspace{-2.7in}
\caption{Normal ({\em a}) and umklapp ({\em b}) contributions to the conductivity.}
\label{fig:nu}
\end{figure}

\subsubsection{Composite scattering, $T\neq 0$}

A natural question is: can composite scattering  of lukewarm fermions also lead to NFL scaling of the  {\em dc} conductivity with temperature?    Unfortunately, this does not seem to be the case because the effective low-energy theory operating with lukewarm fermions does not explicitly contain umklapp processes.\cite{patel:2014,hartnoll:2014} As we discussed in Sec.~\ref{sec:BE}, the main difference between the optical and {\em dc} conductivities is that the dissipative part of the former is finite even in the absence of umklapp scattering but the latter is finite only in the presence of umklapp scattering. Umklapp scattering (by momentum $\bq_\pi$) is subsumed into the composite vertex (Fig.~\ref{fig:comp}{\em d}), and the total momentum of its initial states is conserved in the same way as for normal scattering (cf. Fig.~\ref{fig:nu}{\em a}). This is to be contrasted with true umklapp scattering (cf. Fig.~\ref{fig:nu}{\em b}), in which the momenta of the initial and final states differ by an integer multiple of  the reciprocal lattice vector.  Therefore, composite scattering on its own cannot render the
{\em dc}
conductivity finite,
 which means that the optical conductivity has a delta-function peak at $\Omega=0$.

Suppose however that the momentum-relaxing process  (be it  umklapp or impurity scattering) is much weaker than the momentum-conserving one, i.e., that we are in the hydrodynamic regime.
 As we
 saw
  in  Sec.~\ref{sec:BE}, for frequencies higher than the rate of
 a
 momentum-relaxing process
  ($1/\ti$), the conductivity
takes a quasi-Drude form [cf. Eq.~(\ref{qD})]: although its imaginary part scales as $1/\Omega$ (as if there is no relaxation), the real part behaves in a Drude-like way with the rate of the momentum-conserving process playing the role of the inverse relaxation time. One can ask then how this quasi-Drude form will look like at finite $T$ in a metal at an SDW instability.

 To be on the realistic side, we assume that $T\ll \bar g^2/\e^*F$. For $T\ll\Omega \ll \bar g^2/\e^*F$, $\sigma'(\Omega)$ scales as $1/\Omega^{1/3}$.
 For $\Omega\ll T$, the frequency dependence of the self-energy is replaced by the temperature dependence. Since we are in the 2D regime, it amounts to replacing the $\Omega^2$ factor in Eq.~(\ref{sigma2D}) by $T^2$.  Because $\I\Sigma$ is finite as $\Omega$ goes to zero at finite $T$,
 one cannot expand the denominator of the Drude formula in $\I\Sigma$. Up to an overall factor,
 the conductivity is now given by
 \bea
 \sigma'(\Omega,T)\propto \int d\kp \frac{\I\Sigma_\bk(T)}{(\Omega/Z_\bk)^2+\left[\I\Sigma_\bk(T)\right]^2},\label{sigma_int}
 \eea
 where $\I\Sigma_\bk(T)\sim \bar g^2\e_F^* T^2/(v_F\kp)^4$.
 Assuming that the relevant $\kp$ in this integral are within the lukewarm region and thus the $Z$-factor is given by Eq.~(\ref{Zlw}), we find that the integral is dominated by such $\kp$ that
 the two terms in the denominator of the equation above are of the same order, i.e., $\kp\sim k_T\sim (T^2\e_F^*\bar g/\Omega)^{1/3}/v_F$, and the conductivity is given by
 \bea
 \sigma'(\Omega,T)\sim\sigma_0\frac{(\e_F^*)^{2/3}}{\bar g^{1/3}}\frac{T^{4/3}}{\Omega^{5/3}}.\label{sigma_T1}
 \eea
 This expression matches with the first line of Eq.~(\ref{sigma_Sigma}) at $\Omega\sim T$, as it should.
 As we see, the conductivity depends both on $T$ and $\Omega$ in a NFL way (for a FL, we would have $\sigma'(\Omega,T)\propto T^2/\Omega^2$ in a similar interval of $\Omega$ and $T$).
 Equation (\ref{sigma_T1})
  is valid at not too low frequencies, such that $k_T$ is still smaller
 than the upper boundary of the lukewarm region, $k_2$ in Eq.~(\ref{k12}). At lower frequencies, the integral over $\kp$ in Eq.~(\ref{sigma_int}) is controlled by the cold region, where the $Z$-factor is close to one, while $\I\Sigma\propto T^2$. This gives a conventional, FL form of the conductivity, $\sigma(\Omega,T)\propto T^{-2}$. The crossover between this form and the one in Eq.~(\ref{sigma_T1}) occurs at $\omega\sim T^2\e^*_F/\bar g^2\ll T$. The behavior of $\sigma'(\Omega,T)$ is sketched in Fig.~\ref{fig:13to1} {\em b}. We remind the reader that the analysis above is valid only at frequencies above the
 momentum-relaxation rate ($1/\ti$);
correspondingly, the
 region $\Omega\lesssim 1/\ti$ is masked by a box in the sketch.

 \section{Conclusions}
 In this review, we discussed three
 particular
 aspects
 of
 optical response of
 correlated electron systems.
The
first  one
is the
role of momentum relaxation,
 the
second
one
 is
 $\Omega/T$ scaling of the optical conductivity of a Fermi-liquid metal,
 and the third
 one is
 the optical conductivity of a non-Fermi liquid metal.
 We
 argued
  that,
   in each of these
   three
   aspects, optical response is different
   from
   and
   complementary to
    other probes,
    such as photoemission and {\em dc} transport.
Accordingly, this review is divided into three parts addressing each of the three aspects mentioned above.

In the first part,
 we analyzed
the interplay between the contributions to the conductivity
   from normal and umklapp {\em ee}
    scattering,
    both
     at finite frequency and near the {\em dc} limit.
    We
    discussed
    the
    similarities and differences between
    the
    optical
     and {\em dc}
    conductivities,
    and
    demonstrated that, unlike the {\em dc} conductivity, the optical one has a finite dissipative part in non-Galilean--invariant systems, even if only normal scattering is present.
     As a specific example of a non-Galilean--invariant system,  we re-visited a two-band model
     with momentum-conserving inter-band scattering and momentum-relaxing intra-band scattering.
     A useful lesson from this model is that although its optical conductivity does have a finite dissipative part, it does not obey the Drude form, because the scattering rates of momentum-conserving and momentum-relaxing processes do not add up according to the Matthiessen rule.
    We also discussed how  the Fermi surface geometry affects the behavior of the optical and {\em dc} conductivities.  In particular,
    we reviewed the theoretical predictions that,
    for any convex
    and
    simply-connected
     Fermi surface in 2D,
     the effective scattering rate scales as $\max\{T^4,\Omega^4\}$
    rather than $\max\{T^2,\Omega^2\}$, as it is to be expected for a Fermi liquid.

In the second part, we re-visited
 the Gurzhi formula for
 the
 optical
 conductivity
of a Fermi-liquid
metal,
  $\R\sigma^{-1}(\Omega, T) \propto  \Omega^2+4\pi^2 T^2$, and
showed
that
 a factor of $4\pi^2$
  in front   of
 the
 $T^2$
 term is a manifestation of
 the
 ``first-Matsubara-frequency rule'' for boson response,
  which states
  that
  a combination of
  the
  $T^2$ and $\Omega^2$ terms must vanish
  upon analytic continuation to the first
  boson
   Matsubara frequency,
         $
         \Omega \to \pm
         2
         \pi iT$. We discussed the origin  and the accuracy of
         this rule
   for the single-particle self-energy
   and
          conductivity,
          both for Fermi and non-Fermi liquids.  We then  discussed recent experiments in several materials, which
          showed
          that,
          although the
          conductivity can be fitted to the Gurzhi-like form, $\R\sigma^{-1} (\Omega, T) \propto  \Omega^2+b\pi^2 T^2$,
          the coefficient $b$
           happens to
           deviate from the theoretical value of $4$
            in all cases, except for Sr$_2$RuO$_4$.
  \cite{stricker:2014}
           The discrepancy is especially pronounced in
rare-earth Mott insulators and heavy-fermion materials, where $b$ is in general smaller than $2$
 and
 remarkably close to $1$ in some cases, e.g.,
 in URu$_2$Si$_2$.
 \cite{nagel:2012}
          We
          proposed
           that the deviations from Gurzhi scaling may be due to the presence of
           elastic scattering, which decreases the value of $
           b$
           below
           $4$,
           with $
           b=1$
           corresponding to the limit
           where elastic scattering dominates over inelastic {\em ee}
           one.

          In the third part,
          we considered
         the optical
          conductivity
          of a metal
           near quantum
           phase transitions to nematic
           and
           spin-density-wave
           states with
           nesting
           momentum $(\pi,\pi)$.
           In the last case,
           we reviewed the special role of a
           ``composite'' scattering process, which consists
           of
            two
           consequent
           $(\pi,\pi)$ scatterings.  We
           demonstrated  that this effectively small-momentum scattering gives rise to
           a
           non-Fermi--liquid behavior of the
           optical
            conductivity at the critical point
            and
            at $T=0$.
          We reviewed the results of recent papers,\cite{hartnoll:2011,chubukov:2014} which predict
          that the dissipative part of the conductivity,
      $\sigma' (\Omega)$,
      scales as   $\Omega^{-1/3}$ at asymptotically low frequencies  and as
   $\Omega^{-1}$  at higher frequencies, up to the bandwidth.
  The  $1/\Omega$ scaling of $\R\sigma (\Omega)$  is consistent with the behavior
  observed in the superconducting cuprates.
   We  also argued that composite scattering alone cannot render the {\em dc} conductivity finite--to do so, one needs to invoke
  some momentum-relaxing process (with scattering rate
  $1/\ti$).
 Nevertheless, if $
 1/\ti$ is the slowest
 scattering
 rate in the problem,
  one can discuss $\Omega/T$ scaling of $\sigma'(\Omega,T)$ at finite $T$ and $\Omega\gg
  1/\ti$.
 Within this approximation,  we showed that  $\sigma'(\Omega,T)$  is of
       the Fermi-liquid form, $\sigma'(\Omega,T)\propto T^{-2}$, below some $T$-dependent frequency,
       but scales in a non-Fermi-liquid way,    as $
       T^{4/3}\Omega^{-5/3}$, above that frequency.

\acknowledgements
 We acknowledge stimulating  discussions with
 I. L. Aleiner,
 B. L. Altshuler,
D. Basov,
C. Batista,
K. Behnia,
M. Broun,
G. Blumberg,
D. Dessau,
P. Coleman,
 S. Dodge,
 M. Dressel,
 B. Fauqu{\`e},
 A. Georges,
 L. P. Gor'kov,
 I. V. Gornyi,
 D. Gryaznov,
 K. Haule,
 S. Hartnoll,
K. Ingersent,
 M. Kennett,
Y.-B. Kim,
 G. Kotliar,
 E. A. Kotomin,
P. Kumar,
I. V. Lerner,
 S. Maiti,
 D. van der Marel,
 M. Metlitski,
A. Millis,
  U. Nagel,
  H. Pal,
  I. Paul,
  M. Yu. Reizer,
  T. R{\~o}{\~o}m,
  S. Sachdev,
  M. Scheffler,
  B. I. Shklovskii,
 D. Tanner,
 T. Timusk,
 A.-M. Tremblay,
 D. Vanderbilt,
 and V. I. Yudson.
  This work was supported by the National Science Foundation via grant NSF DMR-1308972 (D.L.M.)
   and NSF-DMR 1523036 (A.V.C.).
  We acknowledge
  hospitality of the Center for Non-Linear Studies, Los Alamos National Laboratory,
  which  both of us visited in
 2015-2016 as Ulam Scholars.

 \appendix
 \section{Diagrams for the optical conductivity}
 \label{app:diags}
 In this Appendix, we show that the diagrams for the optical conductivity do not cancel for a system with broken Galilean invariance, even in the absence of umklapp processes.
 We select five diagrams shown in Fig.~\ref{fig:5diag}. Such a choice can be justified, e.g., within the large-$N$ limit, when each the five diagrams contains a factor of $N^2$ while the diagrams not included in this group are smaller by a factor of $N$. The interaction, $ U_{\bq}$,  is static but otherwise an arbitrary function of the momentum transfer $\bq$.

  First, we consider the self-energy diagrams {\em a} and {\em b}, whose combined contribution to the current-current correlation function reads
 \bea
 {\cal K}_{ab}=-\sum_{P}\bv^2_\bp\left(G_P^2 G_{P+Q} \Sigma_P+G^2_{P+Q} G_P \Sigma_{P+Q}\right).
 \eea
As in the main text, $P=(\bp,p_0)$, $Q=({\bf 0},q_0)$, etc., $\sum_P$ stands for $T\sum_{p_0} \int d^{D}p/(2\pi)^D$, and $\bv_\bp=\boldsymbol{\nabla}\e_\bp$ is the group velocity of Bloch electrons. Notice that $Q$ has only a temporal component. With the help of an identity
\beq
\label{ident}
G(P)G(P+Q)=\frac{1}{iq_0}\left(G_P-G_{P+Q}\right),
\eeq ${\cal K}_{ab}$ can be re-written as
\bea
 {\cal K}_{ab}=\frac{1}{q_0^2}\sum_{P}\bv^2_\bp\left(G_P-G_{P+Q}\right)\left(\Sigma_{P+Q}-\Sigma_P\right).
\eea
Using an explicit form of the self-energy $\Sigma(P)=-\sum_LU^2_{\bl} G_{P+L}\Pi_{L}$, where  $\Pi_L=\sum_{K}G_KG_{K-L}$ is the particle-hole polarization bubble, we re-write ${\cal K}_{ab}$ as
\bea
 {\cal K}_{ab}=\frac{1}{q_0^2}\sum_{P,K,L}\bv^2_\bp U_{\bl}^2\left(G_P- G_{P+Q}\right)\left(G_{P+L}-G_{P+L+Q}\right)G_K G_{K-L}.
 \label{kab}
\eea
Applying idenitity (\ref{ident}) to vertex diagram {\em c} and adding up the result with Eq.~(\ref{kab}), we obtain for the combined contribution of diagrams {\em a-c}
\bea
{\cal K}_{ab}+{\cal K}_c&=&\frac{1}{q_0^2}\sum_{P,K,L} \bv_\bp\cdot\left(\bv_{\bp}-\bv_{\bp+\bl}\right)U_{\bl}^2\left(G_P-G_{P+Q}\right)\left(G_{P+L}-G_{P+L+Q}\right)G_K G_{K-L}\nn\\
&=&\frac{1}{q_0^2}\sum_{P,K,L} \bv_\bp\cdot\left(\bv_{\bp}-\bv_{\bp+\bl}\right)U_{\bl}^2\left(2G_PG_{P+L}-G_PG_{P+L+Q}-G_{P+Q} G_{P+L}\right)G_KG_{K-L},\nn\\
\label{kabc}
\eea
where we relabeled $P+Q\to P$ in the term $G_{P+Q}G_{P+L+Q}$. Since $Q$ has only a temporal component, this transformation affects neither the velocities
nor the interaction.

Now we turn to the Aslamazov-Larkin diagrams, {\em d} and {\em e}. Applying identity (\ref{ident}) twice, we obtain for diagram {\em d}
\bea
{\cal K}_{d}&=&-\frac{1}{q_0^2}\sum_{ P,K,L}\bv_{\bp}\cdot\bv_{\bk-\bl} U_{\bl}^2\left(G_P-G_{P+Q}\right)\left(G_{K-L}-G_{K-L+Q}\right)G_KG_{P+L}\nn\\
&=&-\frac{1}{q_0^2}\sum_{ P,K,L}\bv_{\bp}\cdot\bv_{\bk-\bl} U_{\bl}^2\left(G_PG_{K-L}+G_{P+Q}G_{K-L+Q}-G_PG_{K-L+Q}-G_{P+Q}G_{K-L}\right)G_KG_{P+L}.\nn\\
\eea
Now, we relabel $P+Q\to P$, $L-Q\to L$ in the second term and $L-Q\to L$ in the third one.  This yields the same combination of the Green's functions as in Eq.~(\ref{kabc}):
\bea
{\cal K}_{d}&=&-\frac{1}{q_0^2}\sum_{ P,K,L}\bv_{\bp}\cdot\bv_{\bk-\bl} U_{\bl}^2\left(2G_PG_{P+L}-G_PG_{P+L+Q}-G_{P+Q} G_{P+L}\right)G_KG_{K-L}.
\label{kd}
\eea
Similarly, diagram {\em e} reads
\bea
{\cal K}_{e}&=&-\frac{1}{q_0^2}\sum_{ P,K,L}\bv_{\bp}\cdot\bv_{\bk} U_{\bl}^2\left(G_P-G_{P+Q}\right)\left(G_{K}-G_{K+Q}\right)G_{K-L}G_{P+L+Q}\nn\\
&=&-\frac{1}{q_0^2}\sum_{ P,K,L}\bv_{\bp}\cdot\bv_{\bk} U_{\bl}^2\left(G_PG_K+G_{P+Q}G_{K+Q}-G_PG_{K+Q}-G_{P+Q}G_K\right)G_{K-L}G_{P+L+Q}\nn\\
\eea
Relabeling $K+Q\to K$ and $L+Q\to L$ in the second and third terms, and $P+Q\to P$ in the fourth one, we obtain the same combination of the Green's functions as in Eqs.~(\ref{kabc})
and Eq.~(\ref{kd}) but with an opposite sign:
 \bea
{\cal K}_{e}&=&-\frac{1}{q_0^2}\sum_{ P,K,L}\bv_{\bp}\cdot\bv_{\bk} U_{\bl}^2\left(G_PG_{P+L+Q}+G_{P+Q} G_{P+L}-2G_PG_{P+L}\right)G_KG_{K-L}.
\eea
Collecting all contributions, we obtain the final result for the current-current correlation function
\bea
{\cal K}&=&{\cal K}_{ab}+{\cal K}_c+{\cal K}_d+{\cal K}_e=\frac{1}{q_0^2}\sum_{P,K,L} \bv_\bp\cdot\left(\bv_{\bp}+\bv_\bk-\bv_{\bp+\bl}-\bv_{\bk-\bl}\right)U_{\bl}^2\nn\\
&&\times\left(2G_PG_{P+L}-G_PG_{P+L+Q}-G_{P+Q} G_{P+L}\right)G_KG_{K-L}.
\eea
The combination of the velocities in the equation above vanishes in a Galilean-invariant system, where $\bv_\bk=\bk/m$ but, in general, is non-zero otherwise.

\section{Accuracy of the first-Matsubara-frequency rule}
\label{sec:remainder}
The first-Matsubara-frequency rule is not exact. In this appendix, we discuss the conditions under which the remainder $R(T)$ in Eq.~(\ref{rule1}) is indeed subleading to the first term.

\subsection{Fermi liquid}
\label{sec:RFL}
Equation  (\ref{sum}) was derived under the assumption that momentum transfers along the normal to the FS are small, of order $T/v_F$, whereas the momentum transfers tangential to the FS are determined by the internal scale of the interaction, $\Lambda$. It is the breakdown of this (local) approximation that gives rise to corrections to first Matsubara rule. In what follows, we assume that the linear-in-$T$ term in  the Matsubara self-energy is already singled out. The next term is  obtained by retaining only one dynamic polarization bubble in $\chi(\bq,\Omega)$, which scales as $|\Omega_n|/q_{||}$. The lower limit in the integral over $q_{||}$ is set  by the condition $q_{||}\gg q_{\perp}\sim T/v_F$. For an estimate, we will replace $T\sum_{\Omega_n}\Omega_n$ by $T^2$. Then
\beq
\Sigma(\omega_m,T)-i\lambda \text{sgn}\omega_mT\sim T\sum_{\Omega_n}\Omega_n\int_{T/v_F}^\Lambda dq^{\phantom{D}}_{||} q^{D-3}_{||}\sim T^2\left[\Lambda^{D-2}-(T/v_F)^{D-2}\right].\label{r1}
\eeq
For $D>2$, the $T$-dependent contribution from the lower limit is subleading, and the local approximation indeed works. The next order term contains a square of the dynamic bubble, which yield a correction to Eq.~(\ref{r1}) on the order of
\beq
T\sum_{\Omega_n}\Omega^2\int_{T/v_F}^\Lambda dq^{\phantom{D}}_{||} q^{D-4}_{||}\sim T^3\left(\Lambda^{D-3}-T^{D-3}\right).\label{r2}
\eeq
For $D=3$, the integral diverges logarithmically at the lower limit. This is a well-studied $E^3\ln E$ term ($E=\max\{\omega_m,T\}$) in the self-energy of a 3D FL, which gives rise to a non-analytic, $T^3\ln T$ correction to the specific heat\cite{eliashberg:1960,eliashberg:1963,pethick:1973,
[{For a more complete list of references on the non-analytic behavior of the specific heat, see:} ]chubukov:2006}
observed both in $^3$He (Refs.~\onlinecite{abel:1966,greywall:1983}) and heavy-fermion materials.\cite{stewart:1984} The correction in Eq.~(\ref{r2}) is not nullified at the first Matsubara frequency, and its $T$-dependent part gives an estimate for the remainder
\beq
R={\mathcal O}(T^D).
\label{R}
\eeq

 For $D=2$, the integral Eq.~(\ref{r1}) diverges logarithmically at the lower limit. This gives a familiar $E^2\ln E$ form of the self-energy in  2D FL. Still, this term is nullified at $\omega_m=\pi T$ and the surviving term is of order  $T^2\ll T^2\ln T$. An exact result for the surviving term is\cite{chubukov:2012}
\beq
\Sigma(\pi T,T)-i\pi \lambda T=\frac{AT^2}{2\pi v_F^2}\left(K+\frac{\pi\ln 2}{4}\right),
\eeq
where $A$ is a coupling constant, which is expressed via the charge and spin components of the forward- and backscattering amplitudes, and $K=0.9160$ is the Catalan constant. To summarize, Eq.~(\ref{R}) works for any $D\geq 2$.

\subsection{First-Matsubara rule in non-Fermi liquids: Hertz-Millis-Moriya criticality}
\label{sec:NFL}

FMFR holds only for FLs but also for (NFLs, in a sense that the $\Omega_n=0$ term in the sum (\ref{sigmaM}) gives, under certain conditions, the leading contribution to the result. However, in contrast to Eq.~(\ref{rule1}), the leading term does not scale linearly with $T$ because the coefficient $\lambda$ depends on $T$ itself. This is related to the fact that the effective mass of a NFL depends on the frequency. In addition, the estimate for the remainder term, Eq.~(\ref{R}), changes.

Below, we demonstrate how FMFR  works for a NFL using  a Hertz-Millis-Moriya quantum critical point\cite{hertz:1976,millis:1993,moriya:1985} as a concrete example.
We consider a generic Hertz-Millis-Moriya model with the propagator of critical fluctuations described by
\beq
\chi(q,\Omega_n)=\frac{1}{q^2+\xi^{-2}(T)+\frac{\gamma|\Omega_n|}{q^{z-2}}},\label{chi}
\eeq
where $\xi(T)$ is the correlation length and $z$ is the dynamical scaling exponent. Here, $q$ is measured from the center of the Brillouin zone for $z=3$ (Pomeranchuk transition), and $q=|\bq-\bq_n|$ for $z=2$, where $\bq_n$ is the nesting wavevector of a  spin- or charge-density-wave. Right at the critical point, $\xi$ is finite but temperature-dependent, and it diverges at $T\to 0$. We assume that $\xi\propto T^{-\beta}$. At the tree level, $\beta=1/2$ modulo logarithmic renormalizations.\cite{millis:1993} The derivation for the leading term in FMFR  proceeds in the same way as in Sec.~\ref{sec:1stM}, and we arrive again at Eq.~(\ref{sum1}). However,  $\chi_{\text{loc}}(0)$ in this equation depends now on $T$ via $\xi(T)$:
\beq
\chi_{\text{loc}}(0)\propto \int dq^{\phantom{D}}_{||} q_{||}^{D-2}\frac{1}{q_{||}^2+\xi^{-2}(T)}\propto \xi^{3-D}\propto T^{-\beta(3-D)}.
\eeq
In the local approximation,  the sum in Eq.~(\ref{sum}) vanishes at $\omega=\pm\pi T$, and we obtain instead of Eq.~(\ref{rule1})
\beq
\Sigma(\pm \pi T,T)=\mp i\; \text{const}\times T^{1-\beta(3-D)}+\tilde R(T).
\label{HM}
\eeq
Physically, the modification of the leading term is to due to scattering from static fluctuations of the order parameter. Naturally, the leading term--being entirely static--does not depend on the dynamical exponent $z$. An immediate consequence of Eq.~(\ref{HM}) is that the dHvA amplitude deviates from the Lifshitz-Kosevich form of  Eq.~(\ref{LK}) and is now given by
 \beq
A(T)=\frac{4\pi^2T}{\omega_c}\exp\left(-2\pi^2\frac{T+\text{const} \times T^{1-\beta(3-D)}}{\omega_c}\right).\label{LK1}
\eeq
A non-Lifshitz-Kosevich behavior of this type was observed near a magnetic-field--driven quantum critical point  in CeCoIn$_5$.\cite{mccollam:2005}

As in a FL, the remainder term in Eq.~(\ref{HM}), $\tilde R(T)$, comes from the corrections arising from keeping finite $q_{\perp}$ in the boson propagator.  Since we already singled out the static term, we can put $\xi^{-1}=0$ in Eq.~(\ref{chi}), upon which it is reduced to a scaling form
\beq
\chi(q,\Omega_n)=|\Omega_n|^{-2/z}f\left(\frac{q}{|\Omega_n|^{1/z}}\right)\label{deltachi}
\eeq
with $f(x)=x^{z-2}/(x^z+1)$. Now we expand $q=\sqrt{q_{\perp}^2+q_{||}^2}$
to first order  in $q^2_{\perp}$ and obtain a correction to the propagator
\beq
\delta\chi(q,\omega)=\frac{q_{\perp}^2}{2 q_{||} |\Omega_n|^{3/z}}f'\left(\frac{q_{||}}{|\Omega_n|^{1/z}}\right).\label{delta2}
\eeq
The Green's function is given by
$G=\left[i\omega_m+\Sigma(\omega_m,T)-v_Fq_{\perp}\right]^{-1}$. Since we are in a NFL regime, $\Sigma\gg\omega_m$ and thus typical $q_{\perp}\sim |\Sigma|/v_F$. Replacing $\Omega_n$ by $T$ in Eq.~(\ref{delta2}) and evaluating the corresponding correction to the self-energy, we obtain an estimate for $\tilde R(T)$
\beq
\tilde R(T)\propto T^{1-3/z}|\Sigma|^2\int dq^{\phantom{D}}_\perp q^{D-3}_\perp f'\left(\frac{q_{||}}{T^{1/z}}\right)\sim |\Sigma|^2 T^{(z+D-5)/z},
\label{R2}
\eeq
which itself depends on the self-energy.
The self-energy at $\omega_m\sim T$ but $\omega_m\neq \pi T$ contains two contributions: the static one, $\Sigma_s\propto T^{1-\beta(3-D)}$, and the dynamical one, $\Sigma_d\propto T^{(z+D-3)/z}$. If $\beta>1/z$, $\Sigma_s\gg \Sigma_d$ and $\Sigma$ in the equation above needs to be replaced by $\Sigma_s$ and vice versa
 for $\beta<1/z$.
For $\beta>1/z$,  the ratio of the remainder to the first (static) term in Eq.~(\ref{HM}) scales as
\beq
\tilde R(T)/\Sigma_s\propto \Sigma_sT^{(z+D-5)/z}\propto T^{(2z+D-5)/z-\beta(3-D)}.
\eeq
Limiting our consideration to $D=2,3$ and $z=2,3$ cases, we notice that the only situation when the ratio in the equation above does not necessarily go to zero at $T\to 0$ is $D=z=2$ (a spin/charge density-wave phase transition in 2D). In this case, the right-hand-side scales as $T^{1/2-\beta}$ and therefore diverges if $\beta>1/2$. For $\beta<1/z$,  $\Sigma$ in Eq.~(\ref{R2}) is to be replaced by $\Sigma_s$, which gives for the ratio
\beq
\tilde R(T)/\Sigma_s\propto \Sigma^2_dT^{(z+D-5)/z}/\Sigma_s\propto T^{(2z+3D-11)/z+\beta(3-D)}.
\eeq
Again, the only case when the ratio does not vanish at $T\to 0$, is $D=z=2$.

\section{First-Matsubara-frequency rule for vertex corrections to conductivity}
\label{sec:vertex_sigma}

In this Appendix, we demonstrate how FMFR  for the self-energy transforms into an analogous rule for the vertex corrections to the conductivity. This will be done by employing the Ward-type relations between the partial self-energy
[as defined by Eq.~(\ref{part2}) and (\ref{part3})] and current vertex.
As an example, we will consider a particular self-energy diagram (diagram {\em a} in Fig.~\ref{fig:vertex}):
\bea
\Sigma_{K}=\sum_{K',K''}G_{K'}G_{K''}G_{K-K'+K''}U_{\bk-\bk'}U_{\bk'-\bk''}.
\eea
{\begin{figure}[h]
\includegraphics[width=0.6 \linewidth]{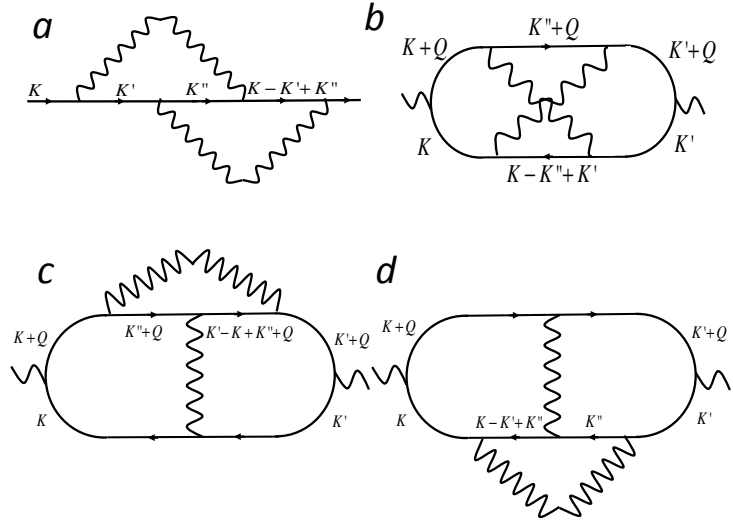}
\caption{An example of the self-energy diagram ({\em a}) and the conductivity diagrams ({\em b-d}), which it generates.\label{fig:vertex}}
\end{figure}
We will actually need the difference between the self-energy in the equation above and the
one in which the frequency of the incoming fermion is shifted by the frequency of the external field:
\bea
\Sigma_{K+Q}=\sum_{K',K''}G_{K'+Q}G_{K''+Q}G_{K-K'+K''+Q}U_{\bk-\bk'}U_{\bk'-\bk''}.
\eea
 In the notations of Appendix \ref{app:diags}, \beq
\Delta\Sigma_{K,Q}\equiv \Sigma_K-\Sigma_{K+Q}
\eeq
with $Q=({\bf 0},q_0)$.
 Following Refs.~\onlinecite{goetze:1972} and \onlinecite{maebashi:1997}, we use an identity
\beq
x_1x_2x_3-y_1y_2y_3=y_1x_3(x_2-y_2)+y_1y_2(x_3-y_3)+x_2x_3(x_1-y_1)
\eeq
to re-write  $\Delta\Sigma_{K,Q}$ as a sum of three terms:
\bea
\Delta\Sigma_{K,Q}=\Delta\Sigma^{(b)}_{K,Q}+\Delta\Sigma^{(c)}_{K,Q}+\Delta\Sigma^{(d)}_{K,Q},
\eea
where
\bse
\bea
\Delta \Sigma^{(b)}_K&=&\sum_{K',K''} G_{K'+Q}G_{K-K'+K''}\left(G_{K''}-G_{K''+Q}\right)U_{\bk-\bk'}U_{\bk'-\bk''},\label{sb}\\
\Delta \Sigma^{(c)}_K&=&\sum_{K',K''} G_{K'+Q}G_{K''+Q}\left(G_{K-K'+K''}-G_{K-K'+K''+Q}\right)U_{\bk-\bk'}U_{\bk'-\bk''},\label{sc}\\
\Delta \Sigma^{(d)}_K&=&\sum_{K',K''} G_{K''}G_{K-K'+K''}\left(G_{K'}-G_{K'+Q}\right)\label{sd}U_{\bk-\bk'}U_{\bk'-\bk''}.\label{sd}
\eea
\ese
In Eq.~(\ref{sb}), we relabel $K'\leftrightarrow K''$. In (\ref{sc}), we also relabel $K'\leftrightarrow K''$ and then $K'+K-K''\to K'$.
Equation (\ref{sd}) is left as is. Then
\bse
\bea
\Delta \Sigma^{(b)}_{K,Q}&=&\sum_{K',K''} G_{K''+Q}G_{K-K''+K'}\left(G_{K'}-G_{K'+Q}\right)U_{\bk-\bk''}U_{\bk'-\bk''}
=\sum_{\hat\bk'}\Delta{\cal S}^{(b)}_{K,Q;\bk'}\label{sb2},\\
\Delta \Sigma^{(c)}_{K,Q}&=&\sum_{K',K''} G_{K''+Q}G_{K-K''+K'+Q}\left(G_{K'}-G_{K'+Q}\right)U_{\bk-\bk'}U_{\bk'-\bk''}=\sum_{\hat\bk'}\Delta{\cal S}^{(c)}_{K,Q;\hat\bk'},\label{sc2}\\
\Delta \Sigma^{(d)}_{K,Q}&=&\sum_{K',K''} G_{K''}G_{K-K'+K''}\left(G_{K'}-G_{K'+Q}\right)\label{sd3}U_{\bk-\bk'}U_{\bk'-\bk''}=\sum_{\hat\bk'}\Delta{\cal S}^{(d)}_{K,Q;\hat\bk'},\label{sd2}
\eea
\ese
where $\Delta S^{j}_{K,Q;\hat\bk'}\equiv S^{j}_{K,\bk'}-S^j_{K+Q,\bk'}$ with $j=b\dots d$ is the difference of the corresponding partial self-energies, defined by Eqs.~(\ref{part2}) and (\ref{part3}), and $\sum_{\hat\bk'}$ is a shorthand for an integral over the FS:  $\sum_{\hat\bk'}\equiv (2\pi)^{-D} \int da_{\bk'}/v_{\bk'}$. The sum $\sum_jS^{j}_{K,\hat\bk'}$
gives the partial self-energy in diagram {\em a}.

Now we observe that the same combinations of the Green's functions and interactions appear in vertex diagrams  {\em b-d}, Fig.~\ref{fig:vertex}.  Indeed, applying identity (\ref{ident}) to the two Green's functions adjacent to the right current vertex, we obtain for the sum of diagrams {\em b-d} (as in the main text, we assume a cubic lattice in the $D$-dimensional space)
\bea
{\cal K}(Q)=-\frac{1}{iq_0 D}\sum_{j=b\dots d}\sum_{K}G_KG_{K+Q} \bv_\bk\cdot {\bf L}_{K,Q}^{(j)},\label{K}
\eea
where
\bse
\bea
{\bf L}_{K,Q}^{(b)}&=&\sum_{K',K''} \bv_{\bk'}G_{K''+Q}G_{K-K''+K'}\left(G_{K'}-G_{K'+Q}\right) U_{\bk-\bk''}U_{\bk'-\bk''},\label{lb}\\
{\bf L}_{K,Q}^{(c)}&=&\sum_{K',K''}\bv_{\bk'}G_{K''+Q}G_{K-K''+K'+Q} \left(G_{K'}-G_{K'+Q}\right) U_{\bk-\bk'}U_{\bk'-\bk''},\label{lc}\\
{\bf L}_{K,Q}^{(d)}&=&\sum_{K',K''}\bv_{\bk'}G_{K''}G_{K-K'+K''}\ \left(G_{K'}-G_{K'+Q}\right) U_{\bk-\bk'}U_{\bk'-\bk''}\label{ld}
\eea
are the renormalized current vertices.
\ese
Comparing Eqs.~(\ref{sb2}-\ref{sd2}) and (\ref{lb}-\ref{ld}), we obtain the relations between the current vertices and partial self-energies:
\bea
{\bf L}_{K,Q}^{(b-d)}=\sum_{\hat\bk'}\bv_{\bk'}\Delta{\cal S}^{(b-d)}_{K,Q;\hat\bk'}\label{lb2}
\eea
and thus
\bea
{\cal K}(Q)=-\frac{1}{iq_0 D}\sum_{K,{\hat\bk'}}G_KG_{K+Q} \bv_\bk\cdot\bv_{\bk'} \Delta{\cal S}_{K,Q;\hat\bk'}.\label{K3}
\eea
Within the local approximation, the current vertices do not depend on the fermion dispersions. Then the product of two Green's functions in Eq.~(\ref{K3})
can be integrated over $\e_{\bk}$ with the  result
\bea
{\cal K}(Q)=\frac{\pi}{iq_0^2 D} T\sum_{k_0}\left[\text{sgn} (k_0+q_0)-\text{sgn} k_0\right]\sum_{\hat\bk,\hat\bk'}
\bv_\bk\cdot\bv_{\bk'}\left({\cal S}_{K,\bk'}-{\cal S}_{K+Q,\bk'}\right)\label{K2}.
\eea
For $q_0=2\pi T$, the Matsubara sum in the equation above contains only one term ($k_0=-\pi T$), and therefore the first and second partial self-energy are evaluated at $k_0=-\pi T$ and $k_0+q_0=\pi T$, correspondingly. Recalling that $K=(\bk,-\pi T)$ and $K+Q=(\bk,\pi T)$, and that the partial self-energy obeys FMFR  [Eq.~(\ref{part})], we find that ${\cal K}(2\pi T)$ is reduced to a $T$-independent constant:
\bea
{\cal K}(2\pi T)=\frac{1}{2\pi i T D}\sum_{\hat\bk,\hat\bk'}\bv_\bk\cdot\bv_{\bk'}\left( {\cal S}_{(\bk,-\pi T),\bk'}-{\cal S}_{(\bk,\pi T),\bk'}\right)
=-\frac{1}{D} \sum_{\hat\bk,\hat\bk'}\bv_\bk\cdot\bv_{\bk'}\mu_{\bk,\bk'}.
\eea
The corresponding Matsubara conductivity $\sigma(2\pi T,T)=-e^2K(2\pi T)/2\pi T$ scales as $1/T$, in agreement with Eq.~(\ref{rule1_s}).

The same procedure can be extended to other vertex-correction diagrams. Namely, one can always find a correspondence between a particular self-energy diagram and a set of vertex diagrams for the conductivity which it generates. FMRF for the conductivity then follows from the analogous rule for the (partial) self-energy.

\bibliography{dm_references}

\end{document}